%% file: lmcs.tex
\documentclass{LMCS}

\usepackage{amsmath}
\usepackage{amssymb}
\usepackage{amsthm}
\usepackage{mathdots}
\usepackage{misc}
\usepackage{latexsym}
\usepackage{color}
\usepackage{bbm}
\usepackage{pifont}
\usepackage{epsfig}
\usepackage{xspace}           
\usepackage{boxedminipage}    
\usepackage{hyperref}
\theoremstyle{plain}

\def\doi{2 (2:5) 2006}
\lmcsheading%
{\doi}
{41}
{}
{}
{Nov.~24, 2005}
{Jun.~22, 2006}
{}

\begin{document}

\title[Modal Logics of Topological Relations]{Modal Logics of Topological Relations}

\author[C.~Lutz]{Carsten Lutz\rsuper a}	
\address{{\lsuper a}Institute of Theoretical Computer Science 
  TU Dresden, Germany} 	
\email{lutz@tcs.inf.tu-dresden.de}  

\author[F.~Wolter]{Frank Wolter\rsuper b}	
\address{{\lsuper b}Department of Computer Science,
University of Liverpool,
United Kingdom}
\email{frank@csc.liv.ac.uk}  



\keywords{Spatial reasoning, topology, modal logic, RCC8, expressive completeness, decidability, axiomatizability}
\subjclass{F4.1, H2.8, I2.4}


\begin{abstract}
  \noindent Logical formalisms for reasoning about relations between spatial
  regions play a fundamental role in geographical information systems,
  spatial and constraint databases, and spatial reasoning in AI. In
  analogy with Halpern and Shoham's modal logic of time intervals
  based on the Allen relations, we introduce a family of modal logics
  equipped with eight modal operators that are interpreted by the
  Egenhofer-Franzosa (or \mn{RCC8}) relations between regions in
  topological spaces such as the real plane.  We investigate the
  expressive power and computational complexity of logics obtained in
  this way. It turns out that our modal logics have the same
  expressive power as the two-variable fragment of first-order logic,
  but are exponentially less succinct. The complexity ranges from
  (undecidable and) recursively enumerable to $\Pi_{1}^{1}$-hard,
  where the recursively enumerable logics are obtained by considering
  substructures of structures induced by topological spaces.  As our
  undecidability results also capture logics based on the real line,
  they improve upon undecidability results for interval temporal
  logics by Halpern and Shoham.  We also analyze modal logics based on
  the five \mn{RCC5} relations, with similar results regarding the
  expressive power, but weaker results regarding the complexity.
\end{abstract}

\maketitle

%
%
%
%
%
%
\section{Introduction}

Reasoning about topological relations between regions in space is
recognized as one of the most important and challenging research areas
within spatial reasoning in artificial intelligence (AI) and
philosophy, spatial and constraint databases, and geographical
information systems (GISs).  Research in this area can be classified
according to the logical apparatus
employed: 

\smallskip
\noindent
-- First-order theories of topological relations between regions, as
studied in AI and philosophy
\cite{Clarke85,Randell92rcc,PrattSchoop98,CohnHazarika}, spatial databases
\cite{Papa96,Schaefer01} and from an algebraic viewpoint in
\cite{dwm_rcc,StellAI2000,dw_rep1};

\smallskip
\noindent
-- Purely existential theories formulated as constraint satisfaction systems
over jointly exhaustive and mutually disjoint sets of topological relations
between regions \cite{Egenhofer94,RenzNebel,Papa95,Schaefer01,Randell92rcc,Bennett94a,CohnHazarika}

\smallskip
\noindent
-- Modal logics of space with operators interpreted by the closure and
interior operator of the underlying topological space and propositions
interpreted as subsets of the topological space, see e.g.,
\cite{McKinseyTarski,Bennett96b,AilloBenthem,Nutt-99,Pratt-Hartmann-02}.

\smallskip
\noindent
A similar classification can be made for temporal reasoning: we have general
first-order theories~\cite{Allen-84}, temporal constraint
systems \cite{Allen-83,Vilain-et-al-90,NebelBu} and modal temporal
logics like Prior's tense logics, LTL, and CTL
\cite{Gabbay-et-al-94,Emerson-90}.
Surprisingly, one of the most natural 
approaches to temporal reasoning has not yet found a fully developed
analogue on the spatial reasoning research agenda: Halpern and
Shoham's modal logic of intervals \cite{HalpernShoham91}, in which
propositions are evaluated at intervals (rather than time points), and
where reference to other intervals is enabled by modal operators
interpreted by Allen's 13 relations between intervals, see also
\cite{vanBenthem83,Galton87}. Despite its bad computational behavior
(undecidable, usually not even r.e.), this framework proved rather
fruitful and influential in temporal reasoning, see e.g.\ 
\cite{Venema-90,Venema92,Artale-Franconi-98,rasmussen99,lodaya00,Lutz-03b}.

In this paper, we consider modal logics in which propositions are
evaluated at the regions of topological spaces, and reference to other
regions is enabled by modal operators interpreted as topological
relations. For defining such logics, the two most important decisions
to be made are choosing an appropriate set of relations and
identifying a suitable notion of a ``region'' in a topological space.

Regarding the relations, in the initially mentioned research areas
there appears to be consensus that the eight Egenhofer-Franzosa (or
$\mn{RCC8}$) relations, which have been independently introduced in
\cite{Randell92rcc} and \cite{EgenhoferFranzosa}, and their coarser
relative $\mn{RCC5}$ consisting of only five relations, are the most
fundamental sets of relations between regions of topological
spaces---both from a theoretical and a practical viewpoint, see
e.g.~\cite{Papa96,Egenhofer94,RenzNebel,Schaefer01,Randell92rcc}.
Therefore, in the current paper we concentrate on these two sets of
relations. We should note that modal logics based on the
Egenhofer-Franzosa relations have been suggested in an early paper by
Cohn \cite{Cohn93a} and further considered in~\cite{Wessel-01}.
However, it proved difficult to analyze the expressive power and
computational behavior of such logics: despite several efforts, to the
best of our knowledge no results have been obtained so far.

Concerning the regions of a topological space, we adopt a rather
relaxed view: we generally assume that regions are non-empty regular
closed subsets of a topological space, but we do not require that
\emph{every} such subset is a region. This view allows us to consider
logical structures, henceforth called \emph{region structures}, that
are based on various kinds of regions. Among others, we consider the
following options:

\smallskip
\noindent
-- Region structures in which the set of regions is  \emph{exactly}
the set of non-empty regular closed subsets of a topological space.

\smallskip
\noindent
-- In the Euclidean space $\Rbbm^{n}$, region structures where regions
are identified with all non-empty convex regular closed sets, or with
all hyper-rectangles.

\smallskip
\noindent
-- Substructures of the above region structures: for example, we may
admit region structures in which only some, but not all
hyper-rectangles of $\Rbbm^n$ are regions. To distinguish this case
from the former two, we call region structures in which all regions of
a particular kind are present \emph{full} region structures.

\smallskip
\noindent
-- Finite substructures of the above region structures. 

\smallskip
\noindent
The rationale behind the latter two choices of structures is that, for
certain applications, it is sufficient to require the presence of only
those regions in region structures that are inhabited by spatial
objects. If it is known that there are only finitely many such
objects, but their exact number is unknown, then finite
substructures are the appropriate choice.

The main purpose of this paper is to \emph{introduce modal logics of
  topological relations in a systematic way, to perform an
  investigation of their expressiveness and relationships, and to
  analyze their computational behavior}.  Regarding expressiveness,
our main result concerns the relationship to first-order theories of
topological relations. The expressive power of our modal logics is
incomparable with that standard theories of this kind since modal
logics offer an infinite supply of propositional variables
corresponding to unary predicates of first-order logic. In contrast,
standard first-order theories of topological relations offer only
eight binary predicates interpreted as topological relations, and no
unary predicates~\cite{Randell92rcc,PrattSchoop98,Papa96,Schaefer01}.
Therefore, we consider the extension of first-order theories of
topological relations with an infinite number of ``free'' unary
predicates. Then, we can show that our logics based on the
Egenhofer-Franzosa or \mn{RCC5} relations has exactly the same
expressive power as the two-variable fragment of first-order logic on
the same set of relations (indeed, this holds for any mutually
disjoint and jointly exhaustive set of topological relations).  We
also show that first-order logic is exponentially more succinct.  We
argue that the availability of unary predicates is essential for a
wide range of application areas: in contrast to describing only purely
topological properties of regions, it allows one to also capture other
properties such as being a country (in a GIS), a ball (for a
soccer-playing robot), or a protected area (in a spatial database). In
our modal logics, we can thus formulate constraints based on 
non-spatial properties such as ``there are no two overlapping regions
that are both countries'' and ``every river is connected to an ocean
or a lake''.

The main results of this paper concern the computational behavior of
modal logics of topological relations. We prove a very general
undecidability result that captures all modal logics of the \mn{RCC8}
relations that are determined by a class of region structures whose
regions are (not necesserily all) non-empty regular closed sets, and
that contains at least one infinite structure. It is interesting to
note that this result also covers logics that are determined by
\emph{substructures} of region structures. In particular, it captures
the substructures of the real line where regions are intervals, and
thus improves upon undecidability results for interval temporal logics
by Halpern and Shoham that do not capture substructures of interval
structures \cite{HalpernShoham91}.  Using a variation of the proof of
our central theorem, we can even show that logics based on finite
substructures of region structures are undecidable. Although our
results show that moving from full region structures to substructures
does not help to regain decidability, there is an improvement in
computational complexity: we show that most logics of \mn{RCC8}
relations based on full region structures are $\Pi^1_1$-hard and thus
not recursively enumerable. In contrast, we also prove that many
logics determined by substructures \emph{are} recursively enumerable.
Finally, we establish the undecidability of a number of modal logics
based on the \mn{RCC5} relations. The result is less general and, for
example, does not cover the substructure case.  Recursive
enumerability of \mn{RCC5}-based logics is left as an open problem.

This paper is organized as follows: in Section~\ref{sect:structures},
we introduce region structures as the semantical basis for modal
logics of topological relations.  The modal language is introduced in
Section~\ref{sect:languages}. In this section, we also compare its
expressiveness to that of first-order logic. Additionally, we show
that our modal logics are strictly more expressive than topological
constraint satisfaction problems. In Section~\ref{sect:logics}, we
introduce a number of natural modal logics based on the
Egenhofer-Franzosa relations that are induced by different notions of
regions, and briefly analyze their relationship.  In
Section~\ref{sect:rcc8undec}, we then prove the central undecidability
result capturing basically all interesting modal logics of \mn{RCC8}
relations determined by sets of region structures containing at least
one infinite structure.  For logics of full region structures, this is
strengthened to a $\Pi^1_1$-hardness proof in
Section~\ref{sect:notre}. We also prove recursive enumerability of
many modal logics based on substructures of region structures. In
Section~\ref{sect:finite}, we prove undecidability of logics
determined by classes of finite region structures. Finally, in
Section~\ref{sect:rcc5} we consider modal logics based on the
\mn{RCC5} relations.

\section{Structures}
\label{sect:structures}

The purpose of the logics considered in this paper is to reason about
regions in topological spaces. In this section, we show how a
topological space together with an appropriate definition of
``region'' induces a logical structure, and establish some basic
properties of the structures obtained in this way.  

Recall that a topological space is a pair ${\mathfrak T}=(U,
\mathbb{I})$, where $U$ is a set and $\mathbb{I}$ is an \emph{interior
  operator} on $U$, i.e., for all $s,t \subseteq U$, we have
$$
  \begin{array}{rclcrcl}
    \mathbb{I}(U) & = & U & \quad &
    \mathbb{I}(s) & \subseteq & s \\[1mm]
    \mathbb{I}(s) \cap \mathbb{I}(t) &=& \mathbb{I}(s \cap t) &&
    \mathbb{I}\mathbb{I}(s) &=& \mathbb{I}(s).
  \end{array}
$$
The closure $\mathbb{C}(s)$ of $s$ is 
$
{\mathbb C}(s) = U-\mathbb{I}(U-s).
$  %
Of particular interest for spatial reasoning are $n$-dimensional
Euclidean spaces $\Rbbm^{n}$ based on Cartesian products of the real
line with the standard topology induced by the Euclidean metric.
Depending on the application domain, different definitions of regions
in topological spaces have been introduced. Almost all of them have in
common that the regions of a topological space $\Tmf=(U, \mathbb{I})$
are identified with some set of non-empty, regular closed subsets of
$U$, where a subset $s \subseteq U$ is called {\em regular closed} if
$\mathbb{C}\mathbb{I}(s)=s$.\footnote{Another possibility is to identify regions
with non-empty regular open sets instead of non-empty regular closed ones. The results presented in
this paper hold for this alternative definition of regions as well.} 
Some popular choices for topological spaces and regions are the following:
\begin{itemize}
  
\item the set ${\mathfrak T}_{\mn{reg}}$ of all non-empty regular
  closed subsets of some topological space ${\mathfrak T}$, in
  particular the topological spaces $\Rbbm^n$ for some $n \geq 1$;
  
\item the set $\Rbbm^{n}_{\mn{conv}}$ of non-empty convex regular
  closed subsets of $\Rbbm^{n}$, for some $n \geq 1$;
  
\item the set $\Rbbm^{n}_{\mn{rect}}$ of closed hyper-rectangular
  subsets of $\Rbbm^{n}$, i.e., regions of the form $\prod_{i=1}^{n}
  C_{i}$, where $C_{1},\ldots,C_{n}$ are non-singleton closed
  intervals in $\Rbbm$, for some $n \geq 1$.

\end{itemize}
Sometimes, regions are required to satisfy additional constraints such
as being connected or homeomorphic to the closed unit disc.

\begin{figure}
  \begin{center}
    \framebox[1\columnwidth]{\input{rcc8.pstex_t}}
    \caption{The eight relations between regions.}
    \label{fig:rcc8}
  \end{center}
\end{figure}
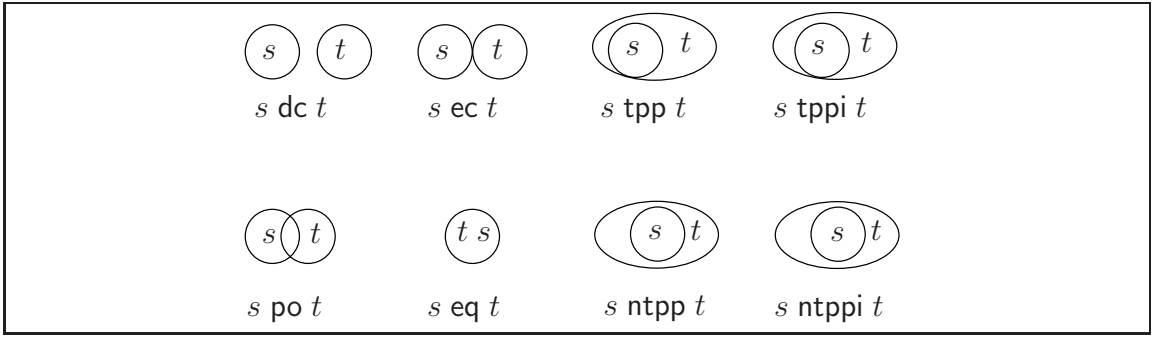

Given a topological space ${\mathfrak T}$ and a set of regions
$U_{\mathfrak T}$, we define the extension of the eight
\emph{Egenhofer-Franzosa (or \mn{RCC8}) relations} $\mn{dc}$
(`disconnected'), $\mn{ec}$ (`externally connected'), $\mn{tpp}$
(`tangential proper part'), $\mn{tppi}$ (`inverse of tangential proper
part'), $\mn{po}$ (`partial overlap'), $\mn{eq}$ (`equal'),
$\mn{ntpp}$ (`non-tangential proper part'), and $\mn{nttpi}$ (`inverse
of non-tangential proper part') as the following subsets of
$U_{\mathfrak T} \times U_{\mathfrak T}$:
  $$
  \begin{array}{rcl}
        (s,t) \in \mn{dc}^\mathfrak{T}
        &\text{ iff }& s \cap t = \emptyset \\
        (s,t) \in \mn{ec}^\mathfrak{T}
        &\text{ iff }& \mathbb{I}(s) \cap \mathbb{I}(t) = \emptyset \; \wedge
        \; s \cap t \neq \emptyset \\
        (s,t) \in \mn{po}^\mathfrak{T}
        &\text{ iff }& \mathbb{I}(s) \cap \mathbb{I}(t) \neq \emptyset \; \wedge
        \; s \not\subseteq t \; \wedge \; t\not\subseteq s  \\
        (s,t) \in \mn{eq}^\mathfrak{T}
        &\text{ iff }& s = t \\
\end{array}
$$
$$
  \begin{array}{rcl}
        (s,t) \in \mn{tpp}^\mathfrak{T}
        &\text{ iff }& s \subseteq t \; \wedge \;
        s \not\subseteq \mathbb{I}(t) \; \wedge \; s\neq t \\
        (s,t) \in \mn{ntpp}^\mathfrak{T}
        &\text{ iff }& s \subseteq \mathbb{I}(t) \; \wedge \; s\neq t \\
        (s,t) \in \mn{tppi}^\mathfrak{T}
        &\text{ iff }&
        (t,s) \in \mn{tpp}^\mathfrak{T} \\
        (s,t) \in \mn{ntppi}^\mathfrak{T}
        &\text{ iff }&
        (t,s) \in \mn{ntpp}^\mathfrak{T}.
\end{array}
$$
Figure~\ref{fig:rcc8} shows examples of the \mn{RCC8} relations in the
real plane $\Rbbm^{2}$.
\psfull
\psdraft
The structure $ {\mathfrak R}({\mathfrak T},U_{\mathfrak T}) := \auf
U_{\mathfrak T}, \mn{dc}^{\mathfrak T},\mn{ec}^{\mathfrak
  T},\mn{po}^{\mathfrak T},\mn{eq}^{\mathfrak T},\mn{tpp}^{\mathfrak
  T},\mn{ntpp}^{\mathfrak T},\mn{tppi}^{\mathfrak
  T},\mn{ntppi}^{\mathfrak T} \zu $ is called the \emph{concrete
  region structure induced by} $({\mathfrak T},U_{\mathfrak T})$.
Observe that concrete region structures do not include a valuation of
propositional letters, and thus correspond to a frame in standard
modal logic.  We will later extend region structures to region models
by augmenting them with valuation functions.

We now develop a first-order characterization of concrete region
structures. This will establish some fundamental properties of
concrete region structures that are used throughout the whole paper,
and will also provide us with an easy proof of the fact that certain
logics considered in this paper are recursively enumerable. We call a
relational structure
$$
{\mathfrak R}= \auf W, \mn{dc}^{\mathfrak R}, \mn{ec}^{\mathfrak R},
\mn{po}^{\mathfrak R},\mn{eq}^{\mathfrak R},\mn{tpp}^{\mathfrak
  R},\mn{ntpp}^{\mathfrak R},\mn{tppi}^{\mathfrak
  R},\mn{ntppi}^{\mathfrak R} \zu
$$
a \emph{general region structure} if $W$ is a non-empty set and the
$\mn{r}^{\mathfrak R}$ are binary relations on $W$ that are mutually
disjoint (i.e., $\mn{r}^{\mathfrak R}\cap \mn{q}^{\mathfrak
  R}=\emptyset$, for $\mn{r}\not=\mn{q}$), jointly exhaustive (i.e.,
the union of all $\mn{r}^{\mathfrak R}$ is $W\times W$), and
satisfy the following:
\begin{itemize}
\item $\mn{eq}$ is interpreted as the identity on $W$, $\mn{dc}^{\mathfrak R}$,
  $\mn{ec}^{\mathfrak R}$, and $\mn{po}^{\mathfrak R}$ are symmetric, and $\mn{tppi}^{\mathfrak R}$ and
  $\mn{ntppi}^{\mathfrak R}$ are the inverse relations of $\mn{ttp}^{\mathfrak R}$ and
  $\mn{ntpp}^{\mathfrak R}$, respectively;
\item the rules of the composition table
(Figure~\ref{fig:comptable}) are satisfied in the sense that, for any entry
$\mn{q}_{1},\ldots,\mn{q}_{k}$ in row $\mn{r}_{1}$ and column
$\mn{r}_{2}$, the first-order sentence
$$
\!\!\!\!\!\forall x\forall y \forall z ((\mn{r}_{1}(x,y) \wedge \mn{r}_{2}(y,z)) \rightarrow 
(\mn{q}_{1}(x,z) \vee \cdots \vee \mn{q}_{k}(x,z))
$$
is valid ($\ast$ is the disjunction over all eight relations).
\end{itemize}
\psfull
\begin{figure}
  \begin{center}
\scriptsize
\setlength{\tabcolsep}{1mm}
\begin{tabular}{|l||c|c|c|c|c|c|c|}
        \hline
        & & & & & & & \\ 
        \multicolumn{1}{|c||}{$\circ$} &
        \mn{dc} & \mn{ec} & \mn{tpp} & \mn{tppi} & \mn{po} & \mn{ntpp} & \mn{ntppi} \\
        & & & & & & & \\ 
        \hline
        \hline
        & & \mn{dc},\mn{ec}, & \mn{dc},\mn{ec}, & & \mn{dc},\mn{ec}, & \mn{dc},\mn{ec}, & \\ 
        \mn{dc} & $*$ & \mn{po},\mn{tpp}, & \mn{po},\mn{tpp}, & \mn{dc} & \mn{po},\mn{tpp}, & \mn{po},\mn{tpp}, & \mn{dc} \\ 
        & & \mn{ntpp} & \mn{ntpp} & & \mn{ntpp} & \mn{ntpp} & \\ 
        \hline
        & \mn{dc},\mn{ec}, & \mn{dc},\mn{ec}, & \mn{ec},\mn{po}, & & \mn{dc},\mn{ec}, & \mn{po}, & \\ 
        \mn{ec} & \mn{po},\mn{tppi}, & \mn{po},\mn{tpp}, & \mn{tpp}, & \mn{dc},\mn{ec} & \mn{po},\mn{tpp}, & \mn{tpp}, & \mn{dc}\\ 
        & \mn{ntppi} & \mn{tppi},\mn{eq} & \mn{ntpp} & & \mn{ntpp} & \mn{ntpp} & \\ 
        \hline
        & & & & \mn{dc},\mn{ec}, & \mn{dc},\mn{ec}, & & \mn{dc},\mn{ec}, \\ 
        \mn{tpp} & \mn{dc} & \mn{dc},\mn{ec} & \mn{tpp},\mn{ntpp} & \mn{po},\mn{tpp}, & \mn{po},\mn{tpp}, & \mn{ntpp} & \mn{po},\mn{tppi}, \\ 
        & & & & \mn{tppi},\mn{eq} & \mn{ntpp} & & \mn{ntppi} \\ 
        \hline
        & \mn{dc},\mn{ec}, & \mn{ec},\mn{po}, & \mn{po},\mn{eq}, & & \mn{po}, & \mn{po}, & \\ 
        \mn{tppi} & \mn{po},\mn{tppi}, & \mn{tppi}, & \mn{tpp}, & \mn{tppi},\mn{ntppi} & \mn{tppi}, & \mn{tpp}, & \mn{ntppi} \\ 
        & \mn{ntppi} & \mn{ntppi} & \mn{tppi} & & \mn{ntppi} & \mn{ntpp} & \\ 
        \hline
        & \mn{dc},\mn{ec}, & \mn{dc},\mn{ec}, & \mn{po}, & \mn{dc},\mn{ec}, & & \mn{po}, & \mn{dc},\mn{ec}, \\ 
        \mn{po} & \mn{po},\mn{tppi}, & \mn{po},\mn{tppi}, & \mn{tpp}, & \mn{po},\mn{tppi}, & $*$ & \mn{tpp}, & \mn{po},\mn{tppi}, \\ 
        & \mn{ntppi} & \mn{ntppi} & \mn{ntpp} & \mn{ntppi} & & \mn{ntpp} & \mn{ntppi} \\ 
        \hline
        & & & & \mn{dc},\mn{ec}, & \mn{dc},\mn{ec}, & & \\ 
        \mn{ntpp} & \mn{dc} & \mn{dc} & \mn{ntpp} & \mn{po},\mn{tpp}, & \mn{po},\mn{tpp}, & \mn{ntpp} & $*$ \\ 
        & & & & \mn{ntpp} & \mn{ntpp} & & \\ 
        \hline
        & \mn{dc},\mn{ec}, & \mn{po}, & \mn{po}, & & \mn{po}, & \mn{po}, \mn{tppi}, & \\ 
        \mn{ntppi} & \mn{po},\mn{tppi}, & \mn{tppi}, & \mn{tppi}, & \mn{ntppi} & \mn{tppi}, & \mn{tpp},\mn{ntpp}, & \mn{ntppi} \\
        & \mn{ntppi} & \mn{ntppi} & \mn{ntppi} & & \mn{ntppi} & \mn{ntppi},\mn{eq} & \\ 
        \hline
\end{tabular}
\normalsize
    \caption{The composition table.}
    \label{fig:comptable}
  \end{center}
\end{figure}
\psdraft
The following theorem shows that, in some sense, concrete region
structures and general region structures are interchangable.  In what
follows, we will thus often only speak of region structures and only
distinguish between general and concrete region structures when
necessary. A proof can be found in
Appendix~\ref{app:represent}.
%
%
\\[3mm]
\begin{minipage}{\columnwidth}
\begin{theorem}[Representation theorem]\label{represent}~\\[-4mm]
\begin{itemize}

\item[(i)] Every concrete region structure is a general region structure;

\item[(ii)] every general region structure is isomorphic to a concrete region structure;
  
\item [(iii)] for every $n>0$, every countable general region
  structure is isomorphic to a concrete region structure of the form
  ${\mathfrak R}(\Rbbm^{n},U_{\Rbbm^{n}})$ (with
  $U_{\Rbbm^{n}}\subseteq \Rbbm^n_\mn{reg}$). 

\end{itemize}
\end{theorem}
\end{minipage}
Note that Points~(ii) and~(iii) of Theorem~\ref{represent} rely on the
fact that we admit \emph{any} non-empty set of non-empty regular closed sets as
a possible choice for the regions of a topological space. This is of
course different from admitting only structures in which, for example,
\emph{all} non-empty regular closed sets are required to be regions,
or \emph{all} closed hyper-rectangles are required to be regions.  The
logics introduced in Section~\ref{sect:logics} will be based on both
kinds of structures.  Quite informally, we shall in the following call
structures of the latter kind \emph{full} concrete region structures.
We introduce some useful classes of region structures:
\begin{itemize}
  
\item $\mathcal{RS}$ is the class of all general
 region structures;
  
\item $\mathcal{TOP}$ denotes the class of all region structures
  ${\mathfrak R}({\mathfrak T},{\mathfrak T}_{\mn{reg}})$.

\end{itemize}
Observe that the structures in $\mathcal{TOP}$ are full concrete region
structures. It is interesting to note that, in contrast to 
$\mathcal{RS}$, $\mathcal{TOP}$ cannot be characterized by means of a recursively
enumerable set of first-order sentences. This follows from the
non-recursive enumerability of the logic of $\mathcal{TOP}$ to be
introduced and investigated later.

\smallskip

We should also note that the region structure
$\Rmf(\Rbbm,\Rbbm_\mn{rect})=\Rmf(\Rbbm,\Rbbm_\mn{conv})$ is an
interval structure. Therefore, topological modal logics interpreted in 
such structures may be viewed as temporal interval logics similar to
the ones defined by Halpern and Shoham in~\cite{HalpernShoham91}.  A
minor technical difference between our interval structure and the ones
considered by Halpern and Shoham is that our requirement of
regular closedness excludes point-intervals, while such intervals are
admitted by Halpern and Shoham.

\section{The Language}
\label{sect:languages}

The modal language ${\mathcal L}_{\mn{RCC8}}$ extends propositional logic
with countably many variables $p_{1},p_2,\ldots$ and the Boolean
connectives $\neg$ and $\wedge$ by means of the unary modal operators
$[\mn{dc}]$, $[\mn{ec}]$, etc. (one for each topological relation).  A
{\em region model} $ {\mathfrak M}= \auf {\mathfrak
  R},p_{1}^{\mathfrak M},p_{2}^{\mathfrak M},\ldots\zu $ for ${\mathcal
  L}_{\mn{RCC8}}$ consists of a region structure ${\mathfrak R}=\auf
W,\mn{dc}^{\mathfrak R},\mn{ec}^{\mathfrak R},\ldots\zu$ and the
interpretation $p_{i}^{\mathfrak M}$ of the variables $p_i$ of ${\mathcal
  L}_{\mn{RCC8}}$ as subsets of~$W$.  A formula $\varphi$ is either
true at a region $s\in W$ (written $\mathfrak{M},s \models \varphi$)
or false at $s$ (written $\mathfrak{M},s \not\models \varphi$), the
inductive definition being as follows:
\begin{enumerate}

  \item if $\varphi$ is a prop.\ variable, 
  then $\mathfrak{M},s \models \varphi$ iff \mbox{$s \in \varphi^{\mathfrak M}$};

  \item $\mathfrak{M},s\models \neg\varphi$ iff $\mathfrak{M},s\not\models\varphi$;

  \item $\mathfrak{M},s \models \varphi_1 \wedge \varphi_2$ iff ${\mathfrak M},s \models \varphi_1$
    and ${\mathfrak M},s \models \varphi_2$;

  \item $\mathfrak{M},s \models [ \mn{r} ] \varphi$ iff, for all $t \in W$, 
  \mbox{$(s,t) \in \mn{r}^{\mathfrak R}$} implies $\mathfrak{M},t \models \varphi$.

\end{enumerate}
We use the usual abbreviations: $\varphi \rightarrow \psi$ for $\neg \varphi
\vee \psi$ and $\auf \mn{r} \zu \varphi$ for $\neg [\mn{r}] \neg \varphi$.


In the remainder of this section, we discuss the expressive power of
the language~$\Lmc_\mn{RCC8}$. The discussion starts with some simple
observations.  
\begin{itemize}

\item First, the {\em difference modality} $\Box_{d} \varphi$, investigated for example in \cite{deRijke92},
has the following semantics:
$$
\Mmf,s \models \Box_d \varphi \text{ iff } \Mmf,t \models \varphi \mbox{ for all } t \in W \mbox{ such that $t\neq s$}.
$$
In $\Lmc_\mn{RCC8}$, it can be expressed as $\bigwedge_{\mn{r} \in \mn{RCC8} - \{\mn{eq}\}} [\mn{r}] \varphi$
since the relations are jointly exhaustive and mutually exclusive.

\item Second, the useful {\em universal box} $\Box_u \varphi$, which is
well-known from modal logic \cite{Goranko-Passy-92}, has the following
semantics:
$$
\Mmf,s \models \Box_u \varphi \text{ iff } \Mmf,t \models \varphi \mbox{ for all } t \in W.
$$
In $\Lmc_\mn{RCC8}$, it can be expressed as $\varphi \wedge \Box_{d} \varphi$.

\item Third, we can express that a formula $\varphi$ holds in precisely one
region (i.e., is {\em a nominal} \cite{GargovGoranko93}) by writing
$$
\mn{nom}(\varphi) = \Diamond_{u}(\varphi \wedge \Box_{d} \neg \varphi),
$$
where $\Diamond_{u}\varphi = \neg \Box_{u}\neg\varphi$.  The
availability of nominals means that we can introduce names for
regions; e.g., the formulas
$$
  \mn{nom}(\text{\em Elbe}), \quad
  \mn{nom}(\text{\em Dresden})
$$
state that ``{\em Elbe}'' (the name of a river) and ``{\em Dresden}''
each apply to exactly one region.  

\item Finally, it is often useful to define operators $[\mn{pp}]$ and
  $[\mn{ppi}]$ as abbreviations:
$$
\begin{array}{lcl}
  [\mn{pp}]\varphi &=& [\mn{tpp}]\varphi \wedge [\mn{nttp}]\varphi \\[1mm]
  [\mn{ppi}]\varphi &=& [\mn{tppi}]\varphi \wedge [\mn{nttpi}]\varphi.
\end{array}
$$
As in the temporal case \cite{HalpernShoham91} and following Cohn
\cite{Cohn93a}, we can use these new operators to classify
formulas $\varphi$ according to whether
\begin{itemize}
  
\item they are homogeneous, i.e.\ they hold continuously throughout
  regions: 
$$\Box_{u}( \varphi \rightarrow [\mn{pp}]\varphi)$$

\item they are anti-homogeneous, i.e.\ they hold only in regions whose
  interiors are mutually disjoint: 
$$\Box_{u}(\varphi \rightarrow
  ([\mn{pp}]\neg\varphi \wedge [\mn{po}]\neg\varphi)$$
  
\end{itemize}
Instances of anti-homogeneous propositions are ``{\em river}'' and
``{\em university campus}'', while ``{\em occupied-by-water}'' is homogeneous. 
\end{itemize}
As this paper concentrates on the investigation of the expressivity
and computational properties of topological modal logics, it is out of
scope to describe potential applications in detail. Therefore, we only
give a few illustrative examples of statements in $\Lmc_\mn{RCC8}$.
The following example describes, in a drastically simplified way, the
relationship of cities, harbours, rivers, and the sea.  Based on this
`background theory', it then describes the relationship of the city of
Dresden and the river Elbe.
$$
\begin{array}{l}
\Box_{u}(\text{\em harbor-city} \leftrightarrow (\text{\em city} \wedge
\auf \mn{ppi} \zu \text{\em harbor}))\\[1mm]
\Box_{u}(\text{\em harbor} \rightarrow (\auf \mn{ec}\zu \text{\em river} \vee \auf \mn{ec} \zu \text{\em sea}))\\[1mm]
\Box_{u}(\text{\em Dresden} \rightarrow \text{\em harbor-city}) \\[1mm]
\Box_{u}(\text{\em Elbe} \rightarrow \text{\em river}) \\[1mm]
\Box_{u}(\text{\em Dresden} \rightarrow \bigwedge_{\mn{r}\in \mn{RCC8}- \{\mn{dc}\}} 
[\mn{r}] \neg\text{\em sea}) \\[1mm]
\Box_{u}(\text{\em Dresden} \rightarrow (\auf \mn{po}\zu \text{\em Elbe} \wedge \bigwedge_{\mn{r}
\in \mn{RCC8}- \{\mn{dc}\}} [\mn{r}] (\text{\em river}
\rightarrow \text{\em Elbe}))
) \\[1mm] 
\end{array}
$$
From these formulas, it follows that Dresden has a part that is a harbor and is 
related via $\mn{ec}$ to the river Elbe. 

The example suggests a scheme for the representation of spatial
knowledge in ${\mathcal L}_{\mn{RCC8}}$ that is known from description
logic \cite{DL03}: a background theory (called TBox in description
logic) represents knowledge about general classes of regions such as
those describing harbors and rivers.  Knowledge about particular
regions is formulated by using nominals and expressing spatial
relations between them. In description logic, knowledge of this latter
kind would be stored in an ABox.


We now relate the expressive power of the modal language ${\mathcal
  L}_{\mn{RCC8}}$ to the expressive power of two standard formalisms
for spatial reasoning: constraint networks and spatial first-order
theories.

\smallskip

\mn{RCC8} constraint networks are a basic, but rather popular
formalism for representing spatial knowledge using the \mn{RCC8}
relations \cite{RenzNebel,Egenhofer94,Papa95,Schaefer01,Randell92rcc}.
In the following, we show that our
modal language $\Lmc_\mn{RCC8}$ can capture constraint networks in a
straightforward way. An \emph{RCC8 constraint network} is a finite set
of constraints $(s \mathrel{\rsf} r)$ with $s,r$ \emph{region
  variables} and $\rsf$ an \mn{RCC8} relation.
Such a network $N$ is \emph{satisfiable} in a topological space \Tmf
with regions $U_\Tmf$ if there exists an assignment $\delta$ of
regions in $U_\Tmf$ to region variables such that $(s \mathrel{\rsf}
r) \in N$ implies $\delta(s) \mathrel{\rsf^\Tmf} \delta(r)$. In our
language $\Lmc_\mn{RCC8}$, we can express a constraint network $N$
that uses region variables $s_1,\dots,s_k$ by writing
$$
  \bigwedge_{(s_i \mathrel{\rsf} s_j) \in N}
  \Diamond_u(p_i \wedge \auf \mn{r}\zu p_j) \quad \wedge \quad
  \bigwedge_{1 \leq i \leq k }\mn{nom}(p_i).
$$
This formula is clearly satisfiable iff $N$ is satisfiable.


Spatial first-order theories are usually formulated in first-order
languages equivalent to the first-order language $\mathcal{FO}_{\mn{RCC8}}$ that has equality,
eight binary predicates for the $\mn{RCC8}$ relations, no function symbols, and
\emph{no} unary predicates
\cite{Papa96,PrattSchoop98,Schaefer01,Randell92rcc}.  Intuitively, we
cannot reduce ${\mathcal L}_{\mn{RCC8}}$ to such languages because they do
not offer a counterpart of ${\mathcal L}_{\mn{RCC8}}$'s propositional
letters.  A formal proof is provided by the following two
observations:
\begin{enumerate}
  
\item $\mathcal{FO}_{\mn{RCC8}}$ is decidable over the region structure
  $\Rmf(\Rbbm^2,\Rbbm^2_\mn{rect})$. Indeed, it is not hard to verify
  that there is a reduction to the first-order theory of $\auf
  \Rbbm,<\zu$ which coincides with the first-order theory
  of $\auf \Qbbm,<\zu$
  and, therefore, is decidable \cite{enderton}.
  Details of the reduction are omitted as it is
  similar to the proof of Theorem~\ref{repos2} given in
  Appendix~\ref{app:rectanglesre} (but simpler).
  
\item In Section~\ref{sect:notre}, we show that ${\mathcal L}_{\mn{RCC8}}$
  is not recursively enumerable over
  $\Rmf(\Rbbm^2,\Rbbm^2_\mn{rect})$.

\end{enumerate}
Thus, the adequate first-order language to compare ${\mathcal
  L}_{\mn{RCC8}}$ with is the monadic extension $\mathcal{FO}^{m}_{\mn{RCC8}}$ of ${\mathcal FO}_{\mn{RCC8}}$ that is obtained by
adding countably many unary predicates $p_{1},p_{2},\ldots$.  By
well-known results from modal correspondence theory \cite{Gabbay-81c},
any $\Lmc_\mn{RCC8}$ formula $\varphi$ can be polynomially translated
into a formula $\varphi^{\ast}$ of $\mathcal{FO}^{m}_{\mn{RCC8}}$ with only two variables
such that, for any region model ${\mathfrak M}$ and any
region $s$,
$$
{\mathfrak M},s \models \varphi \mbox{ iff } {\mathfrak M}\models \varphi^{\ast}[s].
$$
More surprisingly, the converse holds as well: this follows from
recent results of \cite{Lutz-et-al-01c} since the $\mn{RCC8}$
relations are mutually exclusive and jointly exhaustive. A proof
sketch of the following theorem can be found in
Appendix~\ref{exprcompl}.
\begin{theorem}\label{ete}
  For every $\mathcal{FO}^{m}_{\mn{RCC8}}$-formula $\varphi(x)$ with free
  variable $x$ that uses only two variables, one can effectively
  construct a ${\mathcal L}_{\mn{RCC8}}$-formula $\varphi^{\ast}$ of
  length at most exponential in the length of $\varphi(x)$ such that,
  for every region model ${\mathfrak M}$ and any region $s$, $
  {\mathfrak M},s \models \varphi^{\ast} \mbox{ iff } {\mathfrak
    M}\models \varphi[s].  $
\end{theorem}
However, there is also an important difference between
$\Lmc_\mn{RCC8}$ and the two-variable fragment of $\mathcal{FO}^{m}_{\mn{RCC8}}$: the latter is exponentially more succinct than
the former. This can be shown using a formula proposed by Etessami,
Vardi, and Wilke~\cite{Etessami-et-al-02} stating that any two
regions agreeing on $p_0,\dots,p_{n-1}$ also agree on $p_n$. A 
proof can be found in Appendix~\ref{exprcompl}.
\begin{theorem}
\label{theo:succinct}
  For $n \geq 1$, define a $\mathcal{FO}^{m}_{\mn{RCC8}}$ formula
  $$
  \varphi_n := \forall x \forall y \big ( \bigwedge_{i < n}
  (p_i(x) \leftrightarrow p_i(y)) \rightarrow (p_n(x) \leftrightarrow
  p_n(y)) \big )
  $$
  Then every $\Lmc_\mn{RCC8}$-formula $\psi_{n}$ that is equivalent to $\varphi_{n}$ on 
  the class of all region structures
  $\Rmc\Smc$ has length $2^{\Omega(n)}$.\footnote{Following the formulation of Theorem~\ref{ete},
  the formula $\psi_{n}$ is called equivalent to $\varphi_{n}$
  if the following holds: for every region model ${\mathfrak M}$ and any region $s$, $
  {\mathfrak M},s \models \psi_{n} \mbox{ iff } {\mathfrak
    M}\models \varphi_{n}[s]$. As the formula $\varphi_{n}$ does not 
  have a free variable, the right hand side of this equivalence does not depend on $s$.}
\end{theorem}
We believe that this succinctness result also holds on other classes
of region structures such as the singleton $\{
\Rmf(\Rbbm^n,\Rbbm^n_\mn{reg}) \}$, but leave the proof as an open
problem.

\section{Logics}
\label{sect:logics}

In this section, we define a number of topological modal logics by
applying the language $\Lmc_\mn{RCC8}$ to different classes of region
structures. We also establish a number of separation results showing
that logics obtained from different classes of region structures do
not usually coincide.

Let \Smc be a class of region structures. An $\Lmc_\mn{RCC8}$ formula
$\varphi$ is \emph{valid} in \Smc if it is true in all regions of all
models based on region structures from ${\mathcal S}$. We use
$L_{\mn{RCC8}}({\mathcal S})$ to denote the logic of \Smc, i.e., the set
of all ${\mathcal L}_{\mn{RCC8}}$-formulas valid in ${\mathcal S}$. If ${\mathcal
  S}= \{{\mathfrak R}({\mathfrak T},U_{\mathfrak T})\}$ for some
topological space ${\mathfrak T}$ with regions $U_{\mathfrak T}$, then
we abbreviate $L_{\mn{RCC8}}({\mathcal S})$ by writing
$L_{\mn{RCC8}}({\mathfrak T},U_{\mathfrak T})$. The following logics
of \emph{full} concrete region structures (see
Section~\ref{sect:structures}) will play a prominent role in this
paper:
\begin{itemize}
  
\item the logic $L_\mn{RCC8}(\mathcal{TOP})$ of all full concrete
  region structures of regular closed regions
  $\Rmf(\Tmf,\Tmf_\mn{reg})$;
  
\item logics based on the $\Rbbm^n$, for some $n \geq 1$:
  $L_\mn{RCC8}(\Rbbm^n,\Rbbm^n_\mn{reg})$,
  $L_\mn{RCC8}(\Rbbm^n,\Rbbm^n_\mn{conv})$, and
  $L_\mn{RCC8}(\Rbbm^n,\Rbbm^n_\mn{rect})$.

\end{itemize}
%
We will also study
the logic $L_\mn{RCC8}(\Rmc\Smc)$ of all region structures. Note that
the region classes underlying the above logics admit unbounded regions
such as $\Rbbm^{n}$.  However, the technical results proved in this
paper also hold if we consider bounded regions, only.

\smallskip

We now investigate the relationship between the introduced logics.  As
an exhaustive analysis is out of the scope of this paper, 
we only treat some important cases:
\begin{enumerate}
  
\item $L_{\mn{RCC8}}(\mathcal{TOP}) \not\subseteq L_{\mn{RCC8}}({\mathcal
    RS})$ and $L_{\mn{RCC8}}(\Rbbm^{n},\Rbbm^n_\mn{x}) \not\subseteq
  L_{\mn{RCC8}}(\mathcal{RS})$ for $\mn{x} \! \in \! \{ \mn{reg}, \mn{conv},
  \mn{rect} \}$ and $n > 0$ since
$$
(\mn{nom}(p) \wedge \mn{nom}(q)\wedge \Diamond_{u}(p \wedge \auf \mn{dc}\zu q)) \rightarrow 
\Diamond_{u}(\auf \mn{ppi}\zu p \wedge \auf \mn{ppi}\zu q)
$$
is not valid in $\mathcal{RS}$ (it states that any two disconnected
regions are proper parts of a region). The converse inclusions
obviously hold for all $n>0$.

\item $L_\mn{RCC8}(\Rbbm^n,\Rbbm^n_\mn{x}) \not\subseteq
  L_\mn{RCC8}(\mathcal{TOP})$ for $\mn{x} \in \{ \mn{reg}, \mn{conv},
  \mn{rect} \}$ and $n > 0$: $\auf \mn{ppi} \zu \top$ is valid in
  $\Rmf(\Rbbm^n,\Rbbm^n_\mn{x})$, but not in $\mathcal{TOP}$. 
  For the converse direction, we clearly have
  $L_\mn{RCC8}(\mathcal{TOP}) \subseteq
  L_\mn{RCC8}(\Rbbm^n,\Rbbm^n_\mn{reg})$ for all $n>0$.
  
\item For $n,m>0$ and $m'>1$, $L_{\mn{RCC8}}(\Rbbm^{n},\Rbbm^{n}_{\mn{rect}})
  \not\subseteq L$, where $L$ is any logic from  
  $L_{\mn{RCC8}}(\Rbbm^{n+1},\Rbbm_{\mn{rect}}^{n+1})$,
  $L_\mn{RCC8}(\Rbbm^{m'},\Rbbm^{m'}_{\mn{conv}})$,
  $L_{\mn{RCC8}}(\Rbbm^{m},\Rbbm^{m}_{\mn{reg}})$,
  $L_\mn{RCC8}(\mathcal{TOP})$, $L_\mn{RCC8}(\mathcal{RS})$.  To
  see this define, for $k > 0$, an \mn{RCC8} constraint network
  $\mn{ec}[k]$ as follows:
  $$
  \mn{ec}[k]= \{ (x_{i} \mathrel{\mn{ec}} x_{j}) \mid 1 \leq i , j \leq k\}.
  $$
  For $n>0$, $\mn{ec}[2^{n}+1]$ is not satisfiable in ${\mathfrak
    R}(\Rbbm^{n}, \Rbbm^{n}_{\mn{rect}})$, but it is satisfiable in
  the classes of region structures determining the logics $L$. Observe that
  the condition $m'>1$ is required because ${\mathfrak R}(\Rbbm,\Rbbm_{\mn{conv}})= {\mathfrak R}(\Rbbm,\Rbbm_{\mn{rect}})$.
  
\item For $n>0$, $L_{\mn{RCC8}}(\Rbbm^{n},\Rbbm_{\mn{conv}}^{n})
  \not\subseteq L_{\mn{RCC8}} (\Rbbm^{n+1},\Rbbm_{\mn{conv}}^{n+1})$.
  Since
  $L_{\mn{RCC8}}(\Rbbm,\Rbbm_{\mn{conv}})=L_{\mn{RCC8}}(\Rbbm,\Rbbm_{\mn{rect}})$,
  the case $n=1$ follows from the previous item. Regarding the cases
  $n>1$, for simplicity we only consider \mbox{$n=2$} explicitly.
  A generalization is straightforward. Take region variables
  $x_{ij}$, $1 \leq i < j \leq 4$.  Then the constraint network
  obtained as the union of $\mn{ec}[4]$,
$$
\{(x_{i} \mathrel{\mn{pp}} x_{ij}), (x_{j} \mathrel{\mn{pp}} x_{ij}) \mid 1 \leq i<j\leq 4\}
$$
and
$$
\{(x_{ij} \mathrel{\mn{ec}} x_{k}) \mid 1\leq i<j\leq 4, k \in \{1,2,3,4\}-\{i,j\}\}
$$
is satisfiable in ${\mathfrak R}(\Rbbm^{3},\Rbbm_{\mn{conv}}^{3})$
but not in ${\mathfrak R}(\Rbbm^{2},\Rbbm_{\mn{conv}}^{2})$. 

\item For all $n,m>0$, $L_{\mn{RCC8}}(\Rbbm^{n},\Rbbm^{n}_{\mn{reg}})
  \not\subseteq L_{\mn{RCC8}}(\Rbbm^{m},\Rbbm^{m}_{\mn{conv}})$ and
  $L_{\mn{RCC8}}(\Rbbm^{n},\Rbbm^{n}_{\mn{reg}}) \not\subseteq
  L_{\mn{RCC8}}(\Rbbm^{m},\Rbbm^{m}_{\mn{rect}})$: the
  following formula states that, for any three pair-wise disconnected
  regions, there is another region containing only the first two (but
  not the third) as a proper part:
$$
\begin{array}{l}
\big (\displaystyle \bigwedge_{ 1 \leq i \leq 3} \mn{nom}(p_{i}) \wedge \bigwedge_{1 \leq i<j\leq 3} \Diamond_{u}(p_{i}\wedge
\auf \mn{dc} \zu p_{j}) \big )
\rightarrow \\[5mm]
\hspace*{3cm}\Diamond_{u}(\auf \mn{ppi}\zu p_{1} \wedge \auf \mn{ppi}\zu p_{2}
\wedge \neg \auf \mn{ppi}\zu p_{3}).
\end{array}
$$
This formula is valid in ${\mathfrak
  R}(\Rbbm^{n},\Rbbm^{n}_\mn{reg})$, but not in ${\mathfrak
  R}(\Rbbm^{n},\Rbbm^{n}_\mn{conv})$ and ${\mathfrak
  R}(\Rbbm^{n},\Rbbm^{n}_\mn{rect})$.

\end{enumerate}
As these examples show, $\Lmc_\mn{RCC8}$ is powerful enough to
``feel'' the difference between different topological spaces and
different choices of regions.


While full concrete region structures are appropriate for reasoning
about topological spaces themselves, for many applications it is not
adequate to demand that models have to comprise \emph{all} regions of
a particular form (such as the non-empty regular closed ones or the
closed hyper-rectangles). In such applications, models may contain
only \emph{some} such regions---those that are inhabited by spatial
objects that are relevant for the application. This observation gives
rise to another class of topological modal logics: given a class \Smc
of region structures, we use $L_\mn{RCC8}^\Ssf(\Smc)$ to denote the
logic determined by the class of all \emph{substructures} of
structures in \Smc. Note that the class $\Rmc\Smc$ is closed under
substructures by definition, and thus we have
$L_\mn{RCC8}(\Rmc\Smc)=L_\mn{RCC8}^\Ssf(\Rmc\Smc)$.  Taking this idea
one step further, we may even be concerned with applications where the
number of relevant spatial objects is known to be finite, but their
exact number is unknown. Then, we should consider only models
comprising a finite number of regions, without assuming an upper bound
on their number. Thus, we use $L_{\mn{RCC8}}^{\mn{fin}}({\mathcal S})$ to
denote the logic of all finite substructures of structures in ${\mathcal
  S}$.

\smallskip

The inclusion of such substructure logics and their finite versions is
a distinguishing feature of the undecidability results proved in this
paper: the general undecidability theorems presented in
Sections~\ref{sect:rcc8undec} and~\ref{sect:finite} cover all logics
of full concrete region structures introduced in this section, as
well as their substructure variants and finite substructure variants.
In contrast, the undecidability proofs of Halpern and Shoham for
interval temporal logics are not  applicable to the
substructure variants of these logics~\cite{HalpernShoham91}. Moreover,
it will turn out that logics of full concrete region structures are
usually $\Pi^1_1$-hard, while their substructure counterparts are usually
recursively enumerable.

We now continue our investigation of the relationship between
topological modal logics, taking into account substructure logics and
their finite companions.  Some of the new family members turn out to
be already known:

\begin{theorem} 
\label{equality1}
For $n>0$, we have
\begin{enumerate}
  
\item $L_{\mn{RCC8}}(\mathcal{RS}) = L^\Ssf_{\mn{RCC8}}(\mathcal{TOP}) =
  L_{\mn{RCC8}}^{\mn{S}}(\Rbbm^{n},\Rbbm^{n}_{\mn{reg}})$;
  
\item $L_{\mn{RCC8}}^{\mn{fin}}(\mathcal{RS})= L^\mn{fin}_{\mn{RCC8}}(\mathcal{TOP}) =
  L_{\mn{RCC8}}^{\mn{fin}}(\Rbbm^{n},\Rbbm^{n}_{\mn{reg}})$.

\end{enumerate}
\end{theorem}
%
\begin{Proof}
  All the mentioned logics are modal logics determined by classes of
  structures that are closed under substructures. As shown in
  \cite{Wolter97}, Corollary 3.8, such modal logics are determined by
  the at most countable members of those classes.  Thus,
  Theorem~\ref{equality1} is an immediate consequence of
  Theorem~\ref{represent}.
\end{Proof}
 
\begin{figure}
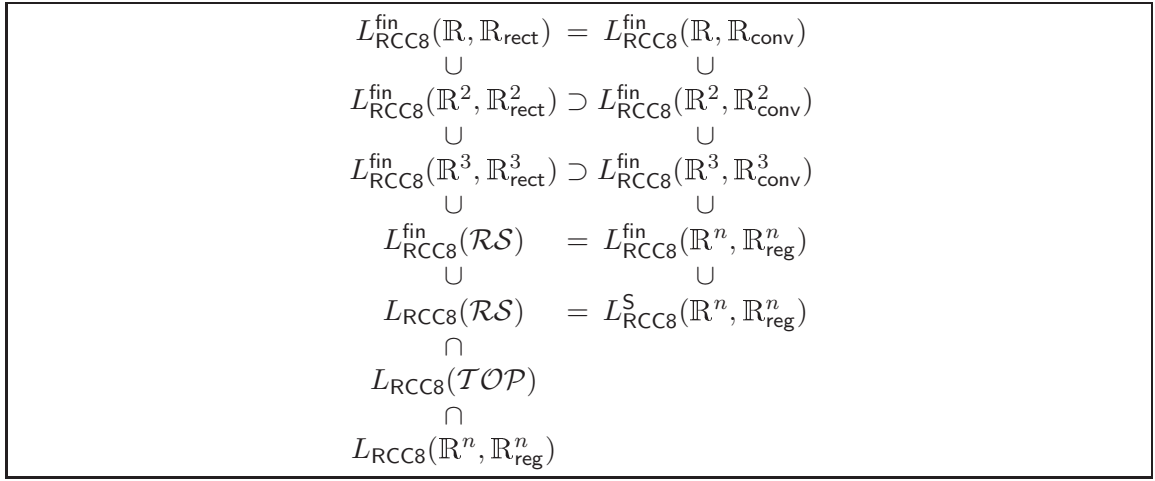

    \framebox[1\columnwidth]{
      \begin{minipage}{1\columnwidth}
$$
\begin{array}{c@{\;}c@{\;}c@{\;}c@{\;}c}
  L^\mn{fin}_\mn{RCC8}(\Rbbm,\Rbbm_\mn{rect}) & = & L^\mn{fin}_\mn{RCC8}(\Rbbm,\Rbbm_\mn{conv}) \\[0mm]
  \cup && \cup \\[0mm]
  L^\mn{fin}_\mn{RCC8}(\Rbbm^2,\Rbbm^2_\mn{rect}) & \supset & L^\mn{fin}_\mn{RCC8}(\Rbbm^2,\Rbbm^2_\mn{conv}) \\[0mm]
  \cup && \cup \\[0mm]
  L^\mn{fin}_\mn{RCC8}(\Rbbm^3,\Rbbm^3_\mn{rect}) & \supset & L^\mn{fin}_\mn{RCC8}(\Rbbm^3,\Rbbm^3_\mn{conv}) \\[0mm]
  \cup && \cup \\[0mm]
  L^\mn{fin}_\mn{RCC8}(\Rmc\Smc) & 
  = & L^\mn{fin}_\mn{RCC8}(\Rbbm^n,\Rbbm^n_\mn{reg}) \\[0mm]
  \cup && \cup   \\[0mm]
  L_\mn{RCC8}(\Rmc\Smc) & 
  = & L^\mn{S}_\mn{RCC8}(\Rbbm^n,\Rbbm^n_\mn{reg}) \\[0mm]
  \cap \\[0mm]
  L_\mn{RCC8}(\mathcal{TOP}) \\[0mm]
  \cap \\[0mm]
  L_\mn{RCC8}(\Rbbm^n,\Rbbm^n_\mn{reg}) 
\end{array}
$$
      \end{minipage}
}
    \caption{Inclusions between logics.}
    \label{fig:inclusions}
\end{figure}

\noindent A few additional interesting observations are the following:

\begin{enumerate}

\item[(6)] The non-inclusions given under Items~3 and 4 above also hold for the
  corresponding substructure and finite
  substructure cases. The proofs are identical. 
  
\item[(7)] The arguments given in Items~1, 2 and 5 do not carry over
  since the given formulas are not valid in the corresponding
  substructures and finite substructures.  Indeed, by
  Theorem~\ref{equality1}, in these cases the first claim of Item~1
  does not hold and the remaining claims of Item~1 and 2 do not hold
  for $\mn{x} = \mn{reg}$. In Item~5, the statement is wrong in the
  substructure case and finite substructure case: it is not hard to
  see that, e.g.,
  $L^\mn{S}_{\mn{RCC8}}(\Rbbm^{n},\Rbbm^{n}_{\mn{reg}}) \subseteq
  L^\mn{S}_{\mn{RCC8}}(\Rbbm^{n},\Rbbm^{n}_{\mn{conv}})$ and
  $L^\mn{S}_{\mn{RCC8}}(\Rbbm^{n},\Rbbm^{n}_{\mn{reg}}) \subseteq
  L^\mn{S}_{\mn{RCC8}}(\Rbbm^{n},\Rbbm^{n}_{\mn{rect}})$ for all
  $n>0$, and analogous claims hold in the finite substructure case.
    
\item[(8)] $L_{\mn{RCC8}}^{\mn{fin}}(\Smc) \not\subseteq L$ for any class
  of region structures $\Smc$ and $L$ among $L_\mn{RCC8}(\Rmc\Smc)$,
  $L_{\mn{RCC8}}(\mathcal{TOP})$ and $L_{\mn{RCC8}}(\Rbbm^{n},U_{n})$
  with $n \geq 1$ and $\Rbbm^n_\mn{rect} \subseteq U_n$: the L\"{o}b-formula
  from modal logic
$$
[\mn{pp}]([\mn{pp}]p \rightarrow p) \rightarrow [\mn{pp}]p.
$$
is valid in a relational structure iff there is no infinite ascending
$\mn{pp}$-chain, see \cite{Gabbayetal}, pages 8-12. Thus, this formula
is valid in all finite region structures, but not in all infinite ones.

\item[(9)] A number of additional inclusions is easily derived such as
  $L^\mn{S}_{\mn{RCC8}}(\Rbbm^{n+1},\Rbbm^{n+1}_{\mn{rect}}) \subseteq
  L^\mn{S}_{\mn{RCC8}}(\Rbbm^{n},\Rbbm^{n}_{\mn{rect}})$, for $n>0$:
  it is easy to convert a substructure of
  $\Rmf(\Rbbm^{n+1},\Rbbm^{n+1}_\mn{rect})$ into an isomorphic
  substructure of $\Rmf(\Rbbm^{n},\Rbbm^{n}_\mn{rect})$.

\end{enumerate}
The derived inclusions are summarized in Figure~\ref{fig:inclusions}.
By Points~1 to~9 above, all listed inclusions are indeed proper. For
the sake of readability, we do not attempt to display all derived
non-inclusions in Figure~\ref{fig:inclusions}.
\psfull
\psdraft

\section{Undecidability}
\label{sect:rcc8undec}

We now establish the central result of this paper: a rather general
undecidability result that covers all logics introduced in the
previous section. The only exceptions are logics based on classes of
finite region structures, whose undecidability will be established in
Section~\ref{sect:finite}. To the best of our knowledge, the
undecidability result proved in this section covers all classes of
region structures that have been considered in the literature and
contain at least one infinite structure.  As the precise formulation
of the result is somewhat technical, we start with a weaker version in
which we require that the class of region structures contains at least
one structure of the form ${\mathfrak R}(\Rbbm^{n},U)$ with
$\Rbbm^{n}_{\mn{rect}}\subseteq U$.  This condition will later be
replaced with a more general one.
\begin{theorem}\label{theo:undec}
  Let ${\mathcal S}\subseteq \mathcal{RS}$ and suppose there exists $n>0$ and
  a set $U \subseteq \Rbbm^n_\mn{reg}$ such that
  $\Rbbm^{n}_{\mn{rect}} \subseteq U$ and ${\mathfrak
    R}(\Rbbm^{n},U)\in {\mathcal S}$. Then $L_{\mn{RCC8}}({\mathcal S})$ is
  undecidable.
\end{theorem}
Concerning the logics introduced in Section~\ref{sect:logics}, we thus
obtain the following:
\begin{corollary}
  The logics $L_\mn{RCC8}(\Smc)$ and $L^\mn{S}_\mn{RCC8}(\Smc)$ are
  undecidable, for \Smc one of $\mathcal{RS}$, $\mathcal{TOP}$,
  $\Rmf(\Rbbm^{n},\Rbbm^{n}_{\mn{reg}})$,
  $\Rmf(\Rbbm^{n},\Rbbm^{n}_{\mn{conv}})$, and
  $\Rmf(\Rbbm^{n},\Rbbm^{n}_{\mn{rect}})$, with $n>0$.
\end{corollary}
We now develop the proof of Theorem~\ref{theo:undec}. As we shall see,
the proof suggests the mentioned generalization of
Theorem~\ref{theo:undec}, which will be stated subsequently. To ease
notation, in the proofs given in this and the following sections we
denote accessibility relations in models simply with \mn{dc}, \mn{ec},
etc., instead of with $\mn{dc}^\Rmf$, $\mn{ec}^\Rmf$, etc.

\smallskip

\begin{figure}[t]
  \begin{center}
    \framebox[1\columnwidth]{\input{dovetail.pstex_t}}
    \caption{Enumerating tile positions.}
    \label{fig:dovetail}
  \end{center}
\end{figure}
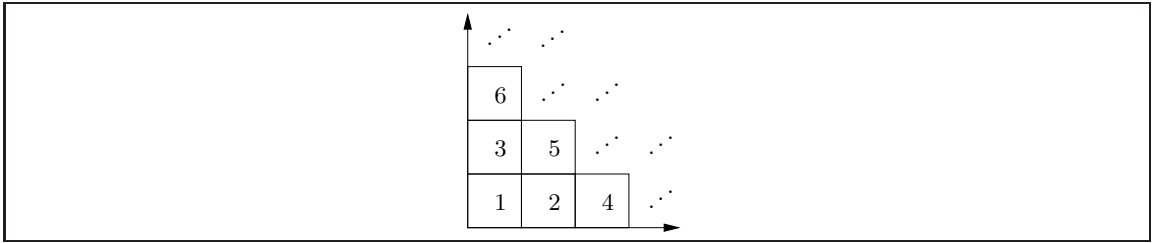

The proof of Theorem~\ref{theo:undec} is by reduction of the domino
problem that requires tiling of the first quadrant of the plane to the
satisfiability of $\Lmc_\mn{RCC8}$ formulas.
%
As usual, a formula $\vp$ is called \emph{satisfiable} in a region
model $\Mmf=\auf W,\mn{dc},\mn{ec},\dots,p_1^\Mmf,p_2^\Mmf,\dots\zu$
if there is an $s \in W$ with $\Mmf,s \models \vp$.
\begin{definition}
\label{domsys}
Let $\mathcal{D} = (T,H,V)$ be a \emphindex{domino system}, where $T$
is a finite set of \emph{tile types} and $H,V \subseteq T \times T$
represent the horizontal and vertical matching conditions.  We say
that $\mathcal{D}$ \emph{tiles the first quadrant of the plane} iff
there exists a mapping $\tau: \Nbbm^2 \to T$ such that, for all
$(x,y) \in \Nbbm^2$:
  \begin{itemize}
 \item if $\tau(x,y) = t$ and $\tau(x + 1,y) = t'$, then $(t,t') \in H$
  \item if $\tau(x,y) = t$ and $\tau(x,y + 1) = t'$, then $(t,t') \in V$
  \end{itemize}
  Such a mapping $\tau$ is called a \emph{solution} for \Dmc.
\end{definition}
For reducing this domino problem to satisfiability in region models
based on \Smc, we fix an enumeration of all the tile positions in the
first quadrant of the plane as indicated in Figure~\ref{fig:dovetail}.
The function $\lambda$ takes positive integers to $\Nbbm \times
\Nbbm$-positions, i.e.\ $\lambda(1)=(0,0)$, $\lambda(2)=(1,0)$,
$\lambda(3)=(1,1)$, etc.
\psfull
\psdraft

\smallskip

The idea of the reduction is to construct a formula $\vp_\Dmc$ that
enforces the existence of a sequence of regions $r_1,r_2,\dots$ such
that $r_i \mathrel{\mn{ntpp}} r_j$ if $i < j$.  Intuitively, each
region $r_i$ corresponds to the position $\lambda(i)$ of the first
quadrant of the plane. We introduce additional regions ``connecting''
each $r_i$ with $r_{i+1}$ to facilitate writing formulas that express
statements such as ``if the current region $r_i$ satisfies $\vp$, then
the next region $r_{i+1}$ satisfies $\psi$'', and likewise for the
previous region. Similarly, we introduce additional regions that
connect each region $r_i$ with the region $r_j$ such that the position
$\lambda(j)$ is to the right of the position $\lambda(i)$ in the first
quadrant of the plane. These latter regions allow statements such as
``if the current region $r_i$ satisfies $\vp$, then the region
representing the position to its right satisfies $\psi$''. Using such
statements, it is obviously easy to enforce the horizontal tiling
condition. By virtue of our enumeration of plane positions, reaching
the position above the current one is simply a matter of going to the
right and then advancing by one in the enumeration. Thus, we can also
enforce the vertical tiling condition. One of the main difficulties of
the proof will be to enforce the existence of the connecting regions
for ``going to the right''. The pursued solution is inspired by
\cite{Marx-Reynolds-99,Reynolds-Zakharyaschev-01}.

Now let $\Dmc=(T,H,V)$ be a domino system. For constructing
$\vp_\Dmc$, we use the following variables:
\begin{itemize}

\item for each tile type $t \in T$, a variable $p_t$;
  
\item variables $a$, $b$, and $c$ that are used to mark
  important regions;
  
\item variables \mn{wall} and \mn{floor} that are used to
  identify regions corresponding to positions from the sets $\{ 0 \}
  \times \Nbbm$ (the \emph{wall}) and $\Nbbm \times \{ 0 \}$ (the
  \emph{floor}), respectively.

\end{itemize}
The reduction formula $\varphi_\Dmc$ is defined as 
$$
a \wedge b \wedge \mn{wall} \wedge \mn{floor} \wedge [\mn{ntppi}]
\neg a \wedge \Box_u \chi,
$$
where $\chi$ is the conjunction of a number of formulas. We list these
formulas together with some intuitive explanations:
  \begin{enumerate}
    
  \item Ensure that the regions $\{ s \in W \mid \Mmf,s \models a\}$
    are ordered by the relation \mn{pp} (i.e.\ the union of \mn{tpp}
    and \mn{ntpp}):
    \begin{equation}
      a \rightarrow ([\mn{dc}] \neg a \wedge [\mn{ec}] \neg
      a \wedge [\mn{po}] \neg a) \\[2mm]
    \end{equation}
    
  \item Enforce that the regions $\{ s \mid \Mmf,s \models a \wedge b
    \}$ are \emph{discretely} ordered by \mn{ntpp}. These regions will
    constitute the sequence $r_1,r_2,\dots$ described above. In order
    to ensure discreteness, we use a sequence of alternating $a \wedge
    b$ and $a \wedge \neg b$ regions as shown in the left part of
    Figure~\ref{fig:ordering}.
\psfull
\begin{figure}[t]
  \begin{center}
      \input{ordering.pstex_t}
    \caption{Left: a discrete ordering in the plane;  Right: the ``going right'' regions.}
    \label{fig:ordering}
  \end{center}
\end{figure}
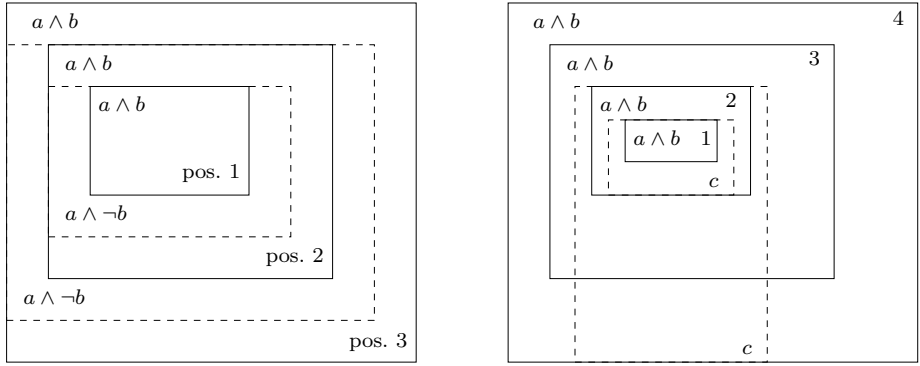
\psdraft
  \begin{eqnarray}
   a \wedge b &\rightarrow& \auf \mn{tpp} \zu (a \wedge \neg b) \\
  a \wedge \neg b &\rightarrow& \auf \mn{tpp} \zu (a\wedge b) \\
  a \wedge \neg b &\rightarrow& [\mn{tpp}] (a \rightarrow b) \\
  a \wedge b &\rightarrow& [\mn{tpp}](a \rightarrow \neg b) 
  \end{eqnarray}
  A formal proof that these formulas work as described is given below
  (Point~5 of Claim~1).  If we are at an $a \wedge b$ region, we can
  access the region corresponding to the next position in the plane
  (w.r.t.\ the fixed enumeration) and to the previous position using
$$
\begin{array}{rcl}
\Diamond^+\varphi &=& \auf \mn{tpp} \zu ( a \wedge \neg b \wedge \auf
\mn{tpp} \zu ( a \wedge b \wedge \varphi))
\\[1mm]
\Diamond^-\varphi &=& \auf \mn{tppi} \zu ( a \wedge  \neg b \wedge \auf
\mn{tppi} \zu ( a \wedge b \wedge \varphi)).
\end{array}
$$

\item The additional regions that will eventually allow us to ``go
  right'' in the plane satisfy the propositional letter $c$ and are
  related to the regions corresponding to plane positions as indicated
  in the right part of Figure~\ref{fig:ordering}. For example,
  Position~2 in the figure is right of Position~1, and Position~4 is
  right of Position~2. We start with stating the following:
  \begin{eqnarray}
  a \wedge b &\rightarrow& \auf \mn{tpp} \zu c \\
  c &\rightarrow& \auf \mn{tpp} \zu (a\wedge b) \\
  c &\rightarrow& ([\mn{dc}] \neg c \wedge [\mn{ec}] \neg c \wedge [\mn{po}] \neg c
  \wedge [\mn{tpp}] \neg c \wedge [\mn{tppi}] \neg c)
  \end{eqnarray}
  These formulas do not yet ensure that the $c$ regions actually bring
  us to the correct position. Roughly spoken, they only help to ensure
  that ``going to the right via regions satisfying~$c$'' is a
  well-defined, monotone, and injective total function.
  
  After further constraining the $c$ regions, we will be able to go to
  the right and upper position with
$$
\begin{array}{rcl}
\Diamond^R\varphi &=& \auf \mn{tpp} \zu ( c \wedge \auf
\mn{tpp} \zu ( a \wedge b \wedge \varphi))
\\[1mm]
\Diamond^U\varphi &=& \Diamond^R\Diamond^+\varphi.
\end{array}
$$
Similarly, we will be able to go to the left and down:
$$
\begin{array}{rcl}
\Diamond^L\varphi &=& \auf \mn{tppi} \zu ( c \wedge \auf
\mn{tppi} \zu ( a \wedge b \wedge \varphi)) \\[1mm]
\Diamond^D\varphi &=& \Diamond^L\Diamond^-\varphi.
\end{array}
$$
%
%


\item Axiomatizing the behavior of tiles on the floor and on the wall
  ensures that our ``going to the right'' relation actually brings us
  to the expected position in the first quadrant of the plane:
  \begin{eqnarray}
       (\mn{floor} \wedge \mn{wall}) &\rightarrow& [\mn{ntppi}] \neg a \\
      \mn{wall} &\rightarrow& \Diamond^+ \mn{floor} \\
      \mn{wall} &\rightarrow& \Diamond^U\mn{wall} \\
      {[}\mn{ntppi}] \neg a \vee (\mn{wall} &\rightarrow& \Diamond^D\mn{wall}) \\
      a \wedge b &\rightarrow& \Diamond^R \neg \mn{wall} \\
      (a \wedge b \wedge \neg\mn{wall}) &\rightarrow& \Diamond^L \top
  \end{eqnarray}
  %









\item Finally, we enforce the tiling:
  \begin{eqnarray}
    \bigwedge_{t,t' \in T} \neg (p_t \wedge p_{t'})\\
    a \wedge b \rightarrow \bigvee_{(t,t') \in H} (p_t \wedge \Diamond^R p_{t'}) \\
    a \wedge b \rightarrow \bigvee_{(t,t') \in V} (p_t \wedge \Diamond^U p_{t'}) 
  \end{eqnarray}

\end{enumerate}
%
%
We now prove two lemmas asserting the correctness of the reduction.
The first one is concerned with constructing solutions for \Dmc from
region models for $\vp_\Dmc$. Observe that this lemma does not
assume anything about the involved region model.
\begin{lemma}
\label{corr1}
If the formula $\varphi_\Dmc$ is satisfiable, then
the domino system \Dmc has a solution.
\end{lemma}
\begin{Proof}
  Let $\Mmf=\auf \Rmf,p_1^\Mmf,p_2^\Mmf,\dots\zu$ be a region model
  of~$\varphi_\Dmc$ with $\Rmf = \auf W , \mn{dc},
  \mn{ec},\dots \zu$.
  \\[2mm]
  {\bf Claim~1}. There exists a sequence $r_1,r_2,\ldots \in W$ such
  that
  \begin{enumerate}

  \item $\Mmf,r_1 \models \varphi_\Dmc$,

  \item $r_1 \mathrel{\mn{ntpp}} r_2 \mathrel{\mn{ntpp}} r_3
  \mathrel{\mn{ntpp}} \cdots$,
  
  \item $\Mmf,r_i \models a \wedge b$ for $i \geq 1$.
    
  \item for each $i \geq 1$, there exists a region $s_i \in
    W$ such that
    \begin{enumerate}

    \item $r_i \mathrel{\mn{tpp}} s_i$,

    \item $\Mmf,s_i \models a \wedge \neg b$,

    \item $s_i \mathrel{\mn{tpp}} r_{i+1}$,

    \item for each region $s$ with $r_i \mathrel{\mn{tpp}} s$ and
      $\Mmf,s \models a \wedge \neg b$, we have $s=s_i$, and

    \item for each region $r$ with $s_i \mathrel{\mn{tpp}} r$ and
      $\Mmf,r \models a \wedge b$, we have $r=r_{i+1}$, 

    \end{enumerate}
    
    
  \item for all $r \in W$ with $\Mmf,r \models a \wedge b$, we have
    that $r=r_i$ for some $i\geq1$ or $r_i \mathrel{\mn{ntpp}} r$ for
    all $i \geq 1$.

  \end{enumerate}
  Proof: We start with inductively constructing a sequence
  $r_1,r_2,\dots \in W$ satisfying Properties~1 to~4. Afterwards, we
  prove that Property~5 is also satisfied.  Since $\Mmf$ is a model of
  $\varphi_\Dmc$, there is a region $r_1$ such that $\Mmf,r_1 \models
  \varphi_\Dmc$. By definition of $\varphi_\Dmc$, Point 3 is
  satisfied.  Due to Formulas~(5.2) and~(5.3), there are regions $s_1$ and
  $r_2$ such that $r_1 \mathrel{\mn{tpp}} s_1$, $\Mmf,s_1 \models a
  \wedge \neg b$, $s_1 \mathrel{\mn{tpp}} r_2$, and $\Mmf,r_2 \models
  a \wedge b$. We show that all necessary Properties are satisfied:
  \begin{itemize}
    
  \item Point~2. Since $r_1 \mathrel{\mn{tpp}} s_1$ and $s_1
    \mathrel{\mn{tpp}} r_2$, we have $r_1 \mathrel{\mn{tpp}} r_2$ or
    $r_1 \mathrel{\mn{ntpp}} r_2$ according to the composition table
    which applies to all region structures by Theorem~\ref{represent}.
    But then, the first possibility is ruled out by Formula~(5.5).
    
  \item Point~4d. Suppose there is an $s \neq s_1$ with $r_1
    \mathrel{\mn{tpp}} s$ and $\Mmf,s \models a \wedge \neg b$.  Since
    $r_1 \mathrel{\mn{tpp}} s_1$, $s_1$ and $s$ are related via one of
    \mn{po}, \mn{tpp}, and \mn{tppi} by the composition
    table. But then, the first option is ruled out by
    Formula~(5.1) and the last two by Formula~(5.4).
    
  \item Point~4e. Analogous to the previous case.

  \end{itemize}
  The induction step is similar: as $\Mmf, r_i \models a \wedge b$, we
  may use Formulas~(5.2) and~(5.3) to find the region $r_{i+1}$, and then
  show in the same way as above that it satisfies all relevant
  properties.  It thus remains to prove Point~5.  Assume that there is
  a region $r$ such that $\Mmf,r \models a \wedge b$, $r \neq r_i$ for
  all $i \geq 1$, and $r_k \mathrel{\mn{ntpp}} r$ does not hold for
  some $k \geq 1$. Since $r_k \mathrel{\mn{ntpp}} r$ does not hold and
  $r_k \neq r$, $r_k$ and $r$ are related by one of \mn{dc}, \mn{ec},
  \mn{po}, \mn{tpp}, \mn{tppi}, and \mn{ntppi}. The first three
  possibilities are ruled out by Formula~(5.1), and \mn{tpp} and
  \mn{tppi} are ruled out by Formula~(5.5). It thus remains to treat the
  case $r_k \mathrel{\mn{ntppi}} r$. Consider the relationship between
  $r_1$ and $r$. Since $r_1 \neq r$ and due to Formulas~(5.1) and~(5.5),
  there are only two possibilities for this relation;
  \begin{itemize}

  \item $r \mathrel{\mn{ntpp}} r_{1}$. Impossible by $\varphi_\Dmc$'s
    subformula $[\mn{ntppi}] \neg a$.
    
  \item $r_{1} \mathrel{\mn{ntpp}} r$. Then we have $r_1
    \mathrel{\mn{ntpp}} r \mathrel{\mn{ntpp}} r_k$. Take the maximal
    $i$ such that $r_i \mathrel{\mn{ntpp}} r$ and the minimal $j$ such
    that $r \mathrel{\mn{ntpp}} r_{j}$. Since $r \neq r_n$ for all $n
    \geq 1$, we have $j=i+1$. By Point~4, there is a region
    $s$ with $r_i \mathrel{\mn{tpp}} s$, $\Mmf,s \models a \wedge \neg b$, 
    and $s \mathrel{\mn{tpp}} r_j$. Then we have
    $r \mathrel{\mn{nttpi}} r_{i} \mathrel{\mn{tpp}} s$.
    By the composition table, $r$ is related to $s$ by \mn{po}, \mn{tppi}, or $\mn{ntppi}$. 
    On the other hand, 
    $r \mathrel{\mn{nttp}}  r_{j} \mathrel{\mn{tppi}} s$.
    By the composition table, we have one of the relations $\mn{dc}$, $\mn{ec}$,
    $\mn{po}$, $\mn{tpp}$, or $\mn{ntpp}$ between $r$ and $s$. Together we obtain $r \mathrel{\mn{po}} s$
    which contradicts Formula~(5.1). 

  \end{itemize}
  The next claim identifies the regions needed for ``going right'' in
  the plane.
  \\[2mm]
  {\bf Claim~2}. For each $i \geq 1$, there exist regions
  $t_i$ and $u_i$ such that
  \begin{enumerate}
      
    \item $r_i \mathrel{\mn{tpp}} t_i$,

    \item $\Mmf,t_i \models c$,

    \item for each region $t$ with $r_i \mathrel{\mn{tpp}} t$ and
      $\Mmf,t \models c$, we have $t=t_i$, 

    \item $t_i \mathrel{\mn{tpp}} u_{i}$,

    \item $\Mmf, u_i \models a \wedge b$,

    \item for each region $u$ with $t_{i} \mathrel{\mn{tpp}} u$ and
      $\Mmf,u \models a \wedge b$, we have $u=u_{i}$.

    \end{enumerate}
    Proof: Let $i \geq 1$. By Formula~(5.6), there is a $t_i$ with $r_i
    \mathrel{\mn{tpp}} t_i$ and $\Mmf, t_i \models c$. Let us show
    that $t_i$ satisfies Property~3. To this end, let $t \neq t_i$
    such that $r_i \mathrel{\mn{tpp}} t$ and $\Mmf,t \models c$. Then
    $t$ and $t_i$ are related via one of \mn{po}, \mn{tpp}, and
    \mn{tppi}. But then, all these options are ruled out by
    Formula~(5.8).
    Now for Points~4 to~6. By Formula~(5.7), there is an $r$ such that
    $t_i \mathrel{\mn{tpp}} r$ and $\Mmf,r \models a \wedge b$.
    Point~6 can now be be proved analogously to Point~3, using
    Formulas~(5.1) and~(5.5) instead of Formula~(5.8).  This finishes the
    proof of Claim~2.

  \noindent
  The next claim states that the regions $u_i$ fixed in Claim~2 
  are ordered by \mn{ntpp}.
  \\[2mm]
  {\bf Claim~3}. Let $i,j \geq 1$ with $i < j$. Then $u_i
  \mathrel{\mn{ntpp}} u_j$.
  \\[2mm]
  Proof: By Claims~1
  and~2, we have (i)~$r_i \mathrel{\mn{ntpp}} r_j$, (ii)~$r_i
  \mathrel{\mn{tpp}} t_i$, and (iii)~$r_j \mathrel{\mn{tpp}} t_j$.  By
  the composition table, (i) and (iii) yield $r_i \mathrel{\mn{ntpp}}
  t_j$, which together with (ii) implies that $t_i$ and $t_j$ are
  related by \mn{po}, \mn{tpp}, or \mn{ntpp}. Since $\Mmf, t_i \models c$ and $\Mmf, t_j
  \models c$ by Claim~2, all but the last possibility are ruled out
  by Formula~(5.8). Therefore $t_{i} \mathrel{\mn{ntpp}} t_{j}$ which together
  with $t_{j} \mathrel{\mn{tpp}} u_{j}$ (Claim 2) implies $t_{i} \mathrel{\mn{ntpp}} u_{j}$.
  By Claim 2 we also have $t_{i} \mathrel{\mn{tpp}} u_{i}$ which by the composition table
  implies that $u_{i}$ and $u_{j}$ are related by $\mn{po}$, $\mn{tpp}$, or $\mn{ntpp}$.
  Again by Claim 2, $\Mmf,u_{i}\models a \wedge b$ and $\Mmf, u_{j}\models a\wedge b$. Hence
  the first two possibilities are ruled out by Formulas~(5.1) and (5.5). It follows
  that $u_{i} \mathrel{\mn{ntpp}} u_{j}$, as required.

  \noindent
  Before proceeding, let us introduce some notation. 
  \begin{itemize}
    
  \item for $i,j > 0$, we write $i \Rightarrow j$ if the tile
    position $\lambda(j)$ can be reached from $\lambda(i)$ by going
    one step to the right. Similarly, we define a relation $i \Uparrow
    j$ for going one step up;
    
  \item for $i,j > 0$ we write $r_i \rightarrow r_j$ if $u_i=r_j$.
    Similarly, we write $r_i \uparrow r_{j}$ if $r_i \rightarrow
    r_{j-1}$.  

  \end{itemize}
  Clearly, the ``$\rightarrow$'' and ``$\uparrow$'' relations are
  partial functions by Claims~1 and~2.  The following claim
  establishes some other important properties of ``$\rightarrow$'':
  first, it moves only ahead in the sequence $r_1,r_2,\dots$, but
  never back. And second, it is monotone and injective.
  \\[2mm]
  {\bf Claim~4}. Let $i,j\geq 1$. Then the following holds:
  \begin{enumerate}

  \item if $r_i \rightarrow r_j$, then $i < j$;

  \item if $i < j$, $r_i \rightarrow r_k$, and $r_j \rightarrow
    r_\ell$, then $k < \ell$;

  \end{enumerate}
  Proof: First for Point~1. Suppose $r_i \rightarrow r_j$ and $i \geq j$.
  Then $u_i=r_j$ and, by Claim~2, $r_i \mathrel{\mn{tpp}} t_i
  \mathrel{\mn{tpp}} r_j$. By the composition table, $r_i$ is related to $r_{j}$ by
  \mn{tpp} or \mn{ntpp}. But by Claim~1, $i \geq j$ implies $r_{i} \mathrel{\mn{eq}} r_{j}$ or
  $r_i \mathrel{\mn{ntppi}} r_j$. We have derived a contradiction. Hence $r_{i} \rightarrow r_{j}$
  implies $i<j$.
  
  Now for Point~2. Assume $i<j$, $r_i \rightarrow r_k$, and $r_j \rightarrow
  r_\ell$. We have $u_{i} = r_{k}$ and $u_{j}=r_{\ell}$. Hence, by Claim 3, 
  $r_{k} \mathrel{\mn{ntpp}} r_{\ell}$. Using Claim~1 and the composition table, 
  we derive $k < \ell$.
  
  \noindent
  The following claim establishes the core part of the proof: the fact
  that the ``$\rightarrow$'' relation ``coincides'' with the
  ``$\Rightarrow$'' relation, and similar for ``$\uparrow$'' and
  ``$\Uparrow$''. More precisely, this follows from Point~3 of the
  following claim. For technical reasons, we simultaneously prove some
  other, technical properties. The proof of this claim follows the
  lines of Marx and Reynolds \cite{Marx-Reynolds-99}.
  \\[2mm]
    {\bf Claim~5}. Let $i \geq 1$ and $i \Rightarrow j$. Then the
    following holds:
  \begin{enumerate}

  \item if $\lambda(j)$ is on the floor, then $\Mmf, r_j \models \mn{floor}$;
    
  \item $\Mmf,r_j \not\models \mn{wall}$;
    
    
  \item $r_i \rightarrow r_j$ and $r_i \uparrow r_{j+1}$.

  \item if $\lambda(j+1)$ is on the wall, then $\Mmf,r_{j+1} \models
    \mn{wall}$

  \end{enumerate}
  Proof: All subclaims are proved simultaneously by induction on $i$.
  First for the induction start. Then we have $i=1$ and $j=2$.
  \begin{enumerate}
    
  \item Clearly, $\lambda(2)$ is on the floor. Since $\Mmf,r_1 \models
    \varphi_\Dmc$, we have $\Mmf, r_1 \models \mn{wall}$. Thus Formula~(5.10)
    yields $\Mmf,r_2 \models \mn{floor}$.
    
  \item We have $1 \Rightarrow 2$. Point~1 gives us $\Mmf,r_2 \models
    \mn{floor}$. Since $r_1 \mathrel{\mn{ntpp}} r_2$, we also have
    $\Mmf,r_2 \not\models [\mn{ntppi}] \neg a$. Thus, Formula~(5.9)
    yields $\Mmf,r_2 \not\models \mn{wall}$.
    
  \item By Point~2, we have $\Mmf,r_2 \not\models \mn{wall}$. By
    Formula~(5.14), there are regions $r,s \in W$ such that $\Mmf,r
    \models a \wedge b$, $r \mathrel{\mn{tpp}} s$, $\Mmf,s \models c$,
    and $s \mathrel{\mn{tpp}} r_2$. By Point~5 of Claim~1, we have
    either $r=r_i$ for some $i \geq 1$ or $r_i \mathrel{\mn{ntpp}} r$
    for all $i\geq1$. In the first case, we have $r_i \rightarrow
    r_2$.  Claim~4.1 yields $i=1$ and we are done. In the second case,
    we have $r_2 \mathrel{\mn{ntpp}} r$: contradiction to $r
    \mathrel{\mn{tpp}} s$ and $s \mathrel{\mn{tpp}} r_2$. Finally,
    $r_1 \uparrow r_3$ is an immediate consequence of $r_1 \rightarrow
    r_2$ and the definition of~``$\uparrow$''.
    
  \item Since $\lambda(3)$ is on the wall, we have to show that $\Mmf,r_3
    \models \mn{wall}$. By Point~3, we have $r_1 \uparrow r_3$. Thus,
    Formula~(5.11) yields the desired result.
    
  \end{enumerate}
  Now for the induction step.
  \begin{enumerate}
    
  \item Suppose that $\lambda(j)$ is on the floor. Since obviously
    $j>1$, $\lambda(j-1)$ is on the wall. Since $i > 1$, there is a
    $k$ with $i-1 \Rightarrow k$. It is readily checked that $j-1 =
    k+1$. Thus, IH (Point~4) yields $\Mmf,r_{j-1} \models \mn{wall}$
    and we can use Formula~(5.10) to conclude that $\Mmf,r_j \models
    \mn{floor}$ as required.
    
  \item First assume that $\lambda(j)$ is on the floor. Since $j>1$, we
    have $\Mmf,r_j \not\models
    [\mn{ntppi}] \neg a$. Thus, Point~1 and Formula~(5.9) yield
    $\Mmf,r_j \not\models \mn{wall}$ as required.
    
    Now assume that $\lambda(j)$ is not on the floor. Suppose, to the
    contrary of what is to be shown, that $\Mmf, r_j \models
    \mn{wall}$.  Since $j>1$, we have $\Mmf,r_j \not\models
    [\mn{ntppi}] \neg a$.  Thus, by Formula~(5.12) we obtain $\Mmf,r_j
    \models \Diamond^D \mn{wall}$.  Since $j$ is not on the floor, $i
    \Rightarrow j$ implies $i-1 \Rightarrow j-1$. Thus, the IH
    (Point~3) yields $r_{i-1} \uparrow r_{j}$. Hence, we can use
    $\Mmf,r_j \models \Diamond^D \mn{wall}$ to derive $\Mmf,r_{i-1}
    \models \mn{wall}$. By IH (Point~2), we cannot have $m \Rightarrow
    i-1$ for any~$m$. Thus, $\lambda(i-1)$ is on the wall
    implying that $\lambda(i)$ is on the floor. We have established a
    contradiction since, with $i \Rightarrow j$, this yields that $j$
    is on the floor.
    
  \item We start with showing $r_i \rightarrow r_j$. To this end, let
    us prove that we have $r_k \rightarrow r_j$ for some $k<j$.  By
    Point~2, we have $\Mmf, r_j \not\models \mn{wall}$.  By
    Formula~(5.14), there are regions $r,s \in W$ such that $\Mmf,r
    \models a \wedge b$, $r \mathrel{\mn{tpp}} s$, $\Mmf,s \models c$,
    and $s \mathrel{\mn{tpp}} r_j$.  By Point~5 of Claim~1, we have
    either $r=r_k$ for some $k \geq 1$ or $r_n \mathrel{\mn{ntpp}} r$
    for all $n\geq 1$. In the first case, Claim~4.1 yields $k<j$ and
    we are done. In the second case, we have $r_j \mathrel{\mn{ntpp}}
    r$: contradiction to $r \mathrel{\mn{tpp}} s$ and $s
    \mathrel{\mn{tpp}} r_j$.
    
    Next, we show that $k=i$. To this end, assume that $k \neq i$. We
    distinguish two cases:
  \begin{itemize}
    
  \item $k < i$.  Let $\ell$ be such that $k \Rightarrow \ell$.  By IH
    (Point~3), we have $r_k \rightarrow r_\ell$. Due to functionality
    of ``$\rightarrow$'' (Claim~2) and since $r_k \rightarrow r_j$, we
    have $\ell=j$.  Due to the injectivity of ``$\Rightarrow$'', we
    get $k=i$, which is a contradiction.
    
  \item $i < k$. By Claim~2, we have $r_i \mathrel{\mn{ntpp}} u_i$ and
    $\Mmf,u_i \models a \wedge b$. By Point~5 of Claim~1, we have
    either (i)~$u_i=r_\ell$ for some $\ell \geq 1$ or (ii)~$r_n
    \mathrel{\mn{ntpp}} u_i$ for all $n\geq1$. In Case~(ii), in
    particular we have $r_j \mathrel{\mn{ntpp}} u_i$. Since $r_k
    \rightarrow r_j$, we have $r_j=u_k$, and thus $u_k
    \mathrel{\mn{ntpp}} u_i$.  As $i<k$, we have obtained a
    contradiction to Claim~3. Thus, Case~(ii) is impossible and we
    conclude $u_i=r_\ell$ for some $\ell \geq 1$. Next, we make a case
    distinction as follows:
    \begin{itemize}
      
    \item $\ell < j$. There are two subcases: the tile position
      $\lambda(\ell)$ may or may not be on the wall. 
      
      First assume that it is not. Then there is an $h < \ell$ with $h
      \Rightarrow \ell$. By definition of the ``$\Rightarrow$''
      function, $i \Rightarrow j$, $h \Rightarrow \ell$, and $ \ell <
      j$ this implies $h < i$.  Thus we can use IH (Point~3) to
      conclude $r_h \rightarrow r_\ell$, a contradiction to the
      injectivity of ``$\rightarrow$'' (Claim 4.2) and the facts that
      $r_i \rightarrow r_\ell$ and $h < i$.
      
      Now assume that $\lambda(\ell)$ is on the wall. Since $1 < i < \ell$,
      there is a $h$ such that $h \uparrow \ell$ and $h \rightarrow
      \ell-1$. Thus, IH (Point~4) yields $\Mmf,r_{\ell} \models \mn{wall}$.
      But then, $r_i \rightarrow r_\ell$ and Formula~(5.13) yield a
      contradiction.

    \item $\ell = j$. Then $r_i \rightarrow r_j$ and $r_k \rightarrow r_j$,
      which is a contradiction to the injectivity of ``$\rightarrow$''
      (Claim~4.2) since $i \neq k$.
      
    \item $\ell > j$. Contradiction to the monotonicity of
      ``$\rightarrow$'' (Claim~4.2).

    \end{itemize}
    
  \end{itemize}
  The second part of Point~3, i.e.\ $r_i \uparrow r_{j+1}$, is now
  an immediate consequence of the fact that $r_i \rightarrow r_j$
  and the definition of~``$\uparrow$''.
     
\item Suppose that $\lambda(j+1)$ is on the wall. Then $\lambda(i)$ is
  also on the wall. Since additionally $i > 1$, there is a $k$ such
  that $k \Uparrow i$ and $k \Rightarrow i-1$. By IH (Point~4), the
  latter yields $\Mmf,r_i \models \mn{wall}$. Since Point~3 yields
  $r_i \uparrow r_{j+1}$, Formula~(5.11) yields $\Mmf,r_{j+1} \models
  \mn{wall}$.
      
  \end{enumerate}
  This finishes the proof of Claim~5. By definition of
  ``$\Rightarrow$'', ``$\Uparrow$'', ``$\rightarrow$'', and
  ``$\uparrow$'', Point~3 of this claim yields the following:
  \begin{equation*}
    i \Rightarrow j \text{ implies } r_i \rightarrow r_j 
    \quad \text { and } \quad
    i \Uparrow j \text{ implies } r_i \uparrow r_j.
  \tag{$*$}
  \end{equation*}
  Using this property, we can finally define the solution of \Dmc: set
  $\tau(i,j)$ to the unique $t \in T$ such that $\Mmf,r_n \models p_t$,
  where $\lambda(n)=(i,j)$. This is well-defined due to Formulas~(5.15)
  and~(5.16).  Thus, it remains to check the matching conditions:
  \begin{itemize}
    
  \item Let $(i,j) \in \Nbbm^2$, $\lambda(n)=(i,j)$, and
    $\lambda(m)=(i+1,j)$.  Then $n \Rightarrow m$. By ($*$), this
    yields $r_n \rightarrow r_m$. By Formula~(5.16), there are $(t,t')
    \in H$ such that $\Mmf,r_n \models p_t$ and $\Mmf,r_m \models p_{t'}$.
    Since this implies $\tau(i,j)=t$ and $\tau(i+1,j)=t'$, the
    horizontal matching condition is satisfied.
    
  \item The vertical matching condition can be verified analogously
    using Formula~(5.17).

  \end{itemize}
\end{Proof}
The second lemma deals with the construction of models for $\vp_\Dmc$
from solutions for \Dmc. Here, we have to make a suitable assumption
on the class of region structures \Smc for the construction to succeed.
One possible such assumption is given in Theorem~\ref{theo:undec}. It
turns out, however, that the following more general condition is also
sufficient.
\begin{definition}[Domino ready]
\label{domready}
Let $\Rmf= \auf W , \mn{dc}, \mn{ec},\dots \zu$ be a region
structure. Then \Rmf is called \emph{domino ready} if it satisfies the
following property: the set $W$ contains sequences $x_1,x_2,\dots$ and
$y_1,y_2,\dots$ such that, for $i,j \geq 1$, we have
    \begin{enumerate}

    \item $x_i \mathrel{\mn{tpp}} x_{i+1}$;
      
    \item $x_i \mathrel{\mn{ntpp}} x_{j}$ if $j>i+1$;

    \item $x_{2i-1} \mathrel{\mn{tpp}} y_i$;
      
    \item $y_i \mathrel{\mn{tpp}} x_{2j-1}$ iff the position
      $\lambda(j)$ can be reached from $\lambda(i)$ by going one step
      to the right;

    \item $y_i \mathrel{\mn{ntpp}} y_j$ if $j>i$.

    \end{enumerate}
\end{definition}
Before discussing this property in some more detail, let us show that
it is indeed suitable for our proof.
\begin{lemma}
\label{corr2}
Let $\Rmf= \auf W , \mn{dc}, \mn{ec},\dots \zu$ be a region structure
that is domino ready.  If the domino system \Dmc has a solution, then
the formula $\varphi_\Dmc$ is satisfiable in a region model based on
\Rmf.
\end{lemma}
\begin{Proof}
  Let \Rmf be a region structure that is domino ready, $\Dmc=(T,H,V)$
  a domino system, and $\tau$ a solution of \Dmc.  We introduce new
  names for the regions listed in Definition~\ref{domready} that are
  closer to the names used in the proof of Lemma~\ref{corr1}:
  \begin{itemize}

  \item $r_i:=x_{2i-1}$ for $i \geq 1$;

  \item $s_i:=x_{2i}$ for $i \geq 1$;

  \item $t_i:=y_i$.

  \end{itemize}
  Now define a region model $\Mmf$ based on \Rmf by interpreting the
  propositional letters as follows:
  \begin{itemize}

  \item $a^\Mmf=\{r_i,s_i \mid i \geq 1\}$;

  \item $b^\Mmf=\{r_i \mid i \geq 1\}$;

  \item $c^\Mmf=\{t_i \mid i \geq 1\}$;

  \item $\mn{wall}^\Mmf = \{ r_i \mid \lambda(i) \mbox{ is on the wall} \}$;

  \item $\mn{floor}^\Mmf = \{ r_i \mid \lambda(i) \mbox{ is on the floor} \}$;

  \item $p_t^\Mmf = \{ r_i \mid \tau(\lambda(i))=t \}$.

  \end{itemize}
  It is now easy to verify that $\chi$ is satisfied by every region
  of \Mmf, and that $\Mmf,r_1 \models \varphi_\Dmc$.
\end{Proof}
We have thus proved the following theorem.
\begin{theorem}
  \label{theo:undecplus}
  Let ${\mathcal S}\subseteq \mathcal{RS}$ such that some $\Rmf \in \Smc$ is
  domino ready.  Then $L_{\mn{RCC8}}({\mathcal S})$ is undecidable.
\end{theorem}

We now show that this theorem implies Theorem~\ref{theo:undec}.

\begin{lemma}
\label{corr3}
Each region structure $\Rmf(\Rbbm^n, U)$ with $n > 0$ and
$\Rbbm^n_{\mn{rect}} \subseteq U$ is domino ready.
\end{lemma}
\begin{Proof}
  We start with $n=1$. Thus, we must exhibit the existence of two
  sequences of convex, closed intervals $x_1,x_2,\dots$ and
  $y_1,y_2,\dots$ satisfying Properties~1 to~5 from
  Definition~\ref{domready}: for $i \geq 1$, set
  \begin{itemize}

  \item $x_i := [-j,j]$ if $i=2j-1$;

  \item $x_i := [-j,j+1]$ if $i=2j$;
    
  \item $y_i := [-i,j]$ if $\lambda(j)$ is the position reached from
    $\lambda(i)$ by going a single step to the right.

  \end{itemize}
  It is readily checked that these sequences of intervals are as required.
  To find sequences for $n>1$, just use the $n$-dimensional products of
  these intervals.
\end{Proof}
Note that we can also prove this lemma if we admit only \emph{bounded}
rectangles of $\Rbbm^n$ as regions: the construction from
Lemma~\ref{corr3} can easily be modified so that the sequence of $a
\wedge b$-rectangles converges against a finite rectangle, rather than
against $\Rbbm^n$.

indeed more general than Theorem~\ref{theo:undec}. For example, region
structures that are obtained by choosing all closed circles or ellipses as regions 
are easily seen to be domino
ready, but they do not satisfy the condition from Theorem~\ref{theo:undec}.

\section{Recursive Enumerability}
\label{sect:notre}

In this section, we discuss the question whether modal logics of
topological relations are recursively enumerable. We start with a
simple observation.
\begin{theorem} 
\label{repos1}
For $n>0$, $L_{\mn{RCC8}}(\mathcal{RS}) = L^\Ssf_{\mn{RCC8}}(\mathcal{TOP}) =
L_{\mn{RCC8}}^{\mn{S}}(\Rbbm^{n},\Rbbm^{n}_{\mn{reg}})$ are
recursively enumerable.
\end{theorem}
\begin{Proof}
  The equality has already been shown in Theorem~\ref{equality1}.
  $L_{\mn{RCC8}}(\mathcal{RS})$ is recursively enumerable since (i)~the
  class of all region structures ${\mathcal RS}$ is first-order definable
  (c.f.\ its definition in Section~\ref{sect:structures}); (ii) it is
  a standard result that ${\mathcal L}_{\mn{RCC8}}$ formulas can be
  translated into equivalent formulas of $\mathcal{FO}^{m}_{\mn{RCC8}}$
  (see Section~\ref{sect:logics}); (iii)~first-order logic is
  recursively enumerable.
\end{Proof}
An alternative proof of Theorem~\ref{repos1} can be obtained by
explicitly giving an axiomatization of $L_{\mn{RCC8}}(\mathcal{RS})$.
Since this is interesting in its own right, in the following we
develop such an axiomatization based on a non-standard rule. Non-standard rules, which are
sometimes called non-orthodox or Gabbay-Burgess style rules, were introduced
in temporal logic in \cite{Burgess80,Gabbay81} and often enable finite axiomatizations
of modal logics for which no finite standard axiomatization 
(using only the rules modus ponens and necessitation) is known. 
For $L_{\mn{RCC8}}(\mathcal{RS})$, we leave it as an open problem whether
a finite standard axiomatization exists. To
guarantee a simple presentation, we develop an axiomatization for
the extension of our language ${\mathcal L}_{\mn{RCC8}}$ with countably
many nominals, i.e.\ a new sort of variables $i,j,k,\ldots$
interpreted in singleton sets. As noted in
Section~\ref{sect:languages}, nominals can be defined in the original
language, but here it is more convenient to treat them as first-class
citizens since this enables the application of general completeness
results from modal logic.\footnote{One could also give a finite non-standard axiomatization without
adding nominals to the language by making use of the definable difference modality $\Box_{d}$ 
and then applying a general completeness result of \cite{Venema92} (Theorem 2.7.7).}
The universal box $\Box_{u}$ is still used
as an abbreviation. Then the logic of all region structures is
axiomatized by the following axiom and rule schemata, where $\varphi$
and $\psi$ range over formulas of ${\mathcal L}_{\mn{RCC8}}$ extended with
nominals, $i$ over the nominals, and $\mn{r}$, $\mn{r}_{1}$,
$\mn{r}_{2}$ over the $\mn{RCC8}$-relations:
\begin{itemize}

\item axioms of propositional logic;
  
\item $[\mn{r}](\varphi \rightarrow \psi) \rightarrow ([\mn{r}]\varphi
  \rightarrow [\mn{r}]\psi)$;
  
\item $\auf\mn{r}_{1}\zu i \rightarrow \neg \auf\mn{r}_{2}\zu i$, for
  $\mn{r}_{1}\not=\mn{r}_{2}$. These axioms ensure that the $\mn{r}$
  are mutually disjoint;
  
\item $\auf\mn{r}_{1}\zu\auf\mn{r}_{2}\zu\varphi \rightarrow
  \auf\mn{q}_{1}\zu\varphi \vee \cdots \vee \auf\mn{q}_{k}\zu\varphi$,
  whenever
$$
\!\!\!\!\!\forall x\forall y \forall z ((\mn{r}_{1}(x,y) \wedge \mn{r}_{2}(y,z)) \rightarrow 
(\mn{q}_{1}(x,z) \vee \cdots \vee \mn{q}_{k}(x,z))
$$
is in the $\mn{RCC8}$-composition table;

\item $\varphi \rightarrow [\mn{r}]\auf \mn{r}\zu\varphi$, whenever
  $\mn{r}$ is symmetric;
  
\item $\varphi \rightarrow [\mn{r}_{1}]\auf \mn{r}_{2}\zu \varphi$ and
  $\varphi \rightarrow [\mn{r}_{2}]\auf \mn{r}_{1}\zu \varphi$,
  whenever $\mn{r}_{1}$ is the inverse of $\mn{r}_{2}$;
  
\item $\Box_{u}\varphi \rightarrow \varphi$, $\Box_{u}\varphi
  \rightarrow \Box_{u}\Box_{u}\varphi$, and $\varphi \rightarrow
  \Box_{u}\Diamond_{u} \varphi$. These axioms ensure that $\Box_{u}$
  is a ${\bf S5}$-modality;

\item $[\mn{eq}]\varphi \leftrightarrow \varphi$;
  
\item $\Diamond_{u}i$. This axiom ensures that the interpretation of
  nominals is non-empty;

\item $\Diamond_{u}(i\wedge \varphi) \rightarrow \Box_{u}(i \rightarrow \varphi)$. This axiom together with
the rule cov below ensures that the interpretation of nominals are at most singleton sets;

\item the rules modus ponens, necessitation, and the non-standard rule cov:
  $$
   \frac{\varphi, \varphi \rightarrow
    \psi}{\psi}
  \qquad
  \frac{\varphi}{\Box_{u}\varphi}
  \qquad
  \frac{i \rightarrow \varphi}{\varphi} \text{ if $i$ not in
  $\varphi$.}
  $$
\end{itemize}
It is straightfoward to prove the soundness of this axiomatization.
Completeness follows from a general completeness result of
\cite{GorankoVakarelov} for logics with nominals and the universal
modality, since all the axioms not involving nominals are Sahlqvist
axioms, and, for each modal operator $[\mn{r}]$, we have an operator
$[\mn{r}^{-1}]$ interpreted by the converse of the accessibility
relation for $[\mn{r}]$.


Returning to our original proof of Theorem~\ref{repos1}, we note that
there is another class of logics for which recursive enumerability can
be proved using first-order logic:
$L_{\mn{RCC8}}^{\mn{S}}(\Rbbm^{n},\Rbbm^{n}_{\mn{rect}})$, $n \geq 1$.
In this case, however, we need a different translation that takes into
account the underlying region structures and the shape of regions.
The proof is similar to the translation of interval temporal logic
into first-order logic given by Halpern and Shoham in
\cite{HalpernShoham91}. The important difference is that Halpern and
Shoham use their translation to prove recursive enumerability of
interval temporal logics determined by \emph{full} interval
structures that are first-order definable, whereas we prove recursive
enumerability of a logic determined by substructures of a structure
that is not first-order definable. The proof can be found in
Appendix~\ref{app:rectanglesre}.
\begin{theorem}\label{repos2}
For $n \geq 1$, $L_{\mn{RCC8}}^{\mn{S}}(\Rbbm^{n},\Rbbm^{n}_{\mn{rect}})$
is recursively enumerable.
\end{theorem}
With the exception of the class of logics
$L_{\mn{RCC8}}^{\mn{S}}(\Rbbm^{n},\Rbbm^{n}_{\mn{conv}})$, whose
recursive enumerability status we have to leave as an open problem, it
thus turns out that all logics introduced in Section~\ref{sect:logics}
that are based on \emph{substructures} of concrete region structures
are recursively enumerable.\footnote{Recall that concrete region
  structures are those region structures induced by topological
  spaces.}  Interestingly, this is not the case for logics based on
full concrete region structures, and thus going from full concrete
region structures to substructures yields a computational benefit. In
the following, we prove that most of the logics introduced in
Section~\ref{sect:logics} based on full concrete region structures
are $\Pi^1_1$-hard, and thus not recursively enumerable.  Note,
however, that the conditions listed in the theorem are much less
general than those from Theorem~\ref{theo:undecplus}.
\begin{theorem}\label{nonax}
  The following logics are $\Pi^1_1$-hard: $L_{\mn{RCC8}}(\mathcal{TOP})$
  and $L_{\mn{RCC8}}(\Rbbm^{n},U_{n})$ with $U_{n}\in
  \{\Rbbm^{n}_{\mn{reg}},\Rbbm^{n}_{\mn{conv}}\}$
  and $n\geq 1$.
\end{theorem}
To prove Theorem~\ref{nonax}, the domino problem of
Definition~\ref{domsys} is modified by requiring that, in solutions, a
distinguished tile $t_0 \in T$ occurs infinitely often in the first
column of the first quadrant, i.e.\ on the wall. It has been shown in
\cite{Harel-85} that this variant of the domino problem is
$\Sigma^1_1$-hard. Since we reduce it to satisfiability as in the
proof of Theorem~\ref{theo:undec}, this yields a $\Pi^1_1$-hardness
bound for validity.

\noindent
As a first step toward reducing this stronger variant of the domino
problem, we extend $\varphi_\Dmc$ with the following conjunct stating
that $\Mmf,s \models \vp_\Dmc$ implies that we find an infinite
sequence of regions $r_1,r_2,\dots$ such that $s=r_1$, $r_i
\mathrel{\mn{ntpp}} r_{i+1}$, and $\Mmf,r_i \models a \wedge b \wedge
\mn{wall} \wedge p_{t_0}$ for all $i \geq 1$:
\begin{equation}
\Box_u (a \wedge b \rightarrow  \auf \mn{ntpp} \zu (a \wedge b \wedge \mn{wall} \wedge p_{t_0}))
\end{equation}
However, this is not yet sufficient: in models of $\varphi_\Dmc$, we
can have not only one discrete ordering of $a \wedge b$ regions, but
rather many such orderings that are ``stacked''. For example, there
could be two sequences of regions $r_1, r_2, \dots$, and
$r'_1,r'_2,\dots$ such that 
$$
r_1 \mathrel{\mn{ntpp}} r_2
\mathrel{\mn{ntpp}} r_3 \cdots, \quad r'_1 \mathrel{\mn{ntpp}} r'_2
\mathrel{\mn{ntpp}} r'_3 \cdots, \quad \text{ and } r_i \mathrel{\mn{ntpp}} r'_j
\text{ for all } i,j \geq 1.
$$
Due to this effect, the above formula does not enforce that the main
ordering (there is only one for which we can ensure a proper ``going
to the right relation'') has infinitely many occurrences of $t_0$.

The obvious solution to this problem is to prevent stacked orderings.
This is done by enforcing that there is only one ``limit region'',
i.e.\ only one region approached by an infinite sequence of
$a$-regions in the limit. We add the following formula
to~$\varphi_\Dmc$:
\begin{eqnarray}
  \Box_u \big ( [\mn{tppi}] \auf\mn{po}\zu a \rightarrow ( \neg a \wedge [\mn{tpp}] \neg a \wedge [\mn{ntpp}] \neg a) \big )
\end{eqnarray}
Let $\varphi'_\Dmc$ be the resulting extension of $\varphi_\Dmc$. The
classes of region structures to which the extended reduction applies
is more restricted than for the original one. We require that they are
concrete, i.e.\ induced by a topological space, and additionally adopt
the following property:
\begin{definition}[Closed under infinite unions]
  Suppose that $\Rmf=\Rmf(\Tmf,U_\Tmf)= \auf W , \mn{dc},
  \mn{ec},\dots \zu$ is a concrete region structure.  Then \Rmf is
  \emph{closed under infinite unions} if, for any sequence
  $r_{1},r_{2},\ldots \in W$ with $r_{1}\; \mn{ntpp}\; r_{2}\;
  \mn{ntpp}\; r_{3} \; \cdots$, we have
  \mbox{$\mathbb{C}\mathbb{I}(\bigcup_{i\in \omega} r_{i}) \in W$}.
\end{definition}
We can now formulate the first part of correctness for the extended
reduction. 
\begin{lemma}
\label{notre1}
Let $\Rmf(\Tmf,U_\Tmf)=\auf W , \mn{dc}, \mn{ec},\dots \zu$ be a
concrete region structure that is closed under infinite unions.  If
the formula $\varphi'_\Dmc$ is satisfiable in a region model based on
\Rmf, then the domino system \Dmc has a solution with $t_0$ occurring
infinitely often on the wall.
\end{lemma}
\begin{Proof}
  Let $\Rmf(\Tmf,U_\Tmf)=\auf W , \mn{dc}, \mn{ec},\dots \zu$ be a
  concrete region structure that is closed under infinite unions,
  $\Mmf=\auf \Rmf,p_1^\Mmf,p_2^\Mmf,\dots\zu$ a region model based on
  $\Rmf(\Tmf,U_\Tmf)$, and $w \in W$ such that $\Mmf,w \models
  \varphi'_\Dmc$. We may establish Claims~1 to~5 as in the proof of
  Lemma~\ref{corr1}, and we will use the same terminology in what
  follows. We first strengthen Point~5 of Claim~1:
  \\[2mm]
  {\bf Claim~1'}. There exists a sequence $r_1,r_2,\dots \in W$ such
  that
  \begin{enumerate}

  \item $\Mmf,r_1 \models \varphi_\Dmc$,

  \item $r_1 \mathrel{\mn{ntpp}} r_2 \mathrel{\mn{ntpp}} r_3
  \mathrel{\mn{ntpp}} \cdots$,
  
  \item $\Mmf,r_i \models a \wedge b$ for $i \geq 1$.
    
  \item for each $i \geq 1$, there exists a region $s_i \in
    W$ such that
    \begin{enumerate}

    \item $r_i \mathrel{\mn{tpp}} s_i$,

    \item $\Mmf,s_i \models a \wedge \neg b$,

    \item $s_i \mathrel{\mn{tpp}} r_{i+1}$,

    \item for each region $s$ with $r_i \mathrel{\mn{tpp}} s$ and
      $\Mmf,s \models a \wedge \neg b$, we have $s=s_i$, and

    \item for each region $r$ with $s_i \mathrel{\mn{tpp}} r$ and
      $\Mmf,r \models a \wedge b$, we have $r=r_{i+1}$, 

    \end{enumerate}
    
  \item[(5')] for all $r \in W$ with $\Mmf,r \models a \wedge b$, we
    have $r=r_i$ for some $i \geq 1$.

  \end{enumerate}
  Proof: We construct the sequence $r_1,r_2,\dots$ as in the proof of Claim~1.
  Since Properties~1 to~4 are satisfied by construction, it remains
  to prove Point~5': as $\Rmf(\Tmf,U_\Tmf)$ is closed under
  infinite unions, we have
  $
    t= \mathbb{C}\mathbb{I}(\bigcup_{i\in\omega} r_{i}) \in W.
  $
  We first show that 
  \begin{equation*}
    t\models [\mn{tppi}] \big \auf\mn{po}\zu a \big.
    \tag{$*$}
  \end{equation*}
  To this end, suppose $t \mathrel{\mn{tppi}} q$. Then we have the
  following:
\begin{enumerate}

\item $q-r_{i} \not=\emptyset$ for all $i>0$.
  
  Since $t \mathrel{\mn{tppi}} q$, there exists $x\in q$ such that
  $x\not\in \mathbb{I}(t)$. Suppose $x\in r_{i}$, for some $r_{i}$.
  Since $r_i \mathrel{\mn{ntpp}} r_{i+1}$, this yields $x \in
  \mathbb{I}(r_{i+1})$.  By definition of $t$, we get $x\in
  \mathbb{I}(t)$ and have a contradiction.
  
\item There exists $n>0$ such that $i\geq n$ implies
  $r_{i}-q \not=\emptyset$.
  
  Suppose $r_{i}\subseteq q$, for all $i>0$. Then $s=\bigcup_{i\in
    \omega}r_{i} \subseteq q$. Since $q \in U_\Tmf$, we have
  $q=\mathbb{CI}(q)$. Thus $t=\mathbb{CI}(s) \subseteq q$, and we have
  a contradiction to $t \mathrel{\mn{tppi}} q$.
  
\item There exists $m>0$ such that $j \geq m$ implies \mbox{$\mathbb{I}(r_{j})
    \cap \mathbb{I}(q)\not=\emptyset$}.
  
  Since $q=\mathbb{CI}(q)$, we have $\mathbb{I}(q) \neq \emptyset$.
  Take any $x\in \mathbb{I}(q)$.  Since $t=
  \mathbb{C}\mathbb{I}(\bigcup_{i\in\omega} r_{i})$ and $t
  \mathrel{\mn{tppi}} q$, this yields $x \in \bigcup_{i\in
    \omega}r_{i}$. Thus there is a $j$ with $x\in r_{j}$. Then $x\in
  \mathbb{I}(r_{j+1})$. Set $m:= j+1$. Since $r_m \mathrel{\mn{ntpp}}
  r_i$ for all $i > m$, we have $x\in \mathbb{I}(q)\cap
  \mathbb{I}(r_{j+1})$ for all $i \geq m$.
\end{enumerate}
Take $k=\max\{n,m\}$. Using the above Points~1 to~3 and the definition
of the \mn{po} relation, it is easily verified that $q\; \mn{po}\;
r_{k}$, thus finishing the proof of ($*$).

\noindent
Now we can establish Point~5'. By Point~5 of the original Claim~1, for
all $r \in W$ with $\Mmf,r \models a \wedge b$, we have that $r=r_i$
for some $i\geq1$ or $r_i \mathrel{\mn{ntpp}} r$ for all $i \geq 1$.
It thus suffices to show that the latter alternative yields a
contradiction. Thus assume $r_i \mathrel{\mn{ntpp}} r$ for all $i \geq
1$. Since $r_1 \mathrel{\mn{ntpp}} r_2 \mathrel{\mn{ntpp}} \cdots$ and
$t= \mathbb{C}\mathbb{I}(\bigcup_{i\in\omega} r_{i})$, it is not hard
to verify that this yields $r=t$, $t \mathrel{\mn{tpp}} r$, or $t
\mathrel{\mn{ntpp}} r$. By ($*$), $t$ satisfies
$[\mn{tppi}]\auf\mn{po}\zu a$. By Formula~(6.2), $t$ thus also
satisfies $\neg a \wedge [\mn{tpp}] \neg a \wedge [\mn{ntpp}] \neg a$:
contradiction since $\Mmf,r \models a$.

Lemma~\ref{corr1}. By Point~5' of Claim~1' and Formula~(6.1), this
solution is such that the tile $t_0$ occurs infinitely often on the
wall.
\end{Proof}
For the second part of correctness, we consider region structures
${\mathfrak R}(\Rbbm^n,U)$ with $\Rbbm^n_\mn{rect} \subseteq U$ as
in Theorem~\ref{theo:undec}. In contrast to the previous section,
it does not suffice to demand that region structures are domino
ready. 
\begin{lemma}
\label{notre2}
If the domino system \Dmc has a solution with $t_0$ occurring
infinitely often on the wall, then the formula $\varphi'_\Dmc$ is
satisfiable in a region model based on
${\mathfrak R}(\Rbbm^n,U)$, for each $n \geq 1$ and each $U$
with $\Rbbm^n_{\mn{rect}} \subseteq U \subseteq \Rbbm^n_{\mn{reg}}$.
\end{lemma}
\begin{Proof}
  Let $\tau$ be a solution of \Dmc with $t_0$ appearing infinitely
  often on the wall. It was shown in the proof of Lemma~\ref{corr3}
  that the region spaces we are considering are domino ready. Thus we
  can use $\tau$ to construct a model \Mmf based on the region space
  $\Rmf(\Rbbm^n,U)$ exactly as in the proof of Lemma~\ref{corr2}. It
  suffices to show that \Mmf satisfies, additionally, Formulas~(6.1)
  and~(6.2). This is easy for Formula~(6.1) since $\tau$ has been chosen
  such that $t_0$ appears infinitely often. Thus, let us concentrate
  on Formula~(6.2).
  
  Let $r_1,r_2,\dots$ be the regions from the construction of \Mmf in
  the proof of Lemma~\ref{corr2}. If
  $$
    t= \mathbb{C}\mathbb{I}(\bigcup_{i\in\omega} r_{i}) =\Rbbm^n \in W,
  $$
  then $t$ satisfies $\neg a \wedge [\mn{tpp}] \neg a \wedge
  [\mn{ntpp}] \neg a$ since, clearly, $t$ is not related via \mn{eq},
  \mn{tpp}, and \mn{ntpp} to any of the $r_i$. To show that
  Formula~(6.2) holds, it thus suffices to prove that, for all $s \in
  W$ such that $s\neq t$, $\Mmf,s \models \neg [\mn{tppi}] \big
  \auf\mn{po}\zu a$.  Hence fix an $s \in W$ and assume that $s \neq
  t$. Since it is a region, $s$ is non-empty and regular closed.
  Therefore, we find a hyper-rectangle $h \in \Rbbm^n_\mn{rect}$
  contained in $s$.  By expanding $h$ until we hit a
  point $x \in s- \mathbb{I}s$, we obtain an $h' \in
  \Rbbm^n_\mn{rect}$ such that $h \subseteq h'$ and $h'$ is a
  tangential proper part of $s$. Now fix an $x \in h' \cap
  (s-\mathbb{I}s)$.  Then, by the construction of the sequence
  $r_{1},r_{2},\dots$, we can find a hyper-rectangle $h'' \subseteq
  h'$ which contains $x$ but is not in the relation $\mn{po}$ with any
  $r_{i}$. In conclusion, $\Mmf, h'' \models [\mn{po}] \neg a$ and,
  therefore, $\Mmf, s \models \auf \mn{tppi}\zu [\mn{po}] \neg a$.
\end{Proof}
\bigskip

\noindent
Note that any region structure ${\mathfrak R}({\mathfrak T},{\mathfrak T}_{\mn{reg}})$,
in particular the structures ${\mathfrak R}(\Rbbm^n,\Rbbm^n_\mn{reg})$, are closed under infinite unions.
This applies as well to
${\mathfrak R}(\Rbbm^{n},\Rbbm^{n}_{\mn{conv}})$.
Since $\Rbbm^n_\mn{rect} \subseteq
\Rbbm^n_\mn{conv} \subseteq \Rbbm^n_\mn{reg}$, Lemmas~\ref{notre1}
and~\ref{notre2} immediately yield Theorem~\ref{nonax}.

It is worth noting that there are a number of interesting region
structures to which this proof method does not apply.  Interesting
examples are the region structure of hyper-rectangles in $\Rbbm^{n}$, $n\geq 2$,
the region structure based on simply connected regions in
$\Rbbm^{2}$ \cite{Schaefer01}, and
the structure of polygons in $\Rbbm^{2}$ \cite{PrattSchoop98}. 
Since these spaces are not closed under infinite
unions, the above proof does not show the non-axiomatizability of the
induced logics.  We believe, however, that slight modifications of
the proof introduced here can be used to prove their
$\Pi^1_1$-hardness as well.
%
%
%
%
%

\section{Finite Region Structures}
\label{sect:finite}

As discussed in Section~\ref{sect:logics}, it can be useful to only
admit models with a finite (but unbounded) number of regions. In this
case, we can again establish a quite general undecidability result.
Moreover, undecidability of a logic $L^\mn{fin}_\mn{RCC8}(\Smc)$
implies that it is not recursively enumerable if \Smc is first-order
definable. We start with proving undecidability.
\begin{theorem}\label{theo:undecfin}
  If ${\mathfrak R}(\Rbbm^{n},\Rbbm^{n}_{\mn{rect}}) \subseteq {\mathcal
    S} \subseteq \mathcal{RS}$ for some $n\geq 1$,
  then $L_{\mn{RCC8}}^\mn{fin}({\mathcal S})$ is undecidable.
\end{theorem}
We obtain the following corollary.
\begin{corollary}
  The following logics are undecidable for $n\geq 1$:
  $L_{\mn{RCC8}}^{\mn{fin}}(\mathcal{RS})$,
  $L_{\mn{RCC8}}^{\mn{fin}}(\mathcal{TOP})$,
  $L_{\mn{RCC8}}^{\mn{fin}}(\Rbbm^{n},\Rbbm^{n}_{\mn{reg}})$,
  $L_{\mn{RCC8}}^{\mn{fin}}(\Rbbm^{n},\Rbbm^{n}_{\mn{conv}})$, and
  $L_{\mn{RCC8}}^{\mn{fin}}(\Rbbm^{n},\Rbbm^{n}_{\mn{rect}})$.
\end{corollary}
To prove this result, we reduce yet another variant of the domino
problem. For $k \in \Nbbm$, the \emph{$k$-triangle} is the set $\{
(i,j) \mid i+j \leq k\} \subseteq \Nbbm^2$. The task of the new domino
problem is, given a domino system $\Dmc=(T,H,V)$, to determine whether
\Dmc tiles an arbitrary $k$-triangle, $k \in \Nbbm$, such that the
position $(0,0)$ is occupied with a distinguished tile $s_0 \in T$,
and some position is occupied with a distinguished tile $f_0 \in T$.
It is shown in Appendix~\ref{dominoundec} that the existence of such a
tiling is undecidable.


Given a domino system \Dmc, the reduction formula $\varphi_\Dmc$ is defined as
\begin{eqnarray*}
a \wedge b \wedge \mn{wall} \wedge \mn{floor} \wedge s_0 \wedge [\mn{ntppi}]
\neg a 
\wedge
\Box_u \chi 
\wedge (f_0 \vee \auf \mn{ntpp}\zu (a \wedge b \wedge f_0)),
\end{eqnarray*}
where $\chi$ is the conjunction of the Formulas~(5.1), (5.3) to~(5.5), and
(5.7) to~(5.17) of Section~\ref{sect:rcc8undec}, and the following
formulas:
\begin{itemize}

\item The first tile that has no tile to the right is on the floor:
  \begin{eqnarray}
    \hspace*{-7mm}\big ( a \wedge b \wedge \neg \Diamond^R \top \wedge [\mn{ntppi}] ((a \wedge b) \rightarrow
    \Diamond^R \top) \big ) \rightarrow \mn{floor}
  \end{eqnarray}
  
\item If a tile has no tile to the right, then the next tile (if
  existent) also has no tile to the right:
  \begin{eqnarray}
    (a \wedge b \wedge \neg \Diamond^R \top) \rightarrow (\neg \Diamond^+ \top \vee
    \Diamond^+ \neg \Diamond^R \top)
  \end{eqnarray}

\item The last tile is on the wall and we have no stacked orderings:
  \begin{equation}
    (a \wedge b \wedge \neg \Diamond^+ \top) \rightarrow (\mn{wall} \wedge [\mn{ntpp}] \neg (a \wedge b)) 
  \end{equation}

\end{itemize}
The proof of the following lemma is now a variation of the proofs of
Lemma~\ref{corr1} and Lemma~\ref{corr2}. Details are left to the reader.
\begin{lemma}
\label{corrfin}
Let \Dmc be a domino system.  Then:

  \noindent
  (i) if the formula $\varphi_\Dmc$ is satisfiable in a finite region
    model, then \Dmc tiles a $k$-triangle for some $k \geq 1$;
    
  \noindent
  (ii) if \Dmc tiles a $k$-triangle for some $k \geq 1$, then
  $\varphi_\Dmc$ is satisfiable in a region model based on a finite
  substructure of $\Rmf(\Rbbm^n,\Rbbm^n_{\mn{rect}})$, for each $n
  \geq 1$.

\end{lemma}
Obviously, Theorem~\ref{theo:undecfin} is an immediate consequence of
Lemma~\ref{corrfin}. 


Since $\mathcal{RS}$ is first-order definable, we can enumerate all
finite region models and also all formulas satisfiable in finite
region models. Similarly, the proof of Theorem~\ref{repos2} shows that
the class of at most countable substructures of ${\mathfrak
  R}(\Rbbm^{n},\Rbbm^{n}_{\mn{rect}})$ is first-order definable
(relative to the class of all at most countable structures), for
$n\geq 1$. Thus, the complements of $L_{\mn{RCC8}}^{\mn{fin}}(\mathcal{RS})$ and
$L_{\mn{RCC8}}^{\mn{fin}}(\Rbbm^{n},\Rbbm^{n}_{\mn{rect}})$ are
recursively enumerable and Theorem~\ref{theo:undecfin} and
Theorem~\ref{equality1} give us the following:
\begin{corollary}
  The following logics are not r.e., for each $n\geq 1$:
  $L_{\mn{RCC8}}^{\mn{fin}}(\mathcal{RS})$,
  $L_{\mn{RCC8}}^{\mn{fin}}(\mathcal{TOP})$,
  $L_{\mn{RCC8}}^{\mn{fin}}(\Rbbm^{n},\Rbbm^{n}_{\mn{reg}})$, and
  $L_{\mn{RCC8}}^{\mn{fin}}(\Rbbm^{n},\Rbbm^{n}_{\mn{rect}})$.
\end{corollary}
%
%
We leave it as an open problem whether the logics
$L_{\mn{RCC8}}^{\mn{fin}}(\Rbbm^{n},\Rbbm^{n}_{\mn{conv}})$, $n\geq 2$,
are recursively enumerable.

\section{The \mn{RCC5} set of Relations}
\label{sect:rcc5}

When selecting a set of relations between regions in topological
spaces, the eight Egen\-hofer-Franzosa relations appear to be the
most popular choice in the spatial reasoning community. However, it is
not the only choice possible.  For example, a refinement of \mn{RCC8}
into 23 relations has been proposed and \mn{RCC5}, a coarsening
into five relations, is also rather popular \cite{Papa95,dwm_rcc,Bennett94a,CohnHazarika}.
Since we have shown that modal logics based on the Egenhofer-Franzosa
relations are undecidable and often even $\Pi^1_1$-complete, a natural
next step for improving the computational behaviour is to consider
modal logics based on a coarser set of relations.  In this section, we
define and investigate modal logics based on the \mn{RCC5} set of
relations.  It turns our that often reasoning is still undecidable, although
different proof methods have to be used that yield less general
theorems.  For example, the recursive enumerability of modal logics determined by
full concrete \mn{RCC5} region structures is left as an open problem.

The \mn{RCC5} set of relations is obtained from \mn{RCC8} by keeping
the relations $\mn{eq}$ and $\mn{po}$, but coarsening (1)~the \mn{tpp}
and \mn{ntpp} relations into a new ``proper-part of'' relation
\mn{pp}; (2) the \mn{tppi} and \mn{ntppi} relations into a new ``has
proper-part'' relation \mn{ppi}; and (3) the \mn{dc} and \mn{ec}
relations into a new disjointness relation \mn{dr}.  Thus, a
\emph{concrete} $\mn{RCC5}$-{\em structure} ${\mathfrak
  R}^5(\Tmf,U_\Tmf)$ induced by a topological space \Tmf and a set of
regions $U_{\mathfrak T} \subseteq \Tmf_\mn{reg}$ is the tuple $\auf
U_{\mathfrak T},\mn{eq}^{\mathfrak R},\mn{po}^{\mathfrak R},
\mn{dr}^{\mathfrak R},\mn{pp}^{\mathfrak R},\mn{ppi}^{\mathfrak R}\zu$
where $\mn{eq}$ and $\mn{po}$ are interpreted as before and
\begin{itemize}
  
\item $\mn{dr}^{\mathfrak R}= \mn{dc}^{\mathfrak R} \cup
  \mn{ec}^{\mathfrak R}$;
  
\item $\mn{pp}^{\mathfrak R}= \mn{tpp}^{\mathfrak R} \cup
  \mn{nttp}^{\mathfrak R}$;
  
\item $\mn{ppi}^{\mathfrak R} = \mn{tppi}^{\mathfrak R} \cup
  \mn{nttpi}^{\mathfrak R}$.

\end{itemize}
It is interesting to note that the \mn{RCC5} relations can be defined
without appealing to the topological notions of interior and closure.
Hence, modal logics based on \mn{RCC5} may also be viewed as modal
logics determined by the following relations between sets:  
`having non-empty intersection', `being disjoint', and `is a subset of'. 
They are thus related to the logics considered in \cite{Vakarelov95a}.

Similarly to concrete region structures induced by the eight
Egenhofer-Franzosa relations, the class of concrete
$\mn{RCC5}$-structures can be characterized by first-order sentences.
Denote by $\mathcal{RS}^{5}$ the class of all \emph{general}
$\mn{RCC5}$-\emph{structures}
$$
\auf W,\mn{dr}^{\mathfrak{R}},\mn{eq}^{\mathfrak{R}},\mn{pp}^{\mathfrak{R}},
\mn{ppi}^{\mathfrak{R}},\mn{po}^{\mathfrak R}\zu
$$
where $W$ is non-empty and the $\mn{r}^{\mathfrak R}$ are mutually
exclusive and jointly exhaustive binary relations on $W$ such that
(1)~$\mn{eq}$ is interpreted as the identity relation on $W$,
(2)~$\mn{po}^{\mathfrak R}$ and $\mn{dr}^{\mathfrak R}$ are symmetric, 
(3)~$\mn{pp}^{\mathfrak R}$ is the
inverse of $\mn{ppi}^{\mathfrak R}$ and (4)~the rules of the $\mn{RCC5}$-composition
table (Figure~\ref{fig:comptable5}) are valid.
\psfull
\begin{figure}
  \begin{center}
\setlength{\tabcolsep}{1mm}
\begin{tabular}{|l||c|c|c|c|}
        \hline
        \multicolumn{1}{|c||}{$\circ$} &
        \mn{dr} & \mn{po} & \mn{pp} & \mn{ppi} \\
        \hline
        \hline
        \mn{dr} & $*$ & \mn{dr},\mn{po},\mn{pp} & \mn{dr},\mn{po},\mn{pp} & \mn{dr} \\ 
        \hline
        \mn{po} & \mn{dr},\mn{po},\mn{ppi}, & $*$ & \mn{po},\mn{pp} & \mn{dr},\mn{po},\mn{ppi} \\ 
        \hline
        \mn{pp} & \mn{dr} & \mn{dr},\mn{po},\mn{pp} & \mn{pp} & $*$  \\ 
        \hline
        \mn{ppi} & \mn{dr},\mn{po},\mn{ppi}, & \mn{po},\mn{pp} & \mn{eq},\mn{po},\mn{pp},\mn{ppi} & \mn{ppi} \\ 
        \hline
\end{tabular}
\normalsize
    \caption{The \mn{RCC5} composition table.}
    \label{fig:comptable5}
  \end{center}
\end{figure}
\psdraft

The following representation theorem is proved by first establishing Point~(ii) 
for finite \mn{RCC5}-structures and then applying the same technique
as in the proof of Theorem~\ref{represent}.
\\[3mm]
\begin{minipage}{\columnwidth}
\begin{theorem}\label{represent1}~\\[-4mm]
  \begin{itemize}
    
  \item [(i)] Every concrete $\mn{RCC5}$-structure is a ageneral
    $\mn{RCC5}$-structure;

   \item[(ii)] every general $\mn{RCC5}$-structure is isomorphic to a concrete $\mn{RCC5}$-structure.
     
   \item[(iii)] for every $n>0$, every countable general
     $\mn{RCC5}$-structure is isomorphic to a concrete
     $\mn{RCC5}$-structure of the form ${\mathfrak R}^5(\Rbbm^{n},
     U_{\Rbbm^{n}})$ (with $U_{\Rbbm^n} \subseteq \Rbbm^n_\mn{reg}$).
  \end{itemize}
\end{theorem}
\end{minipage}
As in the \mn{RCC8} case, we only distinguish between concrete and general
\mn{RCC5}-structures if necessary. \mn{RCC5}-models are defined in the obvious
way by extending $\mn{RCC5}$-structures with a valuation function.


The modal language ${\mathcal L}_{\mn{RCC5}}$ for reasoning about
$\mn{RCC5}$-structures extends propositional logic with unary modal
operators $[\mn{dr}]$, $[\mn{eq}]$, etc. (one for each \mn{RCC5}
relation).  A number of results from our investigation of ${\mathcal
  L}_{\mn{RCC8}}$ have obvious analogues for ${\mathcal L}_{\mn{RCC5}}$.


The results established in Section~\ref{sect:languages} have
counterparts in the \mn{RCC5} case: \mn{RCC5} constraint networks can
be translated into ${\mathcal L}_{\mn{RCC5}}$ in a straightforward way by
defining nominals. Moreover, ${\mathcal L}_{\mn{RCC5}}$ has the same
expressive power as the two-variable fragment of $\mathcal{FL}_{\mn{RCC5}}^{m}$, i.e.\ the first-order language with the five
binary $\mn{RCC5}$-relation symbols and infinitely many unary
predicates. Finally, the two-variable fragment of $\mathcal{FL}_{\mn{RCC5}}^{m}$ is exponentially more succinct on the class of
structures $\Rmc\Smc^5$ than $\Lmc_\mn{RCC5}$. The proofs are
analogous to those from Section~\ref{sect:languages} and
Appendix~\ref{exprcompl}.


Analogous to the \mn{RCC8} case, we define logics of full
\mn{RCC5}-structures, substructure variants, and finite substructure
variants: given a class ${\mathcal S}$ of $\mn{RCC5}$-structures, we
denote with $L_{\mn{RCC5}}({\mathcal S})$ the set of ${\mathcal
  L}_{\mn{RCC5}}$-formulas which are valid in all members of ${\mathcal
  S}$; with $L^\mn{S}_{\mn{RCC5}}({\mathcal S})$ the set of ${\mathcal
  L}_{\mn{RCC5}}$-formulas which are valid in all substructures of
members of ${\mathcal S}$; and with $L^\mn{fin}_{\mn{RCC5}}({\mathcal S})$
the set of ${\mathcal L}_{\mn{RCC5}}$-formulas which are valid in all
finite substructures of members of ${\mathcal S}$. For brevity, we refrain
from developing formulas that separate the different logics obtained
by applying ${\mathcal L}_{\mn{RCC5}}$ to different classes of
\mn{RCC5}-structures. Instead, we only note that there is an obvious
analogue of Theorem~\ref{equality1}.
\\[3mm]
\begin{minipage}{\columnwidth}
\begin{theorem} 
\label{equality2}
For $n>0$, we have
\begin{enumerate}
  
\item $L_{\mn{RCC5}}(\mathcal{RS}) = L^\Ssf_{\mn{RCC5}}(\mathcal{TOP}) =
  L_{\mn{RCC5}}^{\mn{S}}(\Rbbm^{n},\Rbbm^{n}_{\mn{reg}})$;
  
\item $L_{\mn{RCC5}}^{\mn{fin}}(\mathcal{RS})= L^\mn{fin}_{\mn{RCC5}}(\mathcal{TOP}) =
  L_{\mn{RCC5}}^{\mn{fin}}(\Rbbm^{n},\Rbbm^{n}_{\mn{reg}})$.

\end{enumerate}
\end{theorem}
\end{minipage}
We now investigate the computational properties of logics based on
${\mathcal L}_{\mn{RCC5}}$. Analogously to the \mn{RCC8} case, many
natural logics are undecidable. Still, our \mn{RCC5} undecidability
result is considerably less powerful than the one for \mn{RCC8}.
Intuitively, we have to restrict ourselves to $\mn{RCC5}$-structures
with the following property: for any set $S \subseteq W$ of
cardinality two or three, there exists a unique smallest region
$\mn{Sup}(S)$ that covers all regions from $S$.  Formally, we define
the class $\mathcal{RS}^{\exists}$ of \mn{RCC5}-structures $\auf
W,\mn{dr}^{\mathfrak R},\mn{eq}^{\mathfrak R},\dots \zu$ satisfying
the following condition: for every set $S \subseteq W$ of cardinality two or three, there
  exists a region $\mn{Sup}(S) \in W$ such that
\begin{itemize}
  
\item $s \mathrel{\mn{eq}} \mn{Sup}(S)$ or 
$s \mathrel{\mn{pp}} \mn{Sup}(S)$ for each $s \in S$;
  
\item for every region $t \in W$ with $s \;\mn{pp}\; t$ for each $s
\in S$, we have  $\mn{Sup}(S) \mathrel{\mn{eq}} t$ or
$\mn{Sup}(S) \;\mn{pp}\; t$;

\item for every region $t \in W$ with $t\;\mn{dr}\;s$ for each $s \in S$, we
have $t \;\mn{dr}\; \mn{Sup}(S)$.

\end{itemize}
Region structures based on {\em all} non-empty regular closed sets in a topological space
belong to $\mathcal{RS}^{\exists}$. This applies, in particular, to the structures
$\Rmf^5(\Rbbm^n,\Rbbm^n_\mn{reg})$, for $n\geq 1$. However, their substructures
usually do not belong to $\mathcal{RS}^{\exists}$. For example, the structures
$\Rmf^5(\Rbbm^n,\Rbbm^n_\mn{x})$ with $\mn{x} \in \{\mn{conv}, \mn{rect} \}$ and $n \geq 1$,
are not in $\mathcal{RS}^{\exists}$. Our aim is to prove the following theorem:
\begin{theorem}
\label{rcc5undec}
  Suppose ${\mathfrak R}^5(\Rbbm^{n},\Rbbm^{n}_{\mn{reg}})\in {\mathcal S}
  \subseteq \mathcal{RS}^{\exists}$, for some $n\geq 1$.  Then
  $L_{\mn{RCC5}}({\mathcal S})$ is undecidable.
\end{theorem}
This clearly yields the following corollary:
\begin{corollary}
  The following logics are undecidable, for each \mbox{$n\geq 1$:}
  $L_{\mn{RCC5}}(\mathcal{TOP})$ and $L_{\mn{RCC5}}
  (\Rbbm^{n},\Rbbm^{n}_{\mn{reg}})$.
\end{corollary}
The proof of Theorem~\ref{rcc5undec} is by reduction of the
satisfiability problem for the undecidable modal logic $\mn{S5}^{3}$
to satifiability of $\Lmc_\mn{RCC5}$ formulas in \Smc. The original
undecidability proof for $\mn{S5}^{3}$ has been given by Maddux in an
algebraic setting \cite{Maddux80}.  For the reduction, we use the
modal notation of \cite{Gabbayetal}. More precisely, the language
${\mathcal L}_{3}$ is the extension of propositional logic by means of
unary modal operators $\Diamond_{1}$, $\Diamond_{2}$
and~$\Diamond_{3}$. ${\mathcal L}_{3}$ is interpreted in
$\mn{S5}^{3}$-models
$$
{\mathfrak W} = \auf W_{1}\times W_{2} \times W_{3}, p_{1}^{\mathfrak W}, p_{2}^{\mathfrak W},\ldots\zu
$$
where the $W_{i}$ are non-empty sets and $p_{i}^{\mathfrak
  W}\subseteq W_{1}\times W_{2} \times W_{3}$.  The truth-relation
$\models$ between pairs $({\mathfrak W},(w_{1},w_{2},w_{3}))$ with
$w_{i}\in W_{i}$, and ${\mathcal L}_{3}$-formulas $\varphi$ is defined
inductively as follows:
\begin{itemize}
\item ${\mathfrak W},(w_{1},w_{2},w_{3})\models p_{i}$ iff $(w_{1},w_{2},w_{3})\in 
p_{i}^{\mathfrak W}$;
\item ${\mathfrak W},(w_{1},w_{2},w_{3}) \models \neg\varphi$
iff ${\mathfrak W},(w_{1},w_{2},w_{3}) \not\models \varphi$;
\item ${\mathfrak W},(w_{1},w_{2},w_{3}) \models \varphi_{1}\wedge \varphi_{2}$ iff
${\mathfrak W},(w_{1},w_{2},w_{3}) \models \varphi_{1}$ and 
${\mathfrak W},(w_{1},w_{2},w_{3}) \models \varphi_{2}$;
\item ${\mathfrak W},(w_{1},w_{2},w_{3}) \models \Diamond_{1}\varphi$
iff there exists $w_{1}'\in W_{1}$ such that ${\mathfrak W},(w_{1}',w_{2},w_{3}) \models 
\varphi$;
\item ${\mathfrak W},(w_{1},w_{2},w_{3}) \models \Diamond_{2}\varphi$
iff there exists $w_{2}'\in W_{2}$ such that ${\mathfrak W},(w_{1},w_{2}',w_{3}) \models 
\varphi$;
\item ${\mathfrak W},(w_{1},w_{2},w_{3}) \models \Diamond_{3}\varphi$
iff there exists $w_{3}'\in W_{3}$ such that ${\mathfrak W},(w_{1},w_{2},w_{3}') \models 
\varphi$.
\end{itemize}
A formula $\varphi\in {\mathcal L}_{3}$ is called
\emph{$\mn{S5}^{3}$-satisfiable} if there exists an
$\mn{S5}^{3}$-model ${\mathfrak W}$ and a triple $(w_{1},w_{2},w_{3})$
such that ${\mathfrak W},(w_{1},w_{2},w_{3})\models \varphi$.


Now for the reduction. The basic idea is to introduce three variables
$a_1,a_2,a_3$ and then to represent each set $W_i$ of an
$\mathsf{S5}^3$-model by the set of pairwise disconnected regions
$$
  \{ r \in W \mid \Mmf,r \models a_i \}.
$$
The set $W_1 \times W_2 \times W_3$ is then
represented by the set of regions
$$
\{\mn{Sup}(\{w_1,w_2,w_3\}) \mid \Mmf,w_i
\models a_i \text{ for } i \in \{1,2,3\} \}.
$$
The regions in this set will be marked with a variable $d$. To
simulate the modal operators of $\mathsf{S5}^3$, we will additionally
refer to regions $\mn{Sup}(\{w_i,w_j\})$ with $1\leq i < j \leq 3$.
Such regions are marked with the variable $d_{i,j}$.


The details of the reduction are as follows: with every
$\mn{S5}^{3}$-formula $\varphi$, we associate an ${\mathcal
  L}_{\mn{RCC5}}$-formula
\begin{equation*}
\Box_{u}\chi \wedge d \wedge \varphi^{\sharp}
\tag{$*$}
\end{equation*}
where $\vp^\sharp$ is inductively defined below and $\chi$ is the
conjunction of the following formulas:
\begin{enumerate} 
  
\item regions representing elements from $W_1 \cup W_2 \cup W_3$ are
  pairwise disconnected, each such region represents an element from
  $W_i$ for a unique $i$, and the sets $W_i$ are non-empty: for
  $i=1,2,3$, put
\begin{equation}\label{a1}
a_{i} \rightarrow \bigwedge_{j=1,2,3}([\mn{pp}]\neg a_{j} \wedge [\mn{ppi}]\neg a_{j}
\wedge [\mn{po}]\neg a_{j})
\end{equation}
\begin{equation}\label{a2}
a_{1} \rightarrow \neg a_{2}, \;\; a_{1} \rightarrow \neg a_{3}, \;\; a_{2}\rightarrow \neg a_{3},
\end{equation} \\[-4mm]
\begin{equation}\label{a3}
\bigwedge_{i=1,2,3}\Diamond_{u}a_{i}
\end{equation}
\item the variable $d$ identifies regions representing elements of $W_{1}\times W_{2} \times W_{3}$:
\begin{equation}\label{d}
\hspace*{-6mm} d \leftrightarrow(\bigwedge_{i=1,2,3}\auf \mn{ppi}\zu a_{i}) \wedge \neg\auf \mn{ppi}\zu 
(\bigwedge_{i=1,2,3}\auf \mn{ppi}\zu a_{i}) 
\end{equation}
\item $d_{i,j}$ identifies regions representing elements
  of $W_{i}\times W_{j}$: for $1\leq i < j \leq 3$, put
\begin{equation}\label{d12}
\hspace*{-6mm}d_{ij} \leftrightarrow (\bigwedge_{k=i,j}\auf \mn{ppi}\zu a_{k}) \wedge \neg\auf \mn{ppi}\zu 
(\bigwedge_{k=i,j}\auf \mn{ppi}\zu a_{k}).
\end{equation}
\end{enumerate}
Now, we define $\varphi^{\sharp}$ inductively by
$$
\begin{array}{rcl}
  p_{i}^\sharp &:=& p_{i} \\
  (\neg \vp)^\sharp &:=& d \wedge \neg \vp^\sharp \\
  (\vp \wedge \psi)^\sharp &:=& \vp^\sharp \wedge \psi^\sharp\\
  (\Diamond_1 \vp)^\sharp &:=& \auf \mn{ppi} \zu ( d_{23} \wedge \auf \mn{pp} \zu 
  (d \wedge \vp^\sharp)) \\
   (\Diamond_2 \vp)^\sharp &:=& \auf \mn{ppi} \zu ( d_{13} \wedge \auf \mn{pp} \zu 
  (d \wedge \vp^\sharp)) \\
   (\Diamond_3 \vp)^\sharp &:=& \auf \mn{ppi} \zu ( d_{12} \wedge \auf \mn{pp} \zu 
  (d \wedge \vp^\sharp)) 
\end{array}
$$
The following Lemma immediately yields Theorem~\ref{rcc5undec}.
\begin{lemma}
  \label{rcc5lem}
  Suppose ${\mathfrak R}(\Rbbm^{n},\Rbbm^{n}_{\mn{reg}})\in {\mathcal S}
  \subseteq \mathcal{RS}^{\exists}$, for some $n\geq 1$. 
  Then an $\mn{S5}^{3}$-formula $\varphi$ is satisfiable in an
  $\mn{S5}^{3}$-model iff $\Box_{u}\chi \wedge d \wedge
  \varphi^{\sharp}$ is satisfiable in \Smc.
\end{lemma}
\begin{Proof}
  ($\Leftarrow$) Suppose the region model 
$$
{\mathfrak M}=\auf {\mathfrak R},a_{1}^{\mathfrak M},a_{2}^{\mathfrak M},
a_{3}^{\mathfrak M},d^{\mathfrak M},
d_{12}^{\mathfrak M},\ldots,p_{1}^{\mathfrak M},\ldots\zu
$$
satisfies $\Box_{u}\chi \wedge d \wedge \varphi^{\sharp}$, where
${\mathfrak R}=\auf W,\mn{dr}^{\mathfrak R},\mn{eq}^{\mathfrak
  R},\ldots\zu\in \mathcal{RS}^{\exists}$.  Define
$$
{\mathfrak W}=\auf W_{1}\times W_{2}\times W_{3},p_{1}^{\mathfrak
  W},p_{2}^{\mathfrak W},\ldots\zu
$$
by setting
\begin{itemize}

\item $W_{i}= a_{i}^{\mathfrak M}$, for $i=1,2,3$;
  
\item for all $(w_{1},w_{2},w_{3})\in W_{1}\times W_{2}\times W_{3}$
  and $i<\omega$, \\[1mm]
  $(w_{1},w_{2},w_{3}) \in p_{i}^{\mathfrak W}$ iff
  $\mn{Sup}(\{w_{1},w_{2},w_{3}\}) \in p_{i}^{\mathfrak M}$.

\end{itemize}
By Formula (\ref{a3}), the $W_{i}$ are non-empty. Now, the function
$
f: W_{1}\times W_{2} \times W_{3} \rightarrow d^{\mathfrak M},
$
defined by putting
$$
f(w_{1},w_{2},w_{3})= \mn{Sup}\{w_{1}, w_{2}, w_{3}\},
$$
is a well-defined bijection: 
\begin{itemize}
  
\item $f$ is well-defined (i.e., $\mn{Sup}\{w_{1},w_{2},w_{3}\}\in
  d^{\mathfrak M}$) by the properties of $\mn{Sup}(S)$ and by Formula (\ref{d});
  
\item $f$ is injective since, by Formulas (\ref{a1}) and (\ref{a2}),
  we have $w_{1} \;\mn{dr}\; w_{2}$ for distinct $w_{1},w_{2} \in
  W_{1}\cup W_{2}\cup W_{3}$.  By the properties of $\mn{Sup}(S)$, we
  thus get $w\;\mn{dr} \;\mn{Sup}\{w_{1}, w_{2}, w_{3}\}$ for
  every $w\in W_{1}\cup W_{2}\cup W_{3}$ different from
  $w_{1},w_{2},w_{3}$;

\item By Formula (\ref{d}), $f$ is surjective.

\end{itemize} 
Using Formula (\ref{d12}), one can show in the same way that $f_{ij}:
W_{i}\times W_{j}\rightarrow d_{ij}^{\mathfrak M}$, $1\leq i < j\leq
3$, defined by
$$
f_{ij}(w_{i},w_{j})= \mn{Sup}\{w_{i},w_{j}\},
$$
are well-defined bijections. Moreover, for all $(w_{1},w_{2},w_{3})\in
W_{1}\times W_{2}\times W_{3}$ and $u\in W_{i}$, $v\in W_{j}$, $1 \leq
i < j \leq 3$, we obtain $\mn{Sup}\{u,v\} \; \mn{pp} \;
\mn{Sup}\{w_{1},w_{2},w_{3}\}$ iff $u=w_{i}$ and $v=w_{j}$.

Now it is straightforward to show by structural induction that, for
all subformulas $\psi$ of $\varphi$ and all $(w_{1},w_{2},w_{3})\in
W_{1}\times W_{2}\times W_{3}$, we have 
$$
{\mathfrak W},(w_{1},w_{2},w_{3})\models \psi 
\text{ iff } {\mathfrak M},f(w_{1},w_{2},w_{2})\models \psi^{\sharp}.
$$
Take $(w_{1},w_{2},w_{3})\in W_{1}\times W_{2}\times W_{3}$ such that
$f(w_{1},w_{2},w_{3}) \models \varphi^{\sharp}$. Then 
$(w_{1},w_{2},w_{3})\models \varphi$.

\noindent
($\Rightarrow$) By the standard translation of $\mathsf{S5}^3$ into
first-order logic and the theorem of L\"owenheim-Skolem, every
satisfiable $\mathsf{S5}^3$ formula $\vp$ is satisfiable in a
countable model
$$
{\mathfrak W}=\auf W_{1}\times W_{2}\times W_{3},p_{1}^{\mathfrak
  W},p_{2}^{\mathfrak W},\ldots\zu.
$$
We may assume w.l.o.g.\ that the sets $W_{i}$ are mutually disjoint.
Now let $n>0$ and define a model ${\mathfrak M}$ for $\Box_{u}\chi
\wedge d \wedge \varphi^{\sharp}$ based on the structure ${\mathfrak
  R}^5(\Rbbm^{n},\Rbbm_{\mn{reg}}^{n})$ as follows.  Let $f: W_{1}\cup
W_{2}\cup W_{3} \rightarrow \Rbbm_{\mn{reg}}^{n}$ be an injective
mapping such that $f(w) \mathrel{\mn{dr}} f(w')$ if $w \neq w'$, and
set
\begin{itemize}

\item $a_{i}^{\mathfrak M}= \{ f(w) \mid w\in W_{i}\}$, for $i=1,2,3$;
  
\item $d^{\mathfrak M} = \{ f(w_{1})\cup f(w_{2})\cup f(w_{3}) \mid
  (w_{1},w_{2},w_{3}) \in W_{1}\times W_{2} \times W_{3}\}$;

\item $d_{ij}^{\mathfrak M}= \{ f(w_{i})\cup f(w_{j}) \mid
  (w_{i},w_{j}) \in W_{i}\times W_{j}\}$, for $1 \leq i < j \leq 3$;

\item $p_{i}^{\mathfrak M}= \{ f(w_{1})\cup f(w_{2})\cup f(w_{3}) \mid
  (w_{1},w_{2},w_{3})\models p_{i}\}$ for $i<\omega$.

\end{itemize}
It is straightforward to prove that $\chi$ is true in every point of
${\mathfrak M}$.  Moreover, one can easily prove by induction that,
for every subformula $\psi$ of $\varphi$ and every
$(w_{1},w_{2},w_{3})\in W_{1}\times W_{2}\times W_{3}$, we have
$$
{\mathfrak
  W},(w_{1},w_{2},w_{3})\models \psi \text{ iff } {\mathfrak M},f(w_{1})\cup
f(w_{2})\cup f(w_{3}) \models \psi^{\sharp}.
$$
Since $\vp$ is satisfied in \Wmf, we thus obtain that $\Box_{u}\chi
\wedge d \wedge \varphi^{\sharp}$ is satisfied in ${\mathfrak M}$.
\end{Proof}
The decidability of other \mn{RCC5} logics is left as an open problem.
In particular, the decidability status of substructure logics and
their finite companions is one of the most intriguing open problems
suggested by the work presented in this paper.


Concering the recursive enumerability of logics based on
$\Lmc_\mn{RCC5}$, we only note that a counterpart of
Theorem~\ref{repos1} is easily obtained using an analogous proof:
\begin{theorem} 
\label{repos5}
For $n>0$, $L_{\mn{RCC5}}(\mathcal{RS}) = L^\Ssf_{\mn{RCC5}}(\mathcal{TOP})
= L_{\mn{RCC5}}^{\mn{S}}(\Rbbm^{n},\Rbbm^{n}_{\mn{reg}})$ are
recursively enumerable.
\end{theorem}
As already noted, the recursive enumerability of $\mn{RCC5}$ logics
determined by full concrete \mn{RCC5}-structures is left as an open
problem.

\section{Conclusion}
\label{sect:concl}

We first compare our results with Halpern and Shoham results for
interval temporal logic \cite{HalpernShoham91}.  Although one might be
tempted to conjecture that their undecidability proofs can be extended
to logics of region spaces, a close inspection shows that the only
spaces for which this might be possible are the 
logics of hyper-rectangles
$L_{\mn{RCC8}}(\Rbbm^{n},\Rbbm^{n}_{\mn{rect}})$. An extension is not
possible, however, for $L_{\mn{RCC8}}(\mathcal{TOP})$ and
$L_{\mn{RCC8}}(\Rbbm^{n},\Rbbm^{n}_{\mn{reg}})$, and not even for
$L_{\mn{RCC8}}^{\mn{S}}(\Rbbm^{n},\Rbbm^{n}_{\mn{rect}})$.  In fact,
the proof technique developed in this paper is more powerful than that
of \cite{HalpernShoham91}: Theorems~\ref{theo:undec}, \ref{nonax},
and~\ref{theo:undecfin} apply to logics induced by the region space
$\Rmf(\Rbbm,\Rbbm_\mn{conv})$, which is clearly an interval
structure.\footnote{Notice that Halpern and Shoham allow for intervals
  consisting of a single point while our intervals have to be regular
  closed sets and therefore non-singletons. However, as single point
  intervals are definable using the formula $[{\sf pp}]\bot$, all our
  negative results extend to interval structure with single point
  intervals.}  Interestingly, on this interval structure our results
are stronger than those of Halpern and Shoham in two respects: first,
we only need the \mn{RCC8} relations, which can be viewed as a
``coarsening'' of the Allen interval relations used by Halpern and
Shoham.  Second and more interestingly, by Theorem~\ref{theo:undec} we
have also proved undecidability of the \emph{substructure logic}
$L^\mn{S}_\mn{RCC8}(\Rbbm,\Rbbm_\mn{conv})$, which is a natural but
much weaker variant of the full (interval temporal) logic
$L_\mn{RCC8}(\Rbbm,\Rbbm_\mn{conv})$, and not captured by Halpern and
Shoham's undecidability proof.



Several open questions for future research remain.
Similar to the temporal case, the main challenge is to exhibit a
decidable and still useful variant of the logics proposed in this
paper. Perhaps the most interesting candidate is
$L_\mn{RCC5}(\Rmc\Smc)$, which coincides with the logics
$L^\mn{S}_\mn{RCC5}(\Rbbm^n,\Rbbm^n_\mn{reg})$, and to which the
reduction exhibited in Section~\ref{sect:rcc5} does not apply.  Other
candidates could be obtained by modifying the set of relations, e.g.\ 
giving up some of them. It has, for example, been argued that dropping
\mn{po} still results in a useful formalism for applications in geographic information systems.
An interesting step in this direction is \cite{ShapShe},
where a number of decidability and axiomatizability results are proved for modal logics over region
structures with only one modal operator corresponding to certain inclusion relations between regions.
Finally, it is an open problem whether
$L_\mn{RCC5}(\mathcal{RS})$ and
$L_\mn{RCC5}(\Rbbm^n,\Rbbm^n_\mn{reg})$ are recursively enumerable.
Although we believe that they are r.e.\ (in contrast to their \mn{RCC8}
counterparts), a proof is yet lacking.

\section*{Acknowledgement}
The authors wish to acknowledge helpful comments from two anonymous referees.



\newcommand{\etalchar}[1]{$^{#1}$}

\newpage

\appendix

\section{Proof of Representation Theorem}
\label{app:represent}

\noindent
{\bf Theorem~\ref{represent} (Representation theorem).}
\begin{itemize}

\item[(i)] Every concrete region structure is a general region structure;

\item[(ii)] every general region structure is isomorphic to a concrete region structure;
  
\item [(iii)] for every $n>0$, every countable general region
  structure is isomorphic to a concrete region structure of the form
  ${\mathfrak R}(\Rbbm^{n},U_{\Rbbm^{n}})$ (with
  $U_{\Rbbm^{n}}\subseteq \Rbbm^n_\mn{reg}$). 

\end{itemize}
The proof of this theorem refers to \mn{RCC8} constraint networks as
introduced in Section~\ref{sect:languages}, with the only difference
that, in the following, we also admit infinite such networks. For
convenience, we repeat the definition here. An \emph{RCC8 constraint
  network} is a set of constraints $(s \mathrel{\rsf} r)$ with $s,r$
\emph{region variables} and $\rsf$ an \mn{RCC8} relation.  Such a
network $N$ is \emph{satisfiable} in a topological space \Tmf with
regions $U_\Tmf$ if there exists an assignment $\delta$ of regions in
$U_\Tmf$ to region variables such that $(s \mathrel{\rsf} r) \in N$
implies $\delta(s) \mathrel{\rsf^\Tmf} \delta(r)$.
\\[2mm]
\begin{Proof}
  (i) Is easily proved by verifying the conditions formulated for
  general region models. This includes verification of the composition
  table, c.f.\ \cite{Cui-et-al-93}.

\noindent
(ii) is well-known for \emph{finite} general region stuctures,
see \cite{Bennett98constr}. Thus, it remains to extend the result to
infinite structures. We are going to prove this extension with the help of the
compactness theorem for first-order logic. To this end, we reduce
satisfiability of \mn{RCC8} constraint networks in topological spaces to satisfiability
in certain relational structures. Fix a general region structure ${\mathfrak R}
= \auf W,\mn{dc}^{\mathfrak R},\mn{ec}^{\mathfrak R},\ldots\zu$ with
$W$ infinite. An associated \mn{RCC8} constraint network, called the
\emph{diagram} of \Rmf and denoted with $\mn{diag}(\Rmf)$, is defined
by
$$
\mn{diag}({\mathfrak R})= \{ (s_{w} \mathrel{\mn{r}} s_{v}) \mid w,v
\in W \text{ and } {\mathfrak M} \models w \mathrel{\mn{r}} v\},
$$
where the $s_{w}$, $w\in W$, are region variables.
To prove (ii), it suffices to show that $\mn{diag}({\mathfrak R})$ is
satisfiable in some topological space \Tmf with a set $U_\Tmf$ of
non-empty regular closed regions: if this is the case, then
$$
\Rmf(\Tmf, \{\delta(s_w) \mid w \in W \})
$$
is a concrete region structure isomorphic to \Rmf, where $\delta$ is
the assignment witnessing satisfaction of $\mn{diag}(\Rmf)$ in
$(\Tmf,U_\Tmf)$.

Recall that every partial order $(V,R)$ induces a topological space $(V,{\mathbb I}_{R})$
by setting, for $X\subseteq V$,
$$
\mathbb{I}_{R}X = \{ x \in V \mid \forall y \;(xR y \rightarrow y\in
X)\}
$$
(and thus $\mathbb{C}_{R}X = \{ x \in V \mid \exists y \;(xR y
\wedge y\in X)\}$). We call $(V,{\mathbb I}_{R})$ the topological
space induced by $(V,R)$. Of particular interest for us are
topological spaces induced by partial orders that are {\em fork
  frames}: a partial order $(V,R)$ is a \emph{fork frame} if it is the
disjoint union of forks, where a \emph{fork} is a partial order
$(\{x_{b},x_{l},x_{r}\},S)$ such that $S$ is the reflexive closure of
$\{(x_{b},x_{l}),(x_{b},x_{r})\}$. 
For example,
Figure~\ref{fig:forkframe} contains an example fork frame whose
induced topological space satisfies the constraints $(r \; \mn{po} \;
s)$, $(s \; \mn{ec} \; t)$, and $(r \; \mn{dc} \; t)$ if $r$, $s$, and
$t$ are interpreted as regular-closed sets as indicated.
\psfull
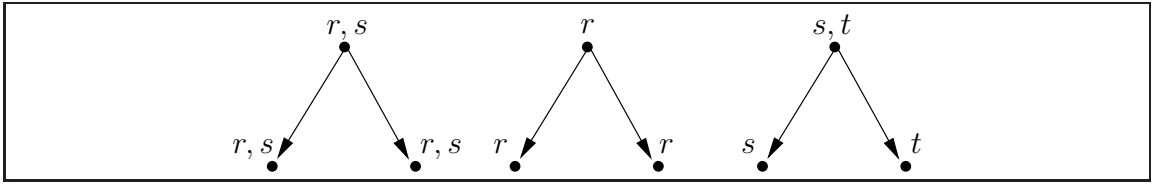
\begin{figure}
  \begin{center}
    \framebox[1\columnwidth]{\input{forkframe.pstex_t}}
    \caption{A fork frame satisfying $(r \; \mn{po} \; s)$, $(s \; \mn{ec} \; t)$, and $(r \; \mn{dc} \; t)$.}
    \label{fig:forkframe}
  \end{center}
\end{figure}
\psdraft
Denote by ${\mathfrak F}$ the class of all 
topological spaces based on fork frames.
It is shown in \cite{Bennett98constr,Renz2002} that every finite
constraint network which is satisfiable in a general region structure
is satisfiable in a topological space $\Tmf\in {\mathfrak F}$ with
regions $\Tmf_\mn{reg}$.  As $\mn{diag}(\Rmf)$ is trivially
satisfiable in the general region structure ${\mathfrak R}$, every
finite subset of $\mn{diag}(\Rmf)$ is satisfiable in a topological
space $\Tmf \in \Fmf$ with regions $\Tmf_\mn{reg}$.

Next, we give a translation of subsets $N$ of the \mn{RCC8} constraint network ${\sf diag}(\Rmf)$
to sets $\Gamma(N)$ of first-order sentences using a binary predicate
$R$ for the partial order in fork frames, and unary predicates
$(P_w)_{w \in W}$ for regions. The translation is such that, for all
$\Tmf \in \Fmf$ based on a fork frame $F=(V,S)$, the following conditions are
equivalent:
\begin{itemize}
\item an assignment $\delta$ witnesses satisfaction of $N$ in \Tmf with regions
$\Tmf_\mn{reg}$;
\item $\Gamma(N)$ is satisfied in the first-order
structure \Mmf with universe $V$ that is obtained by setting $R^\Mmf
:= S$ and $P_w^\Mmf := \delta(s_{w})$ for all region variables $s_{w}$ in $N$.
\end{itemize}
The translation introduces one sentence for each constraint in $N$. We
only treat the case $(s_{w} \mathrel{\mn{ec}} s_{v})$:
$$
\exists x (P_{w}(x) \wedge P_{v}(x)) \wedge \neg \exists x (P_{w}(x) \wedge \forall y (xRy \rightarrow P_{v}(y)))
\wedge  \neg \exists x (P_{v}(x) \wedge \forall y (xRy \rightarrow P_{w}(y))).
$$
The cases for other \mn{RCC8} relations are easily derived from their
semantics and the definition of the topological spaces in \Fmf.
Extend $\Gamma(N)$ to another set of first-order sentences
$\Gamma^*(N)$ by adding the following:
\begin{itemize}

\item ``$P_{w}$ is non-empty and regular closed'', for all $w\in W$:
$$
\exists x P_{w}(x) \wedge \forall x (P_{w}(x) \leftrightarrow \exists y (xRy \wedge \forall z (y R z \rightarrow P_{w}(z)))).
$$

\item ``$R$ is a disjoint union of forks'' (details are left to the
  reader).

\end{itemize}
Clearly, $\Gamma^*(N)$ is satisfiable in an arbitrary first-order
structure iff $\Gamma(N)$ is satisfied in a first-order structure \Mmf
obtained from a fork frame as described above iff $N$ is satisfiable
in a topological space $\Tmf \in \Fmf$ with regions $\Tmf_\mn{reg}$.

Thus, satisfiability of every finite subset of $\mn{diag}(\Rmf)$ in a
topological space $\Tmf \in \Fmf$ with regions $\Tmf_\mn{reg}$ yields that
every finite subset of $\Gamma^*(\mn{diag}(\Rmf))$ is satisfiable. By
compactness of first-order logic, $\Gamma^*(\mn{diag}(\Rmf))$ is also
satisfiable and thus $\mn{diag}(\Rmf)$ is satisfiable in a topological space
$\Tmf \in \Fmf$ with regions $\Tmf_\mn{reg}$.

\noindent
(iii) Suppose that ${\mathfrak R} = \auf W,\mn{dc}^{\mathfrak
  R},\mn{ec}^{\mathfrak R},\ldots\zu$ is at most countable. From the
encoding of constraint networks as sets of first-order sentences to be
interpreted in fork frames  and by L\"owenheim-Skolem, we
obtain that $\mn{diag}({\mathfrak R})$ is satisfiable in a topological
space $\Tmf \in \Fmf$ based on a fork frame $(V,S)$ with $V$
countable. Let $\delta$ be the assignment witnessing this
satisfaction. To satisfy $\mn{diag}({\mathfrak R})$ in $\Rbbm$ with
regions $\Rbbm_\mn{reg}$, assume that we have an enumeration
$(\{x_{b}^{i},x_{l}^{i},x_{r}^{i}\},S_{i})$, $i \in \Nbbm$, of the
forks of $(V,S)$. To define an assignment $\delta'$ in
$\Rbbm_{\mn{reg}}$, consider the sets
$$
W_{i} = \{ w \in W \mid \delta(s_{w}) \supseteq \{x_{b}^{i},x_{r}^{i},x_{l}^{i}\}\}
$$
and take mappings $g_{i}$ from $W_{i}$ into the open interval $(\frac{1}{4},\frac{1}{3})$ such that
\begin{enumerate}
\item $g_{i}(w)\leq g_{i}(v)$ if $\delta(s_{w}) \subseteq \delta(s_{v})$;
\item $g_{i}(w)\not=g_{i}(v)$ if $\delta(s_{w}) \not=\delta(s_{v})$.
\end{enumerate}
Such mappings exist because for each $S_{i}=\{ \delta(s_{w}) \mid w\in W_{i}\}$
the partial order $(S_{i},\subseteq)$ can be extended to a linear order which can then be
embedded into the open interval $(\frac{1}{4},\frac{1}{3})$.
Now set, for $w\in W$,
$$
\delta'(s_{w}) = \bigcup_{i\in \Nbbm, x_{r}^{i}\in \delta(w),x_{l}^{i}\not\in\delta(w)}\hspace*{-4mm}
[i,i+\frac{1}{4}] \cup \bigcup_{i\in \Nbbm, x_{l}^{i}\in \delta(w),x_{r}^{i}\not\in\delta(w)}\hspace*{-4mm}
[i-\frac{1}{4},i]
\cup
\bigcup_{i\in \Nbbm, x_{l}^{i},x_{r}^{i}\in\delta(w)}\hspace*{-4mm}
[i-g_{i}(w),i+g_{i}(w)].
$$
It is not hard to verify that each $\delta'(s_w)$ is non-empty and
regular closed since non-emptyness and regular closedness of
$\delta(s_w)$ implies that \mbox{$x^{i}_{b} \in \delta(s_w)$ iff $\{
x^{i}_{l}, x^{i}_{r} \} \cap \delta(s_w) \neq \emptyset$}. With the
exception of the $\mn{ntpp}$-case, we leave it to the reader to check
that the assignment $\delta'$ witnesses satisfaction of
$\mn{diag}({\mathfrak R})$ in $\Rbbm$ with regions $\Rbbm_\mn{reg}$.
For $\mn{ntpp}$, suppose that $(s_{w} \; \mn{ntpp} \; s_{v}) \in
\mn{diag}({\mathfrak R})$.  Then $\delta(s_{w})$ is in the relation
$\mn{ntpp}$ to $\delta(s_{v})$ in the topological space induced by
$(V,S)$. We show that $\delta'(s_{w})$ is in the relation $\mn{ntpp}$
to $\delta'(s_{v})$ in $\Rbbm$.  Clearly, by Condition 1 for the
functions $g_{i}$, $\delta'(s_{w})$ is a subset of
$\delta'(s_{v})$.
To show that $\delta'(s_{w})$ is included in the interior of $\delta'(s_{v})$ we show that 
$\delta'(s_{w}) \cap [i-\frac{1}{3},i+\frac{1}{3}]$ is included
in the interior of $\delta'(s_{v}) \cap [i-\frac{1}{3},i+\frac{1}{3}]$, for all $i\in \Nbbm$.
Let $i\in \Nbbm$. We distinguish four cases. 
\begin{itemize}

\item $\delta(s_{w}) \supseteq  \{x^{i}_{r},x^{i}_{l}\}$. Then $\delta(s_{v})\supseteq 
\{x^{i}_{r},x^{i}_{l}\}$ and therefore 
$$
\delta'(s_{u}) \cap [i-\frac{1}{3},i+\frac{1}{3}] = [i-g_{i}(u),i+g_{i}(u)],
$$
for $u=w,v$. By Conditions 1 and 2 on the functions $g_{i}$, 
$[i-g_{i}(w),i+g_{i}(w)]$ is included in the interior of $[i-g_{i}(v),i+g_{i}(v)]$.

\item $x^{i}_{l}\in \delta(s_{w})$ and $x^{i}_{r}\not \in \delta(s_{w})$.
Then $\delta(s_{v}) \supseteq \{x^{i}_{r},x^{i}_{l}\}$ (because otherwise $\delta(s_{w})$
would not be included in the interior of $\delta(s_{v})$). But then the claim follows from
the fact that $[i-\frac{1}{4},i]$ is in the interior of $[i-g_{i}(v),i+g_{i}(v)]$.

\item $x^{i}_{r}\in \delta(s_{w})$ and $x^{i}_{l}\not \in \delta(s_{w})$. Dual to the previous case.

\item $\delta(s_{w})\cap \{x^{i}_{r},x^{i}_{r}\}=\emptyset$. Then $\delta'(s_{w})\cap
[i-\frac{1}{3},i+\frac{1}{3}]= \emptyset$ and the claim follows.

\end{itemize}
Assignments witnessing satisfaction of $\mn{diag}({\mathfrak R})$ in
$\Rbbm^{n}$ with regions $\Rbbm^n_\mn{reg}$, $n>1$, can be constructed
similarly using hyper-rectangles.
\end{Proof}

\section{Expressivity and Succinctness}
\label{exprcompl}

\newcommand{\sigmax}{{\sigma_x}}
\newcommand{\sigmay}{{\sigma_y}}

The proof of the following theorem is an adaptation of the proof in
\cite{Etessami-et-al-02}, and a minor variant of the proof in
\cite{Lutz-et-al-01c} that is provided here for convenience.
Throughout this section, we use $2\mathcal{FO}^{m}_{\mn{RCC8}}$ to denote
the two-variable fragment of $\mathcal{FO}^{m}_{\mn{RCC8}}$ and assume
that its two variables are called $x$ and~$y$.

\noindent{\bf Theorem~\ref{ete}.}  For every
$2\mathcal{FO}^{m}_{\mn{RCC8}}$-formula $\varphi(x)$ with free
variable $x$, one can effectively construct a ${\mathcal
L}_{\mn{RCC8}}$-formula $\varphi^{\ast}$ of length at most exponential
in the length of $\varphi(x)$ such that, for every region model
${\mathfrak M}$ and region $s$, we have $ {\mathfrak M},s \models
\varphi^{\ast}$ iff ${\mathfrak M}\models \varphi[s]$.

\begin{Proof}
  A $2\mathcal{FO}^{m}_{\mn{RCC8}}$-formula $\xi$ is called a {\em unary
    atom} if it is of the form $\rsf(x,x)$, $\rsf(y,y)$, $p_{i}(x)$,
  or $p_{i}(y)$.  It is called a {\em binary atom} if it is of the
  form $\rsf(x,y)$, $\rsf(y,x)$, $x=y$, or $y=x$.
%
  W.l.o.g.\ we assume that $2\mathcal{FO}^{m}_{\mn{RCC8}}$-formulas are
  built using the operators $\exists$, $\wedge$, and $\neg$ only. We
  inductively define two mappings $\cdot^\sigmax$ and $\cdot^\sigmay$,
  the former taking each $2\mathcal{FO}^{m}_{\mn{RCC8}}$-formula
  $\varphi(x)$ with free variable $x$ to the corresponding ${\mathcal
    L}_{\mn{RCC8}}$-formula $\varphi^\sigmax$, and the latter
  doing the same for $2\mathcal{FO}^{m}_{\mn{RCC8}}$-formulas
  $\varphi(y)$ with free variable $y$.  We only give the details of
  $\cdot^\sigmax$ since $\cdot^\sigmay$ is defined analogously by
  switching the roles of $x$ and $y$:

  \noindent
  -- If $\varphi(x) = p_{i}(x)$, then put
  $(\varphi(x))^{\sigmax}=p_{i}$.

  \noindent
  -- If $\varphi(x)= \rsf(x,x)$, then put 
  $(\varphi(x))^{\sigmax}= \top$ if $\rsf=\mn{eq}$,
  and $(\varphi(x))^{\sigmax}= \bot$ otherwise.

  \noindent
  -- If $\varphi(x)=\chi_{1}\wedge \chi_{2}$,
  then put $(\varphi(x))^{\sigmax}=\chi_{1}^{\sigmax} \wedge
  \chi_{2}^{\sigmax}$.

  \noindent
  -- If $\varphi(x)=\neg \chi$, then put
  $(\varphi(x))^{\sigmax}=\neg(\chi^{\sigmax})$.

  \noindent
  -- If $\varphi(x) = \exists y \chi(x,y)$, then
  $\chi(x,y)$ can be written as
  $$
  \chi(x,y) = \gamma[\rho_{1},\ldots,\rho_{r},
  \gamma_{1}(x),\ldots,\gamma_{l}(x), \xi_{1}(y),\ldots,\xi_{s}(y)],
  $$

  \noindent i.e.\ as a Boolean combination $\gamma$ of $\rho_{i}$,
  $\gamma_{i}(x)$, and $\xi_{i}(y)$, where the $\rho_{i}$ are binary
  atoms, the $\gamma_{i}(x)$ are unary atoms or of the form $\exists y
  \gamma_{i}'$, and the $\xi_{i}(y)$ are unary atoms or of the form
  $\exists x \xi_{i}'$.  We may assume w.l.o.g.\ that $x$ occurs free
  in $\varphi(x)$.  Our first step is to move all formulas without a
  free variable $y$ out of the scope of $\exists$: obviously,
  $\varphi(x)$ is equivalent to
  \begin{equation*}
  \bigvee_{\auf w_{1},\ldots,w_{\ell}\zu \in
  \{\top,\bot\}^{\ell}}(\bigwedge_{1\leq i\leq \ell}
  (\gamma_{i}\leftrightarrow w_{i}) \wedge \exists y
  \gamma(\rho_{1},\ldots,\rho_{r},
  w_{1},\ldots,w_{l},\xi_{1},\ldots,\xi_{s})).
\tag{$*$}
  \end{equation*}
  Now we ``guess'' a relation \rsf that holds between $x$ and $y$, and then
  replace all binary atoms by either true or false according to the guess. 
  For \rsf an \mn{RCC8} relation and $1 \leq i \leq r$, let
  \begin{itemize}
    
  \item $\rho_{i}^{\rsf}=\top$ if $\rho_i=\rsf(x,y)$;

  \item $\rho_{i}^{\rsf}=\top$ if $\rho_i=\rsf(y,x)$ for $\rsf \in \{ \mn{dc},
    \mn{ec}, \mn{po} \}$;

  \item $\rho_{i}^{\rsf}=\top$ if $\rho_i=\mn{tpp}(y,x)$ and $\rsf=\mn{tppi}$
    or $\rho_i=\mn{ntpp}(y,x)$ and $\rsf=\mn{ntppi}$;

  \item $\rho_{i}^{\rsf}=\top$ if $\rho_i$ is $x=y$ and $\rsf=\mn{eq}$;

  \item $\rho_{i}^{\rsf}=\bot$ otherwise.

  \end{itemize}
  Using this notiation, ($*$) is equivalent to 
  \begin{eqnarray*}
  \begin{array}{l}
  \bigvee_{\auf w_{1},\ldots,w_{\ell}\zu \in
  \{\top,\bot\}^{\ell}}(\bigwedge_{1\leq i\leq
  \ell}(\gamma_{i}\leftrightarrow w_{i}) \wedge 
  {} \\[2mm]
  \hspace*{1cm} 
  \bigvee_{\rsf  \in \mn{RCC8}} \exists y (\rsf(x,y) \wedge
  \gamma(\rho_{1}^{\rsf},\ldots,\rho_{r}^{\rsf},
  w_{1},\ldots,w_{l},\xi_{1},\ldots,\xi_{s}))).
  \end{array}
  \end{eqnarray*}
  Now compute, recursively, $\gamma_{i}^{\sigmax}$ and
  $\xi_{i}^{\sigmay}$, and define $\varphi(x)^{\sigma}$ as
  \begin{eqnarray*}
    \begin{array}{l}
  \bigvee_{\auf w_{1},\ldots,w_{\ell}\zu
  \in \{\top,\bot\}^{\ell}}(\bigwedge_{1\leq i\leq \ell}
  (\gamma_{i}^{\sigmax}\leftrightarrow w_{i})
  \wedge 
   \\[2mm]
   \quad
  \bigvee_{\rsf\in \mn{RCC8}}
  \auf \rsf \zu \gamma(\rho_{1}^{\rsf},\ldots,\rho_{r}^{\rsf},
  w_{1},\ldots,w_{l},\xi_{1}^{\sigmay},\ldots,\xi_{s}^{\sigmay})).
    \end{array}
  \end{eqnarray*}
\end{Proof}

\noindent{\bf Theorem~\ref{theo:succinct}.}
  For $n \geq 1$, define a $\mathcal{FO}^{m}_{\mn{RCC8}}$ formula
  $$
  \varphi_n := \forall x \forall y \big ( \bigwedge_{i < n}
  (p_i(x) \leftrightarrow p_i(y)) \rightarrow (p_n(x) \leftrightarrow
  p_n(y)) \big )
  $$
  Then every $\Lmc_\mn{RCC8}$-formula $\psi_{n}$
  that is equivalent to $\varphi_{n}$ on 
  the class of all region structures
  $\Rmc\Smc$ has length $2^{\Omega(n)}$. 

\begin{Proof}
  Etessami et al.~\cite{Etessami-et-al-02} show that, on
  $\omega$-words, every temporal logic formula equivalent to
  $\varphi_n$ is of length at least~$2^{\Omega(n)}$, where temporal
  logic is assumed to have the operators ``next'', ``previously'',
  ``always in the future'' ($\Box^+ \vp$), and ``always in the
  past'' ($\Box^- \vp$). Assume, to the contrary of what is to be
  shown, that there is an $n \geq 1$ and an $\Lmc_\mn{RCC8}$ formula
  $\psi$ such that $\psi$ is equivalent to $\varphi_n$ on the class of
  structures $\Rmc\Smc$ and the length of $\psi$ is smaller than
  $2^{\Omega(n)}$.  Let $\Rmf=\auf W, \mathsf{dc}^\Rmf,
  \mathsf{ec}^\Rmf,\dots \zu \in \Rmc\Smc$ be such that
  $W=\{s_0,s_1,s_2,\dots\}$ and $s_i \mathrel{\mn{ntpp}^\Rmf} s_j$ if
  $j> i$.  Clearly, $\psi$ is equivalent to $\varphi_n$ on \Rmf.  We
  construct a new formula $\psi^*$ by exhaustively performing the
  following rewritings on (subformulas of) $\psi$:\footnote{Recall
    that $\auf r \zu \vartheta$ is only an abbreviation.}
  \begin{itemize}
    
  \item $[\rsf] \vartheta \rightsquigarrow \top$ if $\rsf \notin \{
    \mn{ntpp}, \mn{ntppi} \}$;

  \item $[\mn{eq}] \vartheta \rightsquigarrow \vartheta$.

  \end{itemize}
  The formula $\psi^*$ is equivalent to $\psi$ (and thus to
  $\varphi_n$) on \Rmf, it only refers to the relations \mn{ntpp} and
  \mn{ntppi}, and it may only be shorter, but not longer than $\psi$.
  We may now convert $\psi^*$ into a temporal logic formula $\psi^t$
  by substituting subformulas $[\mn{ntpp}]\vartheta$ with $\Box^+
  \vartheta$ and subformulas $[\mn{ntppi}]\vartheta$ with $\Box^-
  \vartheta$. It is not hard to see that $\psi^t$ is equivalent to
  $\vp_n$ on $\omega$-words. Thus, we have derived a contradiction to
  the fact that there is no such temporal logic formula of length
  smaller than $2^{\Omega(n)}$.
\end{Proof}

\section{Recursive Enumerability of $L_{\mn{RCC8}}^{\mn{S}}(\Rbbm^{n},\Rbbm^{n}_{\mn{rect}})$}
\label{app:rectanglesre}

\noindent
{\bf Theorem~\ref{repos2}.}
For $n \geq 1$,
$L_{\mn{RCC8}}^{\mn{S}}(\Rbbm^{n},\Rbbm^{n}_{\mn{rect}})$ is
recursively enumerable.

\begin{Proof}
  We show this result for $n=2$. For $n=1$ and $n>2$, the proof is
  similar and left to the reader.  Take the first-order language
  $\mathcal{FL}_{4}$ with one binary relation symbol $<$, infinitely many
  4-ary relation symbols $P_{1}, P_{2}, \ldots$, and one extra 4-ary
  relation symbol ${\sf exists}$.  Define a 4-ary predicate ${\sf
    rect}(x_{1},x_{2},x_{3},x_{4})$ by setting
$$
{\sf rect}(x_{1},x_{2},x_{3},x_{4})= (x_{1}<x_{2}) \wedge (x_{3}<x_{4}).
$$
Clearly, we can identify any vector
$\vec{a}=(a_{1},a_{2},a_{3},a_{4})\in \Rbbm^{2}$ such that $\Rbbm
\models {\sf rect}(\vec{a})$ with the rectangle
$$
[a_{1},a_{2}] \times [a_{3},a_{4}]\in \Rbbm^{2}_{\mn{rect}}.
$$
Moreover, it is easy (but tedious) to find, for every $\mn{RCC8}$
relation $\mn{r}$, a $\mathcal{FL}_{4}$ formula
$\varphi_{\mn{r}}(x_{1},\dots,x_{4},y_{1},\ldots y_{4})$ such that,
for any two rectangles $[a_{1},a_{2}] \times [a_{3},a_{4}]$ and
$[b_{1},b_{2}]\times [b_{3},b_{4}]$, we have
$$
[a_{1},a_{2}] \times [a_{3},a_{4}] \; \mn{r} \; [b_{1},b_{2}]\times [b_{3},b_{4}]
\mbox{ iff } \Rbbm^{2} \models \varphi_{\mn{r}}(\vec{a},\vec{b}).
$$
The details of working out these formulas are left to the reader.
Now fix variables $\vec{x}=x_{1},\ldots,x_{4}$ and $\vec{y} =
y_{1},\ldots,y_{4}$, and define a translation $s$ from ${\mathcal
  L}_{\mn{RCC8}}$ into $\mathcal{FL}_{4}$ by
\begin{eqnarray*}
p_{i}^{s} & = & {\sf rect}(\vec{x}) \wedge {\sf exists}(\vec{x}) \wedge P_{i}(\vec{x})\\
(\psi_{1}\wedge \psi_{2})^{s} & = & \psi_{1}^{s} \wedge \psi_{2}^{s}\\
(\neg \psi)^{s} & = & {\sf rect}(\vec{x}) \wedge {\sf exists}(\vec{x}) \wedge \neg \psi^{s}\\
(\auf r \zu \psi)^{s} & = & {\sf rect}(\vec{x}) \wedge {\sf exists}(\vec{x}) \wedge \exists \vec{y} ( 
\varphi_{\mn{r}}(\vec{x},\vec{y}) \wedge \psi^{s}(\vec{y}/\vec{x})).
\end{eqnarray*}
{\em Claim}. For every formula $\varphi\in {\mathcal L}_{\mn{RCC8}}$,
$\varphi$ is satisfiable in a substructure of ${\mathfrak
  R}(\Rbbm^{2},\Rbbm^{2}_{\mn{rect}})$ iff $\varphi^{s}$ is
satisfiable in a first-order model of the form ${\mathfrak Q}=
(\Qbbm,<,{\sf exists}^{\mathfrak Q}, P_{1}^{\mathfrak Q},
P_{2}^{\mathfrak Q},\ldots)$.
\\[2mm]
($\Rightarrow$) Suppose $\varphi$ is satisfied in a region model \Mmf
based on a substructure of ${\mathfrak
  R}(\Rbbm^{2},\Rbbm^{2}_{\mn{rect}})$.  Then $\varphi^{s}$ is
satisfiable in the first-order model
$$
{\mathfrak R}= (\Rbbm,<,{\sf exists}^{\mathfrak R},
P_{1}^{\mathfrak R}, P_{2}^{\mathfrak R},\ldots)
$$
in which ${\sf exists}$ is interpreted as the set of all rectangles
belonging to the domain of \Mmf and the $P_{i}$ are interpreted as the
set of rectangles in which $p_{i}$ is true in \Mmf.  By
L\"owenheim-Skolem, there exists a countably infinite elementary
substructure of \Rmf in which $\varphi^{s}$ is satisfied (see
\cite{enderton}).  Clearly, this structure is a dense linear order
without endpoints.  As every countable dense linear order without
endpoints is isomorphic to $(\Qbbm,<)$, this structure is of the form
required.

\noindent
($\Leftarrow$) Suppose $\varphi^{s}$ is satisfiable in ${\mathfrak Q}=
(\Qbbm,<,{\sf exists}^{\mathfrak Q}, P_{1}^{\mathfrak
  Q},P_{2}^{\mathfrak Q},\ldots)$.  Define a region model ${\mathfrak
  M}$ based on a substructure of ${\mathfrak
  R}(\Rbbm^{2},\Rbbm^{2}_{\sf rect})$ with domain $U$ and valuation
${\mathfrak V}$ as follows: let $U$ denote the set of rectangles of
the form $[a_{1},a_{2}]\times [a_{3},a_{4}]$ such that ${\mathfrak Q}
\models {\sf rect}(\vec{a}) \wedge {\sf exists}(\vec{a})$.  Let
${\mathfrak V}(p_{i})$ be the set of all rectangles $\vec{a}$ in $U$
such that ${\mathfrak Q}\models P_{i}(\vec{a})$. Then it is readily
checked that ${\mathfrak M}$ satisfies $\varphi$.

\noindent
This finishes the proof of the claim.  Now set $\varphi^{t}= \forall
\vec{x}({\sf rect}(\vec{x}) \wedge {\sf exists}(\vec{x}) \rightarrow
\varphi^{s})$, for every $\varphi\in {\mathcal L}_{\mn{RCC8}}$. Moreover,
let $\Gamma$ be the conjunction of the usual first-order axioms for
dense linear orders without endpoints (see e.g.\ \cite{enderton}).  It
follows from the claim above that $\varphi$ is valid in all
substructures of ${\mathfrak R}(\Rbbm^{2},\Rbbm^{2}_{\sf rect})$ iff
$\Gamma \rightarrow \varphi^{t}$ is a theorem of first-order logic.
Thus, recursive enumerability of
$L_{\mn{RCC8}}^{\mn{S}}(\Rbbm^{2},\Rbbm^{2}_{\mn{rect}})$ is obtained
from recursive enumerability of first-order logic.
\end{Proof}

\section{The Domino Problem for $k$-triangles}
\label{dominoundec}

Recall that, for $k \in \Nbbm$, the \emph{$k$-triangle} is the set
$
  \{(i,j) \mid i+j \leq k\} \subseteq \Nbbm^2.
$
We are going to prove the following undecidability result:
\begin{theorem}
\label{trangleundec}
  Given a domino system $\Dmc=(T,H,V)$, it is undecidable whether \Dmc
  tiles a $k$-triangle, $k \geq 1$, such that the position $(0,0)$ is
  occupied by a distinguished tile $s_0 \in T$ and some position is
  occupied by a distinguished tile $f_0 \in T$.
\end{theorem}
The proof is via a reduction of the halting problem for Turing
machines with a single right-infinite tape that are started on the
empty tape. The basic idea of the proof is to represent a run of the
Turing machine as a sequence of columns of a $k$-triangle, where each
column represents a configuration (with the left-most tape cell at the
bottom of the column).
Let \Amf be a single-tape right-infinite Turing machine with state
space $Q$, initial state $q_0$, halt state $q_f$, tape alphabet
$\Sigma$ ($b \in \Sigma$ stands for blank), and transition relation
$\Delta \subseteq Q \times \Sigma \times Q \times \Sigma \times
\{L,R\}$. W.l.o.g., we assume that Turing machines have the following
properties:
\begin{itemize}
  
\item the initial state $q_0$ is only used at the beginning of computations,
  but not later;

\item the TM comes to a stop only if it reaches $q_f$;
  
\item if the TM halts, its last step is to the right;

\item if the TM halts, then it labels the halting position with
  a special symbol $\# \in \Sigma$ before;

\item the blank symbol is never written.

\end{itemize}
It is easily checked that every TM can be modified to satisfy these
requirements. The configurations of \Amf will be represented by finite
words of one of the forms
\begin{enumerate}

\item $xb^m$,

\item $a_0\cdots a_kxya'_0 \cdots a'_\ell b^m$,

\item $a_0\cdots a_kyxa'_0 \cdots a'_\ell b^m$,

\end{enumerate}
where 
\begin{itemize}


\item $m >0$, 
  
\item all $a_i$ and $a'_i$ are in $\Sigma$, 
  
\item $x \in A := Q \times \Sigma \times \{L,R\}$ represents the
  active tape cell, its content, the current state, and the direction
  to which the TM has moved to reach the current position, and

\item 
$
  y \in A^\dagger:=
\{
\auf q,\sigma,M\zu^\dagger \mid \auf q,\sigma,M\zu \in A \}
$
represents the previously active tape cell, its current content, the
current state, and the direction to which \Amf moved to reach the
current position.

\end{itemize}
Note that the only difference between elements of $A$ and elements of
$A^\dagger$ is that the latter are marked with the symbol
``$\dagger$''.  Intuitively, the elements of $A$ describe the current
head position while the elements of $A^\dagger$ describe the previous
one. For technical reasons, the information whether the last step was
to the left or to the right is stored twice in each column: both in
the $x$ cell and in the $y$ cell. Configurations of Form~1 represent
the initial configuration and thus do not comprise the description of
a previous state.

Given a Turing machine \Amf, we define a domino system $\Dmc_\Amf =
(T,H,V,s_0,f_0)$ as follows:
\begin{itemize}

\item $T := \Sigma \cup A \cup A^\dagger \cup \{ \$ \}$;

\item $s_0 := \auf q_0,b,L \zu$;

\item $f_0 := \auf q_f,\#, R \zu$;

\item ~\\[-3mm] \hspace*{-2mm}$
  \begin{array}{r@{\;}c@{\;}l}
    H &:=&\{ (\sigma,\sigma) \mid \sigma \in \Sigma \} \; \cup \\[1mm]
    && \{ (\auf q,\sigma,M\zu,\auf q',\sigma',M'\zu^\dagger) \mid 
    (q,\sigma,q',\sigma',M') \in \Delta, M \in \{L,R\}  \} \; \cup \\[1mm]
    && \{ (\sigma,\auf q,\sigma,M\zu),(\auf q,\sigma,M\zu^\dagger,\auf q',\sigma,M'\zu) \mid 
    \sigma \in \Sigma, q,q' \in Q, M,M' \in \{L,R\} \} \; \cup \\[1mm]
    && \{ (\auf q,\sigma,M \zu^\dagger,\sigma) \mid q \in Q, \sigma \in \Sigma, M \in \{L,R\}  \}\; \cup \\[1mm]
    && \{(\auf q_f,\#,R\zu,\$),(\$,\$)\} \cup \{ (\sigma,\$) \mid \sigma  \in \Sigma \} \; \cup \\[1mm]
    && \{ (\auf q,\sigma,M \zu^\dagger,\$) \mid q \in Q,  M \in \{L,R\}  \}
  \end{array}
  $

\item ~\\[-3mm] \hspace*{-2mm}$
  \begin{array}{r@{\;}c@{\;}l}
    V &:=& \{ (\sigma,\sigma') \in \Sigma^2 \mid \sigma=b \mbox{ implies } \sigma'=b \} \; \cup \\[1mm]
    && \{ (\sigma,\auf q,\sigma',L\zu ),(\auf q,\sigma',R \zu ,\sigma) \mid
 \sigma,\sigma' \in \Sigma, q \in Q \} \; \cup \\[1mm]
    && \{ (\auf q,\sigma',L\zu^\dagger,\sigma), (\sigma,\auf q,\sigma',R\zu^\dagger) \mid 
    \sigma,\sigma' \in \Sigma, q \in Q \} \; \cup \\[1mm]
    && \{ (\auf q,\sigma,L\zu,\auf q,\sigma',L\zu^\dagger), (\auf q,\sigma',R\zu^\dagger,\auf q,\sigma,R\zu) \mid 
    \sigma,\sigma' \in \Sigma, q \in Q \} \; \cup \\[1mm]
    &&(\$,\$)\}
  \end{array}
  $

\end{itemize}
The tile ``$\$$'' is used for padding purposes: assume that there
exists a terminating computation of \Amc on the empty tape. Then this
computation induces in an obvious way the tiling of a finite
\emph{rectangle} such that $s_0$ is at position $(0,0)$, $f_0$ occurs
somewhere in the right-most column, and the height of the rectangle is
bounded by the width $w$ of the rectangle. We may now perform a
padding of the columns and rows in order to extend this rectangle to a
$2w$-triangle: for extending the height of columns, we may pad with the
blank symbol ``$b$'', and for extending the width of rows, we may pad
with the special symbol ``$\$$''. Since the existence of a tiling of a
$k$-rectangle with $s_0$ at position $(0,0)$ and $f_0$ occurring
somewhere induces a halting computation of \Amc in a straightforward
way, we obtain the following lemma.
\begin{lemma}
\label{trianglelem}
  The Turing machine \Amc halts on the empty tape iff the domino system
  $\Dmc_\Amf$ tiles a $k$-triangle, for some $k \geq 1$, such that 
  position $(0,0)$ is occupied by the tile $s_0$ and some position is 
  occupied by $f_0$.
\end{lemma}
Finally, Theorem~\ref{trangleundec} is an immediate consequence of 
Lemma~\ref{trianglelem}.

\end{document}

%% file: rcc8.pstex_t
\begin{picture}(0,0)%
\includegraphics{rcc8.pstex}%
\end{picture}%
\setlength{\unitlength}{4144sp}%
\begingroup\makeatletter\ifx\SetFigFont\undefined%
\gdef\SetFigFont#1#2#3#4#5{%
  \reset@font\fontsize{#1}{#2pt}%
  \fontfamily{#3}\fontseries{#4}\fontshape{#5}%
  \selectfont}%
\fi\endgroup%
\begin{picture}(3927,1879)(255,-1654)
\put(271,-1596){\makebox(0,0)[lb]{\smash{\SetFigFont{12}{14.4}{\familydefault}{\mddefault}{\updefault}$s \mathrel{\mn{po}} t$}}}
\put(1351,-1596){\makebox(0,0)[lb]{\smash{\SetFigFont{12}{14.4}{\familydefault}{\mddefault}{\updefault}$s \mathrel{\mn{eq}} t$}}}
\put(651,-1161){\makebox(0,0)[lb]{\smash{\SetFigFont{12}{14.4}{\familydefault}{\mddefault}{\updefault}$t$}}}
\put(1646,-1146){\makebox(0,0)[lb]{\smash{\SetFigFont{12}{14.4}{\familydefault}{\mddefault}{\updefault}$s$}}}
\put(3426,-1596){\makebox(0,0)[lb]{\smash{\SetFigFont{12}{14.4}{\familydefault}{\mddefault}{\updefault}$s \mathrel{\mn{ntppi}} t$}}}
\put(2386,-406){\makebox(0,0)[lb]{\smash{\SetFigFont{12}{14.4}{\familydefault}{\mddefault}{\updefault}$s \mathrel{\mn{tpp}} t$}}}
\put(3421,-406){\makebox(0,0)[lb]{\smash{\SetFigFont{12}{14.4}{\familydefault}{\mddefault}{\updefault}$s \mathrel{\mn{tppi}} t$}}}
\put(316,-406){\makebox(0,0)[lb]{\smash{\SetFigFont{12}{14.4}{\familydefault}{\mddefault}{\updefault}$s \mathrel{\mn{dc}} t$}}}
\put(1351,-406){\makebox(0,0)[lb]{\smash{\SetFigFont{12}{14.4}{\familydefault}{\mddefault}{\updefault}$s \mathrel{\mn{ec}} t$}}}
\put(361,-1161){\makebox(0,0)[lb]{\smash{\SetFigFont{12}{14.4}{\familydefault}{\mddefault}{\updefault}$s$}}}
\put(1531,-1146){\makebox(0,0)[lb]{\smash{\SetFigFont{12}{14.4}{\familydefault}{\mddefault}{\updefault}$t$}}}
\put(2536,-36){\makebox(0,0)[lb]{\smash{\SetFigFont{12}{14.4}{\familydefault}{\mddefault}{\updefault}$s$}}}
\put(3646,-31){\makebox(0,0)[lb]{\smash{\SetFigFont{12}{14.4}{\familydefault}{\mddefault}{\updefault}$s$}}}
\put(2671,-1136){\makebox(0,0)[lb]{\smash{\SetFigFont{12}{14.4}{\familydefault}{\mddefault}{\updefault}$s$}}}
\put(3761,-1146){\makebox(0,0)[lb]{\smash{\SetFigFont{12}{14.4}{\familydefault}{\mddefault}{\updefault}$s$}}}
\put(2406,-1591){\makebox(0,0)[lb]{\smash{\SetFigFont{12}{14.4}{\familydefault}{\mddefault}{\updefault}$s \mathrel{\mn{ntpp}} t$}}}
\put(361,-61){\makebox(0,0)[lb]{\smash{\SetFigFont{12}{14.4}{\familydefault}{\mddefault}{\updefault}$s$}}}
\put(801,-61){\makebox(0,0)[lb]{\smash{\SetFigFont{12}{14.4}{\familydefault}{\mddefault}{\updefault}$t$}}}
\put(1396,-61){\makebox(0,0)[lb]{\smash{\SetFigFont{12}{14.4}{\familydefault}{\mddefault}{\updefault}$s$}}}
\put(1741,-61){\makebox(0,0)[lb]{\smash{\SetFigFont{12}{14.4}{\familydefault}{\mddefault}{\updefault}$t$}}}
\put(3946,-16){\makebox(0,0)[lb]{\smash{\SetFigFont{12}{14.4}{\familydefault}{\mddefault}{\updefault}$t$}}}
\put(2866,-16){\makebox(0,0)[lb]{\smash{\SetFigFont{12}{14.4}{\familydefault}{\mddefault}{\updefault}$t$}}}
\put(2926,-1141){\makebox(0,0)[lb]{\smash{\SetFigFont{12}{14.4}{\familydefault}{\mddefault}{\updefault}$t$}}}
\put(4006,-1141){\makebox(0,0)[lb]{\smash{\SetFigFont{12}{14.4}{\familydefault}{\mddefault}{\updefault}$t$}}}
\end{picture}

%% file: dovetail.pstex_t
\begin{picture}(0,0)%
\includegraphics{dovetail.pstex}%
\end{picture}%
\setlength{\unitlength}{2960sp}%
\begingroup\makeatletter\ifx\SetFigFont\undefined%
\gdef\SetFigFont#1#2#3#4#5{%
  \reset@font\fontsize{#1}{#2pt}%
  \fontfamily{#3}\fontseries{#4}\fontshape{#5}%
  \selectfont}%
\fi\endgroup%
\begin{picture}(1835,1854)(559,-2803)
\put(826,-2611){\makebox(0,0)[lb]{\smash{\SetFigFont{9}{10.8}{\familydefault}{\mddefault}{\updefault}{\color[rgb]{0,0,0}1}%
}}}
\put(1276,-2611){\makebox(0,0)[lb]{\smash{\SetFigFont{9}{10.8}{\familydefault}{\mddefault}{\updefault}{\color[rgb]{0,0,0}2}%
}}}
\put(1726,-2611){\makebox(0,0)[lb]{\smash{\SetFigFont{9}{10.8}{\familydefault}{\mddefault}{\updefault}{\color[rgb]{0,0,0}4}%
}}}
\put(1276,-2161){\makebox(0,0)[lb]{\smash{\SetFigFont{9}{10.8}{\familydefault}{\mddefault}{\updefault}{\color[rgb]{0,0,0}5}%
}}}
\put(826,-2161){\makebox(0,0)[lb]{\smash{\SetFigFont{9}{10.8}{\familydefault}{\mddefault}{\updefault}{\color[rgb]{0,0,0}3}%
}}}
\put(826,-1711){\makebox(0,0)[lb]{\smash{\SetFigFont{9}{10.8}{\familydefault}{\mddefault}{\updefault}{\color[rgb]{0,0,0}6}%
}}}
\put(1193,-1269){\makebox(0,0)[lb]{\smash{\SetFigFont{9}{10.8}{\familydefault}{\mddefault}{\updefault}{\color[rgb]{0,0,0}$\iddots$}%
}}}
\put(1651,-1719){\makebox(0,0)[lb]{\smash{\SetFigFont{9}{10.8}{\familydefault}{\mddefault}{\updefault}{\color[rgb]{0,0,0}$\iddots$}%
}}}
\put(2101,-2161){\makebox(0,0)[lb]{\smash{\SetFigFont{9}{10.8}{\familydefault}{\mddefault}{\updefault}{\color[rgb]{0,0,0}$\iddots$}%
}}}
\put(750,-1254){\makebox(0,0)[lb]{\smash{\SetFigFont{9}{10.8}{\familydefault}{\mddefault}{\updefault}{\color[rgb]{0,0,0}$\iddots$}%
}}}
\put(1200,-1719){\makebox(0,0)[lb]{\smash{\SetFigFont{9}{10.8}{\familydefault}{\mddefault}{\updefault}{\color[rgb]{0,0,0}$\iddots$}%
}}}
\put(1650,-2162){\makebox(0,0)[lb]{\smash{\SetFigFont{9}{10.8}{\familydefault}{\mddefault}{\updefault}{\color[rgb]{0,0,0}$\iddots$}%
}}}
\put(2100,-2612){\makebox(0,0)[lb]{\smash{\SetFigFont{9}{10.8}{\familydefault}{\mddefault}{\updefault}{\color[rgb]{0,0,0}$\iddots$}%
}}}
\end{picture}

%% file: ordering.pstex_t
\begin{picture}(0,0)%
\includegraphics{ordering.pstex}%
\end{picture}%
\setlength{\unitlength}{2763sp}%
\begingroup\makeatletter\ifx\SetFigFont\undefined%
\gdef\SetFigFont#1#2#3#4#5{%
  \reset@font\fontsize{#1}{#2pt}%
  \fontfamily{#3}\fontseries{#4}\fontshape{#5}%
  \selectfont}%
\fi\endgroup%
\begin{picture}(8199,3249)(589,-3523)
\put(5326,-511){\makebox(0,0)[lb]{\smash{\SetFigFont{8}{9.6}{\familydefault}{\mddefault}{\updefault}{\color[rgb]{0,0,0}$a \wedge b$}%
}}}
\put(5926,-1261){\makebox(0,0)[lb]{\smash{\SetFigFont{8}{9.6}{\familydefault}{\mddefault}{\updefault}{\color[rgb]{0,0,0}$a \wedge b$}%
}}}
\put(5626,-886){\makebox(0,0)[lb]{\smash{\SetFigFont{8}{9.6}{\familydefault}{\mddefault}{\updefault}{\color[rgb]{0,0,0}$a \wedge b$}%
}}}
\put(6226,-1561){\makebox(0,0)[lb]{\smash{\SetFigFont{8}{9.6}{\familydefault}{\mddefault}{\updefault}{\color[rgb]{0,0,0}$a \wedge b$}%
}}}
\put(6901,-1936){\makebox(0,0)[lb]{\smash{\SetFigFont{8}{9.6}{\familydefault}{\mddefault}{\updefault}{\color[rgb]{0,0,0}$c$}%
}}}
\put(7201,-3436){\makebox(0,0)[lb]{\smash{\SetFigFont{8}{9.6}{\familydefault}{\mddefault}{\updefault}{\color[rgb]{0,0,0}$c$}%
}}}
\put(6826,-1561){\makebox(0,0)[lb]{\smash{\SetFigFont{8}{9.6}{\familydefault}{\mddefault}{\updefault}{\color[rgb]{0,0,0}$1$}%
}}}
\put(7051,-1216){\makebox(0,0)[lb]{\smash{\SetFigFont{8}{9.6}{\familydefault}{\mddefault}{\updefault}{\color[rgb]{0,0,0}$2$}%
}}}
\put(7801,-841){\makebox(0,0)[lb]{\smash{\SetFigFont{8}{9.6}{\familydefault}{\mddefault}{\updefault}{\color[rgb]{0,0,0}$3$}%
}}}
\put(8551,-481){\makebox(0,0)[lb]{\smash{\SetFigFont{8}{9.6}{\familydefault}{\mddefault}{\updefault}{\color[rgb]{0,0,0}$4$}%
}}}
\put(826,-511){\makebox(0,0)[lb]{\smash{\SetFigFont{8}{9.6}{\familydefault}{\mddefault}{\updefault}{\color[rgb]{0,0,0}$a \wedge b$}%
}}}
\put(751,-2986){\makebox(0,0)[lb]{\smash{\SetFigFont{8}{9.6}{\familydefault}{\mddefault}{\updefault}{\color[rgb]{0,0,0}$a \wedge \neg b$}%
}}}
\put(1426,-1261){\makebox(0,0)[lb]{\smash{\SetFigFont{8}{9.6}{\familydefault}{\mddefault}{\updefault}{\color[rgb]{0,0,0}$a \wedge b$}%
}}}
\put(1126,-2236){\makebox(0,0)[lb]{\smash{\SetFigFont{8}{9.6}{\familydefault}{\mddefault}{\updefault}{\color[rgb]{0,0,0}$a \wedge \neg b$}%
}}}
\put(1126,-886){\makebox(0,0)[lb]{\smash{\SetFigFont{8}{9.6}{\familydefault}{\mddefault}{\updefault}{\color[rgb]{0,0,0}$a \wedge b$}%
}}}
\put(2176,-1861){\makebox(0,0)[lb]{\smash{\SetFigFont{8}{9.6}{\familydefault}{\mddefault}{\updefault}{\color[rgb]{0,0,0}pos.\ 1}%
}}}
\put(2926,-2611){\makebox(0,0)[lb]{\smash{\SetFigFont{8}{9.6}{\familydefault}{\mddefault}{\updefault}{\color[rgb]{0,0,0}pos.\ 2}%
}}}
\put(3676,-3376){\makebox(0,0)[lb]{\smash{\SetFigFont{8}{9.6}{\familydefault}{\mddefault}{\updefault}{\color[rgb]{0,0,0}pos.\ 3}%
}}}
\end{picture}

%% file: forkframe.pstex_t
\begin{picture}(0,0)%
\includegraphics{forkframe.pstex}%
\end{picture}%
\setlength{\unitlength}{3947sp}%
\begingroup\makeatletter\ifx\SetFigFont\undefined%
\gdef\SetFigFont#1#2#3#4#5{%
  \reset@font\fontsize{#1}{#2pt}%
  \fontfamily{#3}\fontseries{#4}\fontshape{#5}%
  \selectfont}%
\fi\endgroup%
\begin{picture}(4267,1011)(1191,-1448)
\put(4386,-1326){\makebox(0,0)[lb]{\smash{\SetFigFont{12}{14.4}{\familydefault}{\mddefault}{\updefault}{\color[rgb]{0,0,0}$s$}%
}}}
\put(5451,-1326){\makebox(0,0)[lb]{\smash{\SetFigFont{12}{14.4}{\familydefault}{\mddefault}{\updefault}{\color[rgb]{0,0,0}$t$}%
}}}
\put(4831,-581){\makebox(0,0)[lb]{\smash{\SetFigFont{12}{14.4}{\familydefault}{\mddefault}{\updefault}{\color[rgb]{0,0,0}$s,t$}%
}}}
\put(2371,-1326){\makebox(0,0)[lb]{\smash{\SetFigFont{12}{14.4}{\familydefault}{\mddefault}{\updefault}{\color[rgb]{0,0,0}$r,s$}%
}}}
\put(1781,-581){\makebox(0,0)[lb]{\smash{\SetFigFont{12}{14.4}{\familydefault}{\mddefault}{\updefault}{\color[rgb]{0,0,0}$r,s$}%
}}}
\put(1191,-1326){\makebox(0,0)[lb]{\smash{\SetFigFont{12}{14.4}{\familydefault}{\mddefault}{\updefault}{\color[rgb]{0,0,0}$r,s$}%
}}}
\put(2831,-1326){\makebox(0,0)[lb]{\smash{\SetFigFont{12}{14.4}{\familydefault}{\mddefault}{\updefault}{\color[rgb]{0,0,0}$r$}%
}}}
\put(3871,-1326){\makebox(0,0)[lb]{\smash{\SetFigFont{12}{14.4}{\familydefault}{\mddefault}{\updefault}{\color[rgb]{0,0,0}$r$}%
}}}
\put(3376,-571){\makebox(0,0)[lb]{\smash{\SetFigFont{12}{14.4}{\familydefault}{\mddefault}{\updefault}{\color[rgb]{0,0,0}$r$}%
}}}
\end{picture}